\documentclass[twocolumn]{aastex63}
\usepackage{amsmath}
\usepackage{multirow}
\usepackage{graphicx}
\usepackage[caption=false]{subfig}
\usepackage{subfig}
\usepackage[figuresright]{rotating}
\DeclareCaptionFormat{cont}{#1 (cont.)#2#3\par}

\shorttitle{SOMA Survey III: From Intermediate- to High-Mass Protostars}
\shortauthors{Liu et al.}

\begin{document}

\title{The SOFIA Massive (SOMA) Star Formation Survey. III.\\
From Intermediate- to High-Mass Protostars}

\author{Mengyao Liu}
\affil{Dept. of Astronomy, University of Virginia, Charlottesville, Virginia 22904, USA}
\author{Jonathan C. Tan}
\affil{Dept. of Space, Earth \& Environment, Chalmers University of Technology, 412 93 Gothenburg, Sweden}
\affil{Dept. of Astronomy, University of Virginia, Charlottesville, Virginia 22904, USA}
\author{James M. De Buizer}
\affil{SOFIA-USRA, NASA Ames Research Center, MS 232-12, Moffett Field, CA 94035, USA}
\author{Yichen Zhang}
\affil{Star and Planet Formation Laboratory, RIKEN Cluster for Pioneering Research, Wako, Saitama 351-0198, Japan}
\author{Emily Moser}
\affil{Dept. of Astronomy, Cornell University, Ithaca, NY 14853, USA}
\author{Maria T. Beltr\'{a}n}
\affil{INAF-Osservatorio Astrofisico di Arcetri, Largo E. Fermi 5, I-50125 Firenze, Italy}
\author{Jan E. Staff} 
\affil{College of Science and Math, University of Virgin Islands, St. Thomas, United States Virgin Islands 00802}
\author{Kei E. I. Tanaka}
\affil{ALMA Project, National Astronomical Observatory of Japan, Mitaka, Tokyo 181-8588, Japan}
\author{Barbara Whitney}
\affil{Dept. of Astronomy, University of Wisconsin-Madison, 475 N. Charter St, Madison, WI 53706, USA}
\author{Viviana Rosero}
\affil{National Radio Astronomy Observatory, 1003 L{\'o}pezville Rd., Socorro, NM 87801, USA}
\author{Yao-Lun Yang}
\affil{Dept. of Astronomy, University of Virginia, Charlottesville, Virginia 22904, USA}
\author{Rub\'{e}n Fedriani}
\affil{Dept. of Space, Earth \& Environment, Chalmers University of Technology, 412 93 Gothenburg, Sweden}

\begin{abstract}
We present $\sim10-40\,\mu$m {\it SOFIA}-FORCAST images of 14
intermediate-mass protostar candidates as part of the {\it SOFIA} Massive (SOMA)
Star Formation Survey. We build spectral energy distributions (SEDs),
also utilizing archival {\it Spitzer}, {\it Herschel} and {\it IRAS}
data. We then fit the SEDs with radiative transfer (RT) models of
Zhang \& Tan (2018), based on Turbulent Core Accretion theory, to
estimate key protostellar properties. With the addition of these
intermediate-mass sources, SOMA protostars span luminosities from
$\sim10^{2}-10^{6}\:L_{\odot}$, current protostellar masses from
$\sim0.5-30\:M_{\odot}$ and ambient clump mass surface densities,
$\Sigma_{\rm cl}$ from $0.1-3\:{\rm{g\:cm}^{-2}}$. A wide range of
evolutionary states of the individual protostars and of the
protocluster environments are also probed.
We have also considered about 50 protostars
identified in Infrared Dark Clouds and expected to be at the earliest
stages of their evolution. With this global sample, most of the
evolutionary stages of high- and intermediate-mass protostars are
probed. From the best fitting models, there is no evidence of a
threshold value of protocluster clump mass surface density being
needed to form protostars up to $\sim25\:M_\odot$. However, to form
more massive protostars, there is tentative evidence that
$\Sigma_{\rm{cl}}$ needs to be $\gtrsim1\:{\rm{g\,cm}}^{-2}$. We
discuss how this is consistent with expectations from core accretion
models that include internal feedback from the forming massive star.
\end{abstract}

\keywords{ISM: jets and outflows --- dust --- stars: formation ---
  stars: winds, outflows --- infrared radiation --- ISM: individual
  (S235, IRAS~22198+6336, NGC~2071, Cepheus E, L1206, IRAS~22172+5549,
  IRAS~21391+5802)}

\section{Introduction}

\begin{deluxetable*}{ccccccccc}[t!]
%\tabletypesize{\scriptsize}
\tablecaption{{\it SOFIA} FORCAST Observations: Observation Dates \& Exposure Times (seconds)\label{tab:sofia_obs}}
\tablewidth{14pt}
\tablehead{
\colhead{Source} & \colhead{R.A.(J2000)} & \colhead{Decl.(J2000)} & \colhead{$d$ (kpc)} & \colhead{Obs. Date} & \colhead{7.7$\:{\rm \mu m}$} & \colhead{19.7$\:{\rm \mu m}$} & \colhead{31.5$\:{\rm \mu m}$} & \colhead{37.1$\:{\rm \mu m}$}
}
\startdata
S235 & 05$^h$40$^m$52$\fs$4 & $+$35$\arcdeg$41$\arcmin$30$\arcsec$ & 1.8 & 2016 Sep 20 & 404 & 779 & 642 & 1504  \\
IRAS\,22198+6336 & 22$^h$21$^m$26$\fs$68 & $+$63$\arcdeg$51$\arcmin$38$\farcs$2 & 0.764 & 2015 Jun 05 & 278 & 701 & 482 & 743 \\
NGC\,2071 & 05$^h$47$^m$04$\fs$741 & $+$00$\arcdeg$21$\arcmin$42$\farcs$96 & 0.39 & 2018 Sep 08 & 492 & 1319 & 825 & 2020 \\
Cepheus E & 23$^h$03$^m$12$\fs$8 & $+$61$\arcdeg$42$\arcmin$26$\arcsec$ & 0.73 & 2015 Nov 04 & 281 & 899 & 818 & 281 \\
L1206 & 22$^h$28$^m$51$\fs$41 & $+$64$\arcdeg$13$\arcmin$41$\farcs$1 & 0.776 & 2015 Nov 20 & 116 & 308 & 162 & 630 \\
IRAS\,22172+5549 & 22$^h$19$^m$09$\fs$478 & $+$56$\arcdeg$05$\arcmin$00$\farcs$370 & 2.4 & 2015 Jun 03 & 337 & 664 & 386 & 466 \\
IRAS\,21391+5802 & 21$^h$40$^m$41$\fs$90 & $+$58$\arcdeg$16$\arcmin$12$\farcs$3 & 0.75 & 2015 Nov 06 & 334 & 806 & 488 & 1512 \\
\enddata
\tablecomments{
The source positions listed here are the same as the positions of the
black crosses denoting the radio continuum peak (mm continuum peak in
Cep E and L1206 A, and MIR peak in IRAS22172 MIR2) in each source in Figures 1-7.
%jct5 - this sentence is unclear, so commenting out... maybe it can be added back in refereeing, if it can be explained better
%The ordering of the sources from top to bottom is based on
%the number of sources analyzed in that region (from single to multiple) and
%their isotropic luminosity estimate (from high to low, presented later).
Source distances are from the literature, as discussed below.}
%jct3 - this comment about the ordering is confusing... we should just order by luminosity.
\end{deluxetable*}

Intermediate-mass (IM) protostars are important as representatives of
the transition between the extremes of low- (i.e., $\lesssim
2\:M_\odot$) and high- (i.e., $\gtrsim 8\:M_\odot$) mass star
formation. These objects are relatively rare compared to their
low-mass counterparts and tend to be located at greater
distances. They are precursors of Herbig Ae and Be stars. The
immediate environments of IM protostars can appear quite complex,
with extended emission often resolved into multiple sources when
observed at high resolution (e.g., G173.58+2.45, Shepherd \& Watson
2002). However, there are also examples with relatively simpler, more
isolated morphologies (e.g., Cep E, Moro-Mart\'{i}n et al. 2001).
%As expected, molecular outflows driven by IM protostars are more
%energetic than those from low-mass sources.
Observations of IM protostars indicate that they share some similar
physical properties as low-mass protostars, such as circumstellar
disks (e.g., Zapata et al. 2007; S{\'a}nchez-Monge et al. 2010; van
Kempen et al. 2012; Takahashi et al. 2012) and collimated molecular
outflows (e.g., Gueth et al. 2001; Beltr\'an et al. 2008, 2009; Palau
et al. 2010; Velusamy et al. 2011), but with the latter being more
powerful when driven from IM protostars. Furthermore, IM protostars
%young stellar objects (YSOs)
also share many characteristics with their higher-mass counterparts,
such as correlations between the outflow kinematics and the properties
of their driving sources (e.g., Cabrit \& Bertout 1992; Bontemps et
al. 1996; Wu et al. 2004; Hatchell et al. 2007; Beltr\'an et
al. 2008), and hot core chemistry (e.g., Fuente et al. 2005; Neri et
al. 2007; S\'{a}nchez-Monge et al. 2010). Thus, the observational
evidence suggests that intermediate-mass protostars form in a similar
way as low-mass protostars, and that this formation mechanism is also
shared with at least early B-type or late O-type protostars
(Beltr\'{a}n 2015).

In this paper, we study a sample of 14 IM protostars selected from the
{\it SOFIA} Massive (SOMA) Star Formation Survey (PI: Tan), which aims
to characterize a sample of $\ga$ 50 high- and intermediate-mass
protostars over a range of evolutionary stages and environments with
their $\sim$ 10 to 40\,$\mu$m images observed with the
\textit{SOFIA}-Faint Object infraRed CAmera for the \textit{SOFIA}
Telescope (FORCAST) instrument. In Paper I of the survey (De Buizer et
al. 2017), the first eight sources were presented, which were mostly
massive protostars. In Paper II (Liu et al. 2019), seven especially
luminous sources were presented, corresponding to some of the most
massive protostars in the survey. Thus the IM sample presented here,
which consists of 7 new target regions from which 12 protostars have
been studied plus 2 more protostars extracted as secondary sources
from Papers I and II target regions, serves to extend the luminosity
and mass range of the survey sample down to lower values.

Our approach is to follow the same methods developed in Papers I and
II to build the spectral energy distributions (SEDs) of the
sources. As before, we then fit these SEDs with the Zhang \& Tan
(2018, hereafter ZT18) protostellar radiative transfer (RT) models to
estimate intrinsic source properties. In this way, all the protostars
are analyzed in an uniform way. Finally, we search for trends in
source properties among the overall SOMA sample of 29 sources that
have been so far analyzed in Papers I, II and III.

The observations and data utilized in this paper are described in \S2.
The analysis methods are summarized in \S3. We present the MIR imaging
and SED fitting results in \S4 and discuss these results and their
implications in \S5. A summary is given in \S6.

\section{Observations}

The following seven target regions
%listed in order of decreasing isotropic bolometric luminosity of their primary target, 
were observed by \textit{SOFIA}\footnote{\textit{SOFIA} is jointly operated by the
  Universities Space Research Association, Inc. (USRA), under NASA
  contract NAS2-97001, and the Deutsches SOFIA Institute (DSI) under
  DLR contract 50 OK 0901 to the University of Stuttgart.}  (Young et
al. 2012) with the FORCAST instrument (Herter et al. 2013) (see
Table~\ref{tab:sofia_obs}): S235, IRAS~22198+6336, NGC~2071, Cep E,
L1206 (A and B), IRAS22172+5549 (MIR 1, MIR 2, and MIR 3),
IRAS~21391+5802 (BIMA 2, BIMA 3, and MIR 48). The angular resolutions
of the {\it SOFIA}-FORCAST images are 2.7\arcsec\ at 7~$\rm \mu m$,
2.9\arcsec\ at 11~$\rm \mu m$, 3.3\arcsec\ at 19~$\rm \mu m$,
3.4\arcsec\ at 31~$\rm \mu m$, and 3.5\arcsec\ at 37~$\rm \mu m$. We
also fit the SEDs of two more sources G305.20+0.21 A (hereafter, G305
A) and IRAS 16562-3959 N (hereafter, IRAS 16562 N), which are
secondary sources near primary targets of Paper II. Thus a total of 14
protostars will be analyzed here for the first time as SOMA Survey
sources.

In addition to \textit{SOFIA} observations, for all objects, we also
retrieve publicly-available images of \textit{Spitzer}/IRAC (Fazio et
al. 2004) at 3.6, 4.5, 5.8 and 8.0\,$\mu$m from the \textit{Spitzer}
Heritage Archive, \textit{Herschel}/PACS and SPIRE (Griffin et
al. 2010) at 70, 160, 250, 350 and 500\,$\mu$m from the
\textit{Herschel} Science Archive, and Higher Resolution IRAS Images
({\it
  HIRES})\footnote{https://irsa.ipac.caltech.edu/applications/Hires/}
(Neugebauer et al. 1984) at 60, 100\,$\mu$m from the NASA/IPAC
Infrared Science Archive.

The calibration and astrometry methods are the same as those of Paper
II, except that for Cep E and IRAS 21391 we use the \textit{SOFIA}
19\,$\mu$m image instead of 7\,$\mu$m to calibrate the other
\textit{SOFIA} images and the \textit{Herschel} images given the high
noise level in their 7\,$\mu$m images. For \textit{SOFIA} observations
the calibration error is estimated to be in the range $\sim$ 3\% -
7\%. The astrometric precision is about 0.1\arcsec \ for the
\textit{SOFIA} 7\,$\mu$m image, 0.4\arcsec\ for longer wavelength
\textit{SOFIA} images, and 1$\arcsec$\ for \textit{Herschel}
images. Note that we use HIRES results of the \textit{IRAS} data to
achieve a resolution $\sim$ 1\arcmin. The astrometric precision is
about 20 - 30\arcsec. Fluxes measured from HIRES agree with those of
the Point Source Catalog (PSC2) to within 20\% and ringing (a ring of lower
level flux may appear around a point source) can contribute up to
another 10\% uncertainty in the measurement of the background
subtracted flux of the source. Thus the total uncertainty, summing in
quadrature, is 23\%. Near-Infrared (NIR) images from the Wide Field
Camera (WFC)/ UKIRT InfraRed Deep Sky Survey (UKIDSS) (Lawrence et
al. 2007) surveys and the Two Micron All Sky Survey (2MASS) Atlas
images (Skrutskie et al. 2006) are also used to investigate the
environments of the protostellar sources and look for association with
the MIR counterparts.

\section{Methods}

We follow the methods described in Papers I and II to construct the
SEDs (see \S3 of Papers I and II for more detailed discussion). In
summary, fixed circular aperture, background-subtracted photometry is
estimated from MIR to FIR wavelengths for the sources. The aperture
radius is chosen with reference to the 70~$\rm \mu m$ {\it
  Herschel}-PACS source morphology, when available (else the 37~$\rm
\mu m$ {\it SOFIA}-FORCAST source morphology), with the goal of
enclosing the majority of the flux, while avoiding contamination from
surrounding sources.

We also follow the methods of Papers I and II to fit the SEDs with
ZT18 protostellar radiative transfer models. For IRAS~22198, NGC~2071,
Cep~E, G305~A, IRAS~16562~N, which have {\it Herschel} data, we do not
use {\it IRAS} data for the SED fitting. For L1206, our \textit{SOFIA}
images
%(see Figure~\ref{fig:L1206})
show that L1206 A is much brighter than L1206 B at long wavelengths:
e.g., at 37 $\mu$m L1206 A contributes 96\% of the total flux. Thus we
assume L1206 A is the main source at wavelengths longer than 37 $\mu$m
and use the IRAS flux densities at 60 $\mu$m and 100 $\mu$m as a
normal data point for the SED fitting of L1206 A and upper limits for
the SED fitting of L1206 B. For the other sources, IRAS data are used
as upper limits given its resolution and aperture size.

%We set the flux uncertainty to 23\% for the SED fitting of IRAS data.

There are a few special cases for the SED fitting. For G305 A, at
wavelengths shorter than 8 $\mu$m there is hardly any emission and the
local noise leads to a negative flux measurement at 7 $\mu$m. Thus we
use the non-background subtracted fluxes as upper limits at 3.6, 4.5,
5.8 and 8.0 $\mu$m. In the IRAS 16562 region, the flux densities at
wavelengths longer than 250 $\mu$m are dominated by the main source in
Paper II, thus the background subtracted flux for IRAS 16562 N is
negative at these wavelengths because of the contamination of the main
source. Thus we use the non-background subtracted fluxes as upper
limits at 250, 350 and 500 $\mu$m.

\section{Results}

Table~\ref{tab:flux} lists the types of multi-wavelength data
available for each source, the flux densities derived, and the
aperture sizes adopted. $F_{\lambda, \rm fix}$ is the flux density
derived with a fixed aperture size and $F_{\lambda, \rm var}$ is the
flux density derived with a variable aperture size. The value of flux
density listed in the upper row of each source is derived with
background subtraction, while that derived without background
subtraction is listed in parentheses in the lower row. The
\textit{SOFIA} images for each source are presented in \S4.1. General
results of the \textit{SOFIA} imaging are summarized in \S4.2. The
SEDs and fitting results are presented in \S4.3.

%\clearpage
\begin{rotatetable*}% Landscape page
\centering
\setlength{\tabcolsep}{1pt}
%\renewcommand{\arraystretch}{0.8}
%\vspace{-3in}
\begin{deluxetable*}{cc|ccc|ccc|ccc|ccc|ccc|ccc}
%\tablenum{2}
\tabletypesize{\scriptsize}
\tablecaption{Integrated Flux Densities\label{tab:flux}}
\tablewidth{20pt}
\tablehead{
\colhead{Facility} &\colhead{$\lambda$} &\colhead{F$_{\lambda,\rm fix}$}\tablenotemark{a} &\colhead{F$_{\lambda,\rm var}$} \tablenotemark{b} & \colhead{R$_{\rm ap}$} \tablenotemark{c} &\colhead{F$_{\lambda,\rm fix}$} &\colhead{F$_{\lambda,\rm var}$} & \colhead{R$_{\rm ap}$}&\colhead{F$_{\lambda,\rm fix}$} &\colhead{F$_{\lambda,\rm var}$} & \colhead{R$_{\rm ap}$}&\colhead{F$_{\lambda,\rm fix}$} &\colhead{F$_{\lambda,\rm var}$} & \colhead{R$_{\rm ap}$}&\colhead{F$_{\lambda,\rm fix}$} &\colhead{F$_{\lambda,\rm var}$} & \colhead{R$_{\rm ap}$}&\colhead{F$_{\lambda,\rm fix}$} &\colhead{F$_{\lambda,\rm var}$} & \colhead{R$_{\rm ap}$} \\
\colhead{} &\colhead{($\mu$m)} &\colhead{(Jy)} &\colhead{(Jy)} & \colhead{($\arcsec$)} &\colhead{(Jy)} &\colhead{(Jy)} & \colhead{($\arcsec$)} &\colhead{(Jy)} &\colhead{(Jy)} & \colhead{($\arcsec$)} &\colhead{(Jy)} &\colhead{(Jy)} & \colhead{($\arcsec$)} &\colhead{(Jy)} &\colhead{(Jy)} & \colhead{($\arcsec$)} &\colhead{(Jy)} &\colhead{(Jy)} & \colhead{($\arcsec$)} 
}
\startdata
 &  & \multicolumn{3}{c}{S235} &  \multicolumn{3}{c}{IRAS 22198} &  \multicolumn{3}{c}{NGC2071} &  \multicolumn{3}{c}{Cep E}  &  \multicolumn{3}{c}{L1206 A}  &  \multicolumn{3}{c}{L1206 B} \\
\hline
\multirow{2}{*}{{\it Spitzer}/IRAC} & \multirow{2}{*}{3.6} & 0.50 & 0.48 & \multirow{2}{*}{9.0} & 0.05 & 0.01 & \multirow{2}{*}{6.6} & 0.34 & 0.12 & \multirow{2}{*}{3.6} & 0.05 & 0.06 & \multirow{2}{*}{28.0} & 0.11 & 0.13 & \multirow{2}{*}{12.0} & \multirow{2}{*}{...} & \multirow{2}{*}{...} & \multirow{2}{*}{...} \\
&   & (0.54) & (0.51) &   & (0.08) & (0.01) &   & (0.38) & (0.14) &   & (0.06) & (0.08) &   & (0.12) & (0.15) &   &   &   &   \\
\hline
\multirow{2}{*}{{\it Spitzer}/IRAC} & \multirow{2}{*}{4.5} & 0.46 & 0.44 & \multirow{2}{*}{9.0} & 0.13 & 0.03 & \multirow{2}{*}{4.8} & 1.24 & 0.54 & \multirow{2}{*}{3.6} & 0.17 & 0.24 & \multirow{2}{*}{28.0} & 0.25 & 0.30 & \multirow{2}{*}{12.0} & \multirow{2}{*}{...} & \multirow{2}{*}{...} & \multirow{2}{*}{...} \\
&   & (0.51) & (0.47) &   & (0.15) & (0.04) &   & (1.32) & (0.63) &   & (0.18) & (0.25) &   & (0.27) & (0.33) &   &   &   &   \\
\hline
\multirow{2}{*}{{\it Spitzer}/IRAC} & \multirow{2}{*}{5.8} & 1.99 & 1.90 & \multirow{2}{*}{9.0} & 0.20 & 0.08 & \multirow{2}{*}{5.4} & 2.54 & 1.59 & \multirow{2}{*}{3.6} & 0.23 & 0.27 & \multirow{2}{*}{28.0} & 0.28 & 0.33 & \multirow{2}{*}{12.0} & 2.10 & 2.10 & \multirow{2}{*}{10.0} \\
&   & (2.24) & (2.06) &   & (0.43) & (0.10) &   & (2.78) & (1.71) &   & (0.31) & (0.38) &   & (0.33) & (0.41) &   & (2.19) & (2.19) &   \\
\hline
\multirow{2}{*}{{\it SOFIA}/FORCAST} & \multirow{2}{*}{7.7} & 6.39 & 6.22 & \multirow{2}{*}{9.0} & 0.85 & 0.20 & \multirow{2}{*}{5.4} & 5.58 & 4.04 & \multirow{2}{*}{3.8} & 0.17 & 0.23 & \multirow{2}{*}{6.0} & 0.12 & 0.30 & \multirow{2}{*}{4.6} & 2.21 & 1.53 & \multirow{2}{*}{5.0} \\
&   & (6.24) & (6.13) &   & (1.41) & (0.29) &   & (5.53) & (4.32) &   & (0.19) & (0.20) &   & (0.19) & (0.27) &   & (2.04) & (1.65) &   \\
\hline
\multirow{2}{*}{{\it Spitzer}/IRAC} & \multirow{2}{*}{8.0} & 6.12 & 5.83 & \multirow{2}{*}{9.0} & 0.23 & 0.15 & \multirow{2}{*}{6.6} & 6.08 & 4.09 & \multirow{2}{*}{3.8} & 0.31 & 0.34 & \multirow{2}{*}{28.0} & 0.21 & 0.24 & \multirow{2}{*}{12.0} & 3.45 & 3.45 & \multirow{2}{*}{10.0} \\
&   & (6.76) & (6.25) &   & (0.84) & (0.20) &   & (6.48) & (4.34) &   & (0.56) & (0.70) &   & (0.29) & (0.37) &   & (3.59) & (3.59) &   \\
\hline
\multirow{2}{*}{{\it SOFIA}/FORCAST} & \multirow{2}{*}{19.7} & 33.66 & 32.64 & \multirow{2}{*}{9.0} & 10.35 & 5.40 & \multirow{2}{*}{7.0} & 86.65 & 63.79 & \multirow{2}{*}{3.8} & 1.41 & 1.69 & \multirow{2}{*}{6.0} & 2.11 & 1.82 & \multirow{2}{*}{6.2} & 4.70 & 4.32 & \multirow{2}{*}{8.0} \\
&   & (34.25) & (33.28) &   & (15.13) & (6.17) &   & (86.97) & (66.50) &   & (1.43) & (1.56) &   & (2.42) & (2.09) &   & (4.08) & (4.05) &   \\
\hline
\multirow{2}{*}{{\it SOFIA}/FORCAST} & \multirow{2}{*}{31.5} & 70.87 & 70.87 & \multirow{2}{*}{12.0} & 91.08 & 77.83 & \multirow{2}{*}{9.2} & 310 & 169 & \multirow{2}{*}{3.8} & 20.50 & 16.73 & \multirow{2}{*}{7.7} & 65.17 & 63.01 & \multirow{2}{*}{7.7} & 5.50 & 5.50 & \multirow{2}{*}{10.0} \\
&   & (72.92) & (72.92) &   & (90.51) & (80.47) &   & (318) & (190) &   & (21.61) & (17.47) &   & (67.06) & (65.13) &   & (3.37) & (3.37) &   \\
\hline
\multirow{2}{*}{{\it SOFIA}/FORCAST} & \multirow{2}{*}{37.1} & 84.95 & 84.95 & \multirow{2}{*}{12.0} & 132 & 111 & \multirow{2}{*}{9.2} & 375 & 176 & \multirow{2}{*}{3.8} & 25.89 & 23.56 & \multirow{2}{*}{7.7} & 116 & 116 & \multirow{2}{*}{9.0} & 7.20 & 7.20 & \multirow{2}{*}{10.0} \\
&   & (88.39) & (88.39) &   & (130) & (115) &   & (382) & (205) &   & (25.40) & (24.02) &   & (117) & (117) &   & (5.09) & (5.09) &   \\
\hline
\multirow{2}{*}{{\it IRAS}} & \multirow{2}{*}{60.0} & ... & 2281 & \multirow{2}{*}{186.6} & ... & 224 & \multirow{2}{*}{109.2} & ... & 1146 & \multirow{2}{*}{200.0} & ... & 66.13 & \multirow{2}{*}{141.0} & ... & 432 & \multirow{2}{*}{125.2} & ... & 432 & \multirow{2}{*}{125.2} \\
&   & ... & (2386) &   & ... & (235) &   & ... & (1213) &   & ... & (63.58) &   & ... & (445) &   & ... & (445) &   \\
\hline
\multirow{2}{*}{{\it Herschel}/PACS} & \multirow{2}{*}{70.0} & \multirow{2}{*}{...} & \multirow{2}{*}{...} & \multirow{2}{*}{...} & 449 & 449 & \multirow{2}{*}{25.6} & 694 & 694 & \multirow{2}{*}{9.6} & 99 & 99 & \multirow{2}{*}{23.0} & \multirow{2}{*}{...} & \multirow{2}{*}{...} & \multirow{2}{*}{...} & \multirow{2}{*}{...} & \multirow{2}{*}{...} & \multirow{2}{*}{...} \\
&   &   &   &   & (471) & (471) &   & (753) & (753) &   & (103) & (103) &   &   &   &   &   &   &   \\
\hline
\multirow{2}{*}{{\it IRAS}} & \multirow{2}{*}{100.0} & ... & 2897 & \multirow{2}{*}{244.5} & ... & 525 & \multirow{2}{*}{180.0} & ... & 2559 & \multirow{2}{*}{205.1} & ... & 152 & \multirow{2}{*}{177.5} & ... & 880 & \multirow{2}{*}{215.7} & ... & 880 & \multirow{2}{*}{215.7} \\
&   & ... & (3255) &   & ... & (666) &   & ... & (2879) &   & ... & (137) &   & ... & (947) &   & ... & (947) &   \\
\hline
\multirow{2}{*}{{\it Herschel}/PACS} & \multirow{2}{*}{160.0} & \multirow{2}{*}{...} & \multirow{2}{*}{...} & \multirow{2}{*}{...} & 360 & 360 & \multirow{2}{*}{25.6} & 421 & 421 & \multirow{2}{*}{9.6} & 127 & 127 & \multirow{2}{*}{23.0} & \multirow{2}{*}{...} & \multirow{2}{*}{...} & \multirow{2}{*}{...} & \multirow{2}{*}{...} & \multirow{2}{*}{...} & \multirow{2}{*}{...} \\
&   &   &   &   & (401) & (401) &   & (572) & (572) &   & (143) & (143) &   &   &   &   &   &   &   \\
\hline
\multirow{2}{*}{{\it Herschel}/SPIRE} & \multirow{2}{*}{250.0} & \multirow{2}{*}{...} & \multirow{2}{*}{...} & \multirow{2}{*}{...} & 190 & 190 & \multirow{2}{*}{25.6} & \multirow{2}{*}{...} & \multirow{2}{*}{...} & \multirow{2}{*}{...} & 71.43 & 71.43 & \multirow{2}{*}{23.0} & \multirow{2}{*}{...} & \multirow{2}{*}{...} & \multirow{2}{*}{...} & \multirow{2}{*}{...} & \multirow{2}{*}{...} & \multirow{2}{*}{...} \\
&   &   &   &   & (217) & (217) &   &   &   &   & (87.60) & (87.60) &   &   &   &   &   &   &   \\
\hline
\multirow{2}{*}{{\it Herschel}/SPIRE} & \multirow{2}{*}{350.0} & \multirow{2}{*}{...} & \multirow{2}{*}{...} & \multirow{2}{*}{...} & 93 & 93 & \multirow{2}{*}{25.6} & \multirow{2}{*}{...} & \multirow{2}{*}{...} & \multirow{2}{*}{...} & 29.35 & 29.35 & \multirow{2}{*}{23.0} & \multirow{2}{*}{...} & \multirow{2}{*}{...} & \multirow{2}{*}{...} & \multirow{2}{*}{...} & \multirow{2}{*}{...} & \multirow{2}{*}{...} \\
&   &   &   &   & (107) & (107) &   &   &   &   & (38.37) & (38.37) &   &   &   &   &   &   &   \\
\hline
\multirow{2}{*}{{\it Herschel}/SPIRE} & \multirow{2}{*}{500.0} & \multirow{2}{*}{...} & \multirow{2}{*}{...} & \multirow{2}{*}{...} & 35.06 & 35.06 & \multirow{2}{*}{25.6} & \multirow{2}{*}{...} & \multirow{2}{*}{...} & \multirow{2}{*}{...} & 7.61 & 7.61 & \multirow{2}{*}{23.0} & \multirow{2}{*}{...} & \multirow{2}{*}{...} & \multirow{2}{*}{...} & \multirow{2}{*}{...} & \multirow{2}{*}{...} & \multirow{2}{*}{...} \\
&   &   &   &   & (40.65) & (40.65) &   &   &   &   & (12.45) & (12.45) &   &   &   &   &   &   &   \\
\enddata
\tablenotetext{a}{Flux density derived with a fixed aperture size of the 70\,$\mu$m data.}
\tablenotetext{b}{Flux density derived with various aperture sizes.}
\tablenotetext{c}{Aperture radius.}
\tablecomments{The value of flux density in the upper row is derived with background subtraction. The value in parentheses in the lower line is the flux density derived without background subtraction.\\ The center of the aperture used for photometry of the IRAS images is not the same as those used at other wavelengths, but is determined based on the emission of the image alone. See more details in Papers I \& II.}
\end{deluxetable*}
\end{rotatetable*}

\begin{rotatetable*}% Landscape page
\centering
\setlength{\tabcolsep}{1pt}
\renewcommand{\arraystretch}{0.8}
%\vspace{-3in}
\begin{deluxetable*}{cc|ccc|ccc|ccc|ccc|ccc|ccc|ccc|ccc}
\centerwidetable
%\movetableright=-3in
\tablenum{2}
\tabletypesize{\scriptsize}
\tablecaption{Integrated Flux Densities (continued)}
\tablewidth{20pt}
\tablehead{
\colhead{Facility} &\colhead{$\lambda$} &\colhead{F$_{\lambda,\rm fix}$} &\colhead{F$_{\lambda,\rm var}$} & \colhead{R$_{\rm ap}$} &\colhead{F$_{\lambda,\rm fix}$} &\colhead{F$_{\lambda,\rm var}$} & \colhead{R$_{\rm ap}$}&\colhead{F$_{\lambda,\rm fix}$} &\colhead{F$_{\lambda,\rm var}$} & \colhead{R$_{\rm ap}$} &\colhead{F$_{\lambda,\rm fix}$} &\colhead{F$_{\lambda,\rm var}$} & \colhead{R$_{\rm ap}$}&\colhead{F$_{\lambda,\rm fix}$} &\colhead{F$_{\lambda,\rm var}$} & \colhead{R$_{\rm ap}$}&\colhead{F$_{\lambda,\rm fix}$} &\colhead{F$_{\lambda,\rm var}$} & \colhead{R$_{\rm ap}$} &\colhead{F$_{\lambda,\rm fix}$} &\colhead{F$_{\lambda,\rm var}$} & \colhead{R$_{\rm ap}$} &\colhead{F$_{\lambda,\rm fix}$} &\colhead{F$_{\lambda,\rm var}$} & \colhead{R$_{\rm ap}$} \\
\colhead{} &\colhead{($\mu$m)} &\colhead{(Jy)} &\colhead{(Jy)} & \colhead{($\arcsec$)} &\colhead{(Jy)} &\colhead{(Jy)} & \colhead{($\arcsec$)} &\colhead{(Jy)} &\colhead{(Jy)} & \colhead{($\arcsec$)} &\colhead{(Jy)} &\colhead{(Jy)} & \colhead{($\arcsec$)} &\colhead{(Jy)} &\colhead{(Jy)} & \colhead{($\arcsec$)} &\colhead{(Jy)} &\colhead{(Jy)} & \colhead{($\arcsec$)} &\colhead{(Jy)} &\colhead{(Jy)} & \colhead{($\arcsec$)} &\colhead{(Jy)} &\colhead{(Jy)} & \colhead{($\arcsec$)} 
}
\startdata
 &  & \multicolumn{3}{c}{IRAS22172 MIR2} &  \multicolumn{3}{c}{IRAS22172 MIR1} &  \multicolumn{3}{c}{IRAS22172 MIR3} &  \multicolumn{3}{c}{IRAS21391 BIMA2}  &  \multicolumn{3}{c}{IRAS21391 BIMA3}  &  \multicolumn{3}{c}{IRAS21391 MIR48} &  \multicolumn{3}{c}{IRAS16562 N} \tablenotemark{d} &  \multicolumn{3}{c}{G305 A}\tablenotemark{e} \\
\hline
\multirow{2}{*}{{\it Spitzer}/IRAC} & \multirow{2}{*}{3.6} & 0.15 & 0.09 & \multirow{2}{*}{2.4} & 0.07 & 0.07 & \multirow{2}{*}{3.6} & 0.02 & 0.02 & \multirow{2}{*}{3.6} & 0.01 & 0.01 & \multirow{2}{*}{4.6} & 0.02 & 0.01 & \multirow{2}{*}{3.6} & 0.05 & 0.04 & \multirow{2}{*}{3.6} & 0.06 & 0.04 & \multirow{2}{*}{6.0} & ... & ... & \multirow{2}{*}{12.0} \\
&   & (0.17) & (0.11) &   & (0.09) & (0.08) &   & (0.03) & (0.03) &   & (0.02) & (0.01) &   & (0.03) & (0.02) &   & (0.05) & (0.04) &   & (0.13) & (0.09) &   & (0.16) & (0.16) &   \\
\hline
\multirow{2}{*}{{\it Spitzer}/IRAC} & \multirow{2}{*}{4.5} & 0.43 & 0.29 & \multirow{2}{*}{2.4} & 0.07 & 0.05 & \multirow{2}{*}{2.4} & 0.02 & 0.01 & \multirow{2}{*}{2.4} & 0.07 & 0.05 & \multirow{2}{*}{4.6} & 0.06 & 0.05 & \multirow{2}{*}{4.5} & 0.09 & 0.08 & \multirow{2}{*}{4.5} & 0.08 & 0.06 & \multirow{2}{*}{6.0} & ... & ... & \multirow{2}{*}{12.0} \\
&   & (0.46) & (0.31) &   & (0.09) & (0.06) &   & (0.03) & (0.01) &   & (0.09) & (0.06) &   & (0.08) & (0.06) &   & (0.09) & (0.09) &   & (0.16) & (0.11) &   & (0.19) & (0.19) &   \\
\hline
\multirow{2}{*}{{\it Spitzer}/IRAC} & \multirow{2}{*}{5.8} & 0.71 & 0.43 & \multirow{2}{*}{2.4} & 0.17 & 0.07 & \multirow{2}{*}{2.4} & 0.16 & 0.12 & \multirow{2}{*}{3.6} & 0.11 & 0.10 & \multirow{2}{*}{4.6} & 0.10 & 0.09 & \multirow{2}{*}{4.5} & 0.15 & 0.14 & \multirow{2}{*}{5.4} & 0.20 & 0.07 & \multirow{2}{*}{4.5} & ... & ... & \multirow{2}{*}{12.0} \\
&   & (0.77) & (0.50) &   & (0.25) & (0.12) &   & (0.20) & (0.15) &   & (0.17) & (0.12) &   & (0.15) & (0.11) &   & (0.18) & (0.16) &   & (0.75) & (0.30) &   & (1.13) & (1.13) &   \\
\hline
\multirow{2}{*}{{\it SOFIA}/FORCAST} & \multirow{2}{*}{7.7} & 0.95 & 0.69 & \multirow{2}{*}{2.7} & 0.43 & 0.58 & \multirow{2}{*}{5.4} & 0.54 & 0.54 & \multirow{2}{*}{4.6} & 0.29 & 0.13 & \multirow{2}{*}{4.6} & 0.17 & 0.10 & \multirow{2}{*}{6.4} & 0.21 & 0.26 & \multirow{2}{*}{6.4} & 0.73 & 0.16 & \multirow{2}{*}{4.5} & \multirow{2}{*}{...} & \multirow{2}{*}{...} & \multirow{2}{*}{...} \\
&   & (1.10) & (0.85) &   & (0.71) & (0.88) &   & (0.70) & (0.70) &   & (0.39) & (0.22) &   & (0.37) & (0.24) &   & (0.27) & (0.27) &   & (1.80) & (0.73) &   &   &   &   \\
\hline
\multirow{2}{*}{{\it Spitzer}/IRAC} & \multirow{2}{*}{8.0} & 1.03 & 0.66 & \multirow{2}{*}{2.7} & 0.38 & 0.40 & \multirow{2}{*}{4.8} & 0.40 & 0.42 & \multirow{2}{*}{4.8} & 0.12 & 0.11 & \multirow{2}{*}{4.6} & 0.10 & 0.11 & \multirow{2}{*}{6.4} & 0.21 & 0.21 & \multirow{2}{*}{7.2} & 0.52 & 0.27 & \multirow{2}{*}{6.0} & ... & ... & \multirow{2}{*}{12.0} \\
&   & (1.18) & (0.83) &   & (0.60) & (0.63) &   & (0.53) & (0.55) &   & (0.21) & (0.14) &   & (0.20) & (0.16) &   & (0.26) & (0.25) &   & (1.77) & (1.16) &   & (2.71) & (2.71) &   \\
\hline
\multirow{2}{*}{{\it SOFIA}/FORCAST} & \multirow{2}{*}{19.7} & 3.64 & 3.49 & \multirow{2}{*}{3.6} & 0.64 & 0.26 & \multirow{2}{*}{2.3} & 0.85 & 0.49 & \multirow{2}{*}{2.3} & 0.45 & 0.46 & \multirow{2}{*}{5.4} & 0.28 & 0.49 & \multirow{2}{*}{5.4} & 0.93 & 0.92 & \multirow{2}{*}{7.6} & 3.83 & 3.83 & \multirow{2}{*}{7.7} & 1.60 & 1.58 & \multirow{2}{*}{4.6} \\
&   & (3.83) & (3.70) &   & (0.72) & (0.38) &   & (0.92) & (0.59) &   & (0.52) & (0.49) &   & (0.20) & (0.38) &   & (0.92) & (0.90) &   & (3.94) & (3.94) &   & (0.02) & (1.20) &   \\
\hline
\multirow{2}{*}{{\it SOFIA}/FORCAST} & \multirow{2}{*}{31.5} & 4.99 & 4.99 & \multirow{2}{*}{3.8} & 1.96 & 1.63 & \multirow{2}{*}{3.8} & 3.30 & 3.30 & \multirow{2}{*}{4.6} & 6.81 & 6.26 & \multirow{2}{*}{6.2} & 8.30 & 8.26 & \multirow{2}{*}{8.1} & 3.32 & 3.34 & \multirow{2}{*}{7.6} & 11.06 & 11.06 & \multirow{2}{*}{7.7} & 91 & 87 & \multirow{2}{*}{7.7} \\
&   & (5.76) & (5.76) &   & (2.47) & (2.08) &   & (3.77) & (3.77) &   & (7.09) & (6.63) &   & (8.77) & (8.69) &   & (3.07) & (3.10) &   & (14.82) & (14.82) &   & (104) & (92) &   \\
\hline
\multirow{2}{*}{{\it SOFIA}/FORCAST} & \multirow{2}{*}{37.1} & 6.15 & 6.15 & \multirow{2}{*}{3.8} & 3.18 & 3.18 & \multirow{2}{*}{4.6} & 4.22 & 4.22 & \multirow{2}{*}{4.6} & 11.27 & 11.27 & \multirow{2}{*}{7.7} & 12.94 & 12.94 & \multirow{2}{*}{8.5} & 4.66 & 4.66 & \multirow{2}{*}{8.0} & 13.68 & 13.68 & \multirow{2}{*}{7.7} & 153 & 149 & \multirow{2}{*}{9.2} \\
&   & (7.10) & (7.10) &   & (4.25) & (4.25) &   & (4.66) & (4.66) &   & (11.83) & (11.83) &   & (13.76) & (13.76) &   & (3.91) & (3.91) &   & (20.70) & (20.70) &   & (172) & (158) &   \\
\hline
\multirow{2}{*}{{\it IRAS}} & \multirow{2}{*}{60.0} & ... & 134 & \multirow{2}{*}{94.7} & ... & 134 & \multirow{2}{*}{94.7} & ... & 134 & \multirow{2}{*}{94.7} & ... & 163 & \multirow{2}{*}{77.0} & ... & 163 & \multirow{2}{*}{77.0} & ... & 163 & \multirow{2}{*}{77.0} & \multirow{2}{*}{...} & \multirow{2}{*}{...} & \multirow{2}{*}{...} & \multirow{2}{*}{...} & \multirow{2}{*}{...} & \multirow{2}{*}{...} \\
&   & ... & (220) &   & ... & (220) &   & ... & (220) &   & ... & (178) &   & ... & (178) &   & ... & (178) &   &   &   &   &   &   &   \\
\hline
\multirow{2}{*}{{\it Herschel}/PACS} & \multirow{2}{*}{70.0} & \multirow{2}{*}{...} & \multirow{2}{*}{...} & \multirow{2}{*}{...} & \multirow{2}{*}{...} & \multirow{2}{*}{...} & \multirow{2}{*}{...} & \multirow{2}{*}{...} & \multirow{2}{*}{...} & \multirow{2}{*}{...} & \multirow{2}{*}{...} & \multirow{2}{*}{...} & \multirow{2}{*}{...} & \multirow{2}{*}{...} & \multirow{2}{*}{...} & \multirow{2}{*}{...} & \multirow{2}{*}{...} & \multirow{2}{*}{...} & \multirow{2}{*}{...} & \multirow{2}{*}{...} & \multirow{2}{*}{...} & \multirow{2}{*}{...} & 968 & 968 & \multirow{2}{*}{12.0} \\
&   &   &   &   &   &   &   &   &   &   &   &   &   &   &   &   &   &   &   &   &   &   & (1287) & (1287) &   \\
\hline
\multirow{2}{*}{{\it IRAS}} & \multirow{2}{*}{100.0} & ... & 501 & \multirow{2}{*}{180.9} & ... & 501 & \multirow{2}{*}{180.9} & ... & 501 & \multirow{2}{*}{180.9} & ... & 363 & \multirow{2}{*}{100.0} & ... & 363 & \multirow{2}{*}{100.0} & ... & 363 & \multirow{2}{*}{100.0} & \multirow{2}{*}{...} & \multirow{2}{*}{...} & \multirow{2}{*}{...} & \multirow{2}{*}{...} & \multirow{2}{*}{...} & \multirow{2}{*}{...} \\
&   & ... & (937) &   & ... & (937) &   & ... & (937) &   & ... & (430) &   & ... & (430) &   & ... & (430) &   &   &   &   &   &   &   \\
\hline
\multirow{2}{*}{{\it Herschel}/PACS} & \multirow{2}{*}{160.0} & \multirow{2}{*}{...} & \multirow{2}{*}{...} & \multirow{2}{*}{...} & \multirow{2}{*}{...} & \multirow{2}{*}{...} & \multirow{2}{*}{...} & \multirow{2}{*}{...} & \multirow{2}{*}{...} & \multirow{2}{*}{...} & \multirow{2}{*}{...} & \multirow{2}{*}{...} & \multirow{2}{*}{...} & \multirow{2}{*}{...} & \multirow{2}{*}{...} & \multirow{2}{*}{...} & \multirow{2}{*}{...} & \multirow{2}{*}{...} & \multirow{2}{*}{...} & \multirow{2}{*}{...} & \multirow{2}{*}{...} & \multirow{2}{*}{...} & 668 & 668 & \multirow{2}{*}{12.0} \\
&   &   &   &   &   &   &   &   &   &   &   &   &   &   &   &   &   &   &   &   &   &   & (1160) & (1160) &   \\
\hline
\multirow{2}{*}{{\it Herschel}/SPIRE} & \multirow{2}{*}{250.0} & \multirow{2}{*}{...} & \multirow{2}{*}{...} & \multirow{2}{*}{...} & \multirow{2}{*}{...} & \multirow{2}{*}{...} & \multirow{2}{*}{...} & \multirow{2}{*}{...} & \multirow{2}{*}{...} & \multirow{2}{*}{...} & \multirow{2}{*}{...} & \multirow{2}{*}{...} & \multirow{2}{*}{...} & \multirow{2}{*}{...} & \multirow{2}{*}{...} & \multirow{2}{*}{...} & \multirow{2}{*}{...} & \multirow{2}{*}{...} & \multirow{2}{*}{...} & ... & ... & \multirow{2}{*}{7.7} & \multirow{2}{*}{...} & \multirow{2}{*}{...} & \multirow{2}{*}{...} \\
&   &   &   &   &   &   &   &   &   &   &   &   &   &   &   &   &   &   &   & (73.55) & (73.55) &   &   &   &   \\
\hline
\multirow{2}{*}{{\it Herschel}/SPIRE} & \multirow{2}{*}{350.0} & \multirow{2}{*}{...} & \multirow{2}{*}{...} & \multirow{2}{*}{...} & \multirow{2}{*}{...} & \multirow{2}{*}{...} & \multirow{2}{*}{...} & \multirow{2}{*}{...} & \multirow{2}{*}{...} & \multirow{2}{*}{...} & \multirow{2}{*}{...} & \multirow{2}{*}{...} & \multirow{2}{*}{...} & \multirow{2}{*}{...} & \multirow{2}{*}{...} & \multirow{2}{*}{...} & \multirow{2}{*}{...} & \multirow{2}{*}{...} & \multirow{2}{*}{...} & ... & ... & \multirow{2}{*}{7.7} & 74 & 74 & \multirow{2}{*}{12.0} \\
&   &   &   &   &   &   &   &   &   &   &   &   &   &   &   &   &   &   &   & (29.09) & (29.09) &   & (187) & (187) &   \\
\hline
\multirow{2}{*}{{\it Herschel}/SPIRE} & \multirow{2}{*}{500.0} & \multirow{2}{*}{...} & \multirow{2}{*}{...} & \multirow{2}{*}{...} & \multirow{2}{*}{...} & \multirow{2}{*}{...} & \multirow{2}{*}{...} & \multirow{2}{*}{...} & \multirow{2}{*}{...} & \multirow{2}{*}{...} & \multirow{2}{*}{...} & \multirow{2}{*}{...} & \multirow{2}{*}{...} & \multirow{2}{*}{...} & \multirow{2}{*}{...} & \multirow{2}{*}{...} & \multirow{2}{*}{...} & \multirow{2}{*}{...} & \multirow{2}{*}{...} & ... & ... & \multirow{2}{*}{7.7} & 9.65 & 9.65 & \multirow{2}{*}{12.0} \\
&   &   &   &   &   &   &   &   &   &   &   &   &   &   &   &   &   &   &   & (8.59) & (8.59) &   & (44.62) & (44.62) &   \\
\enddata
\tablenotetext{d}{Aperture is centered at the 37 $\mu$m peak at R.A. (J2000) = 16$^h$59$^m$43$\fs$010, Decl.(J2000) = $-$40$\arcdeg$03$\arcmin$11$\farcs$560.}
\tablenotetext{e}{Aperture is centered at the position of the 6.7\,GHz methanol maser at R.A. (J2000) =  13$^h$11$^m$13$\fs$795, Decl.(J2000) = $-$62$\arcdeg$34$\arcmin$41$\farcs$741 (Norris et al. 1993).}
\tablecomments{The background subtracted flux density of IRAS16562 N at 250, 350, 500 $\mu$m is negative due to contamination of the IRAS 16562 main source presented in Paper II.\\ So we use the non-background subtracted flux density at these wavelengths as upper limits for the SED fitting of IRAS16562 N.\\ There is no emission at wavelengths shorter than 8 $\mu$m from G305 A. So we use the non-background subtracted flux density at these wavelengths as upper limits for the SED fitting of G305 A.}
\end{deluxetable*}
\end{rotatetable*}

\subsection{Description of Individual Sources}\label{S:indiv}

\subsubsection{S235}
\begin{figure*}
\epsscale{1.2}
\plotone{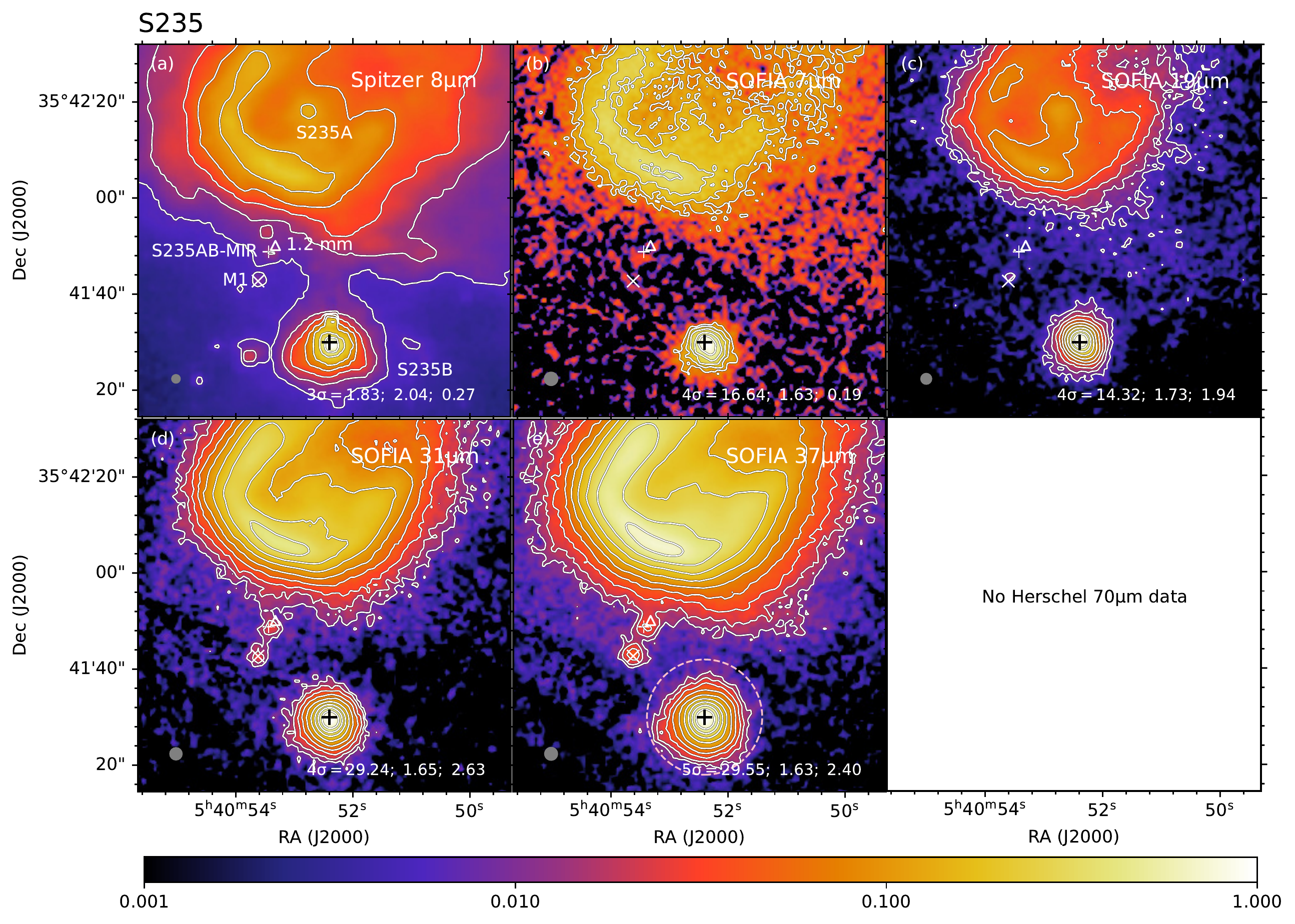}
\caption{
Multi-wavelength images of S235 with facility and wavelength given in
the upper right corner of each panel.  Contour level information is
given in the lower right: lowest contour level in number of $\sigma$
above the background noise and corresponding value in mJy per square
arcsec; then step size between each contour in log$_{10}$ mJy per
square arcsec, then peak flux in Jy per square arcsec. The color map
indicates the relative flux intensity compared to that of the peak
flux in each image panel.
%but only showing the signal above 3$\sigma$. 
The pink dashed circle shown in (e) denotes the aperture used for the
fiducial photometry. Gray circles in the lower left show the
resolution of each image. The black cross in all panels denotes the
position of the radio source VLA-2 of Felli et al. (2006) at
R.A.(J2000) = 05$^h$40$^m$52$\fs$40, Decl.(J2000) =
$+$35$\arcdeg$41$\arcmin$30$\arcsec$. The triangle sign marks the
position of the 1.2 mm core. The small white cross marks the position
of S235AB-MIR. The $\times$ sign marks the position of the NIR K-band
source M1 as well as VLA-1.\label{fig:S235}}
\end{figure*}

%The S235 A-B region is a star-forming site where different stages of
%stellar evolution appear to be present within a small region.

Estimates of the distance to the S235 A-B region vary from 1.6 - 2.5
kpc (e.g., Israel \& Felli 1978; Burns et al. 2015). We adopt 1.8 kpc,
following Evans \& Blair (1981), Dewangan et al. (2016) and Shimoikura
et al. (2016).
%, which is an intermediate value of the published distance range.
High-resolution mm line and continuum and radio continuum observations
towards S235 A-B were reported by Felli et al. (2004,
2006). Shimoikura et al. (2016) carried out observations of C$^{18}$O
emission toward S235 A-B and revealed that the clump has an elliptical
shape, with a mass of $\sim 1000\:M_{\odot}$ and an average radius of
$\sim0.5\:$pc.
%The Position-Velocity (PV) diagram indicates infalling
%and rotating motions of the cores.
Two compact HII regions, called S235 A and S235 B (e.g., Felli et
al. 1997; Klein et al. 2005; Saito et al. 2007) are located in this
clump, along with a mm continuum core with HCO$^+$(1-0) outflows
in-between, which is thought to be an embedded, earlier-stage YSO (Felli et
al. 2004).
%Felli et al. (2006) suggested that the expanding HII region from
%S235~A could have triggered the star formation in the dense cluster
%between:
The mm core has a MIR counterpart S235 AB-MIR and several water masers
and methanol masers nearby (Kurtz et al. 2004). From their estimate of
a luminosity of $\sim 10^3 L_{\odot}$ of the source, Felli et
al. (2004) suggested that S235 AB-MIR is an intermediate-mass YSO
driving the molecular outflows and supplying the energy for the -60 km
s$^{-1}$ water maser nearby. On the other hand, Dewangan \& Anandarao
(2011) concluded from SED fitting that S235 AB-MIR is the most massive
protostar in the region with $m_{*} \sim 11\:M_{\odot}$
%located at the clump center, which is
and still actively accreting and so not yet able to excite an HII
region. However, they were cautious about the reliability of these
results due to the limited number of data points (three in the MIR
from IRAC bands and two in the sub-mm-continuum from Felli et
al. 2004).

Another NIR K-band source with the largest infrared excess, M1, is
reported to be associated with the radio source VLA-1 by Felli et
al. (2006) and they suggested that it could be a B2-B3 star with an
UCHII region, while Dewangan \& Anandarao et al. (2011) suggested that
it is a low-mass star, relatively young in its evolution. Both S235
AB-MIR (counterpart of the 1.2mm core) and M1 can be seen in our
\textit{SOFIA} images in Figure~\ref{fig:S235}. However, due to their
weak MIR emission,
%, which is not ideal for the SED fitting especially our models,
we do not focus on them in this paper.

Our analysis is focussed on the MIR source S235 B, which is associated
with the radio source VLA-2 (Felli et al. 2006). S235 B is the
brightest object in the S235 A-B cluster in all broad-bands from U to
K, and thus may be a massive YSO (Boley et al. 2009).
%jct - is it bright in the U band? That would be unusual, indicating little foreground extinction, which we can compare against our model results.
Krassner et al. (1982) detected hydrogen recombination lines and
polycyclic aromatic hydrocarbon (PAH) emission features at 3.3, 8.7
and 11.3 $\mu$m. However, no 3.3 mm or 1.2 mm continuum or molecular
lines are detected associated with S235 B (Felli et al. 2004).  While
there is large-scale $^{12}$CO, $^{13}$CO and C$^{18}$O emission in
the whole S235 region (Shimoikura et al. 2016; Dewangan \& Ojha 2017),
smaller-scale outflows specifically associated with S235 B have not
yet been reported. For example, even in the high-resolution
HCO$^+$(1-0) map of Felli et al. (2004), whose field of view covers
S235 B, there is no sign of HCO$^+$(1-0) outflows emerging from S235 B. 
%jct - is this component just to the north of the + or to the east?  - We are only sure that the north one is a ghost.
Boley et al. (2009) classified the central star of S235 B as an
early-type (B1V) Herbig Be star surrounded by an accretion disk based
on its spectrum from 3800-7200 {\AA}, its location in a region of
active star formation, the presence of the nearby nebulosity, the
Balmer emission lines in the stellar spectrum, and the large H-K
excess. Furthermore, its spectrum shows that the S235 B nebulosity is
reflective in nature, with the central YSO in S235 B as the
illuminating source. Given the mass inferred from the spectral type
($> 10 M_{\odot}$), Boley et al. suggested S235 B is likely to already
be on the main sequence. 

In our \textit{SOFIA} images as shown in Figure~\ref{fig:S235},
S235 B is much brighter than S235 AB-MIR and M1. The weak second
component to the north of the radio source in the \textit{Spitzer} 8
$\mu$m image is likely to be produced by a ghosting effect of the
primary source, since it is not seen in the other IRAC images, the {\it
  SOFIA} images or the \textit{UKIDSS} JHK band images.

\subsubsection{IRAS 22198+6336}
\begin{figure*}
\epsscale{1.2}
\plotone{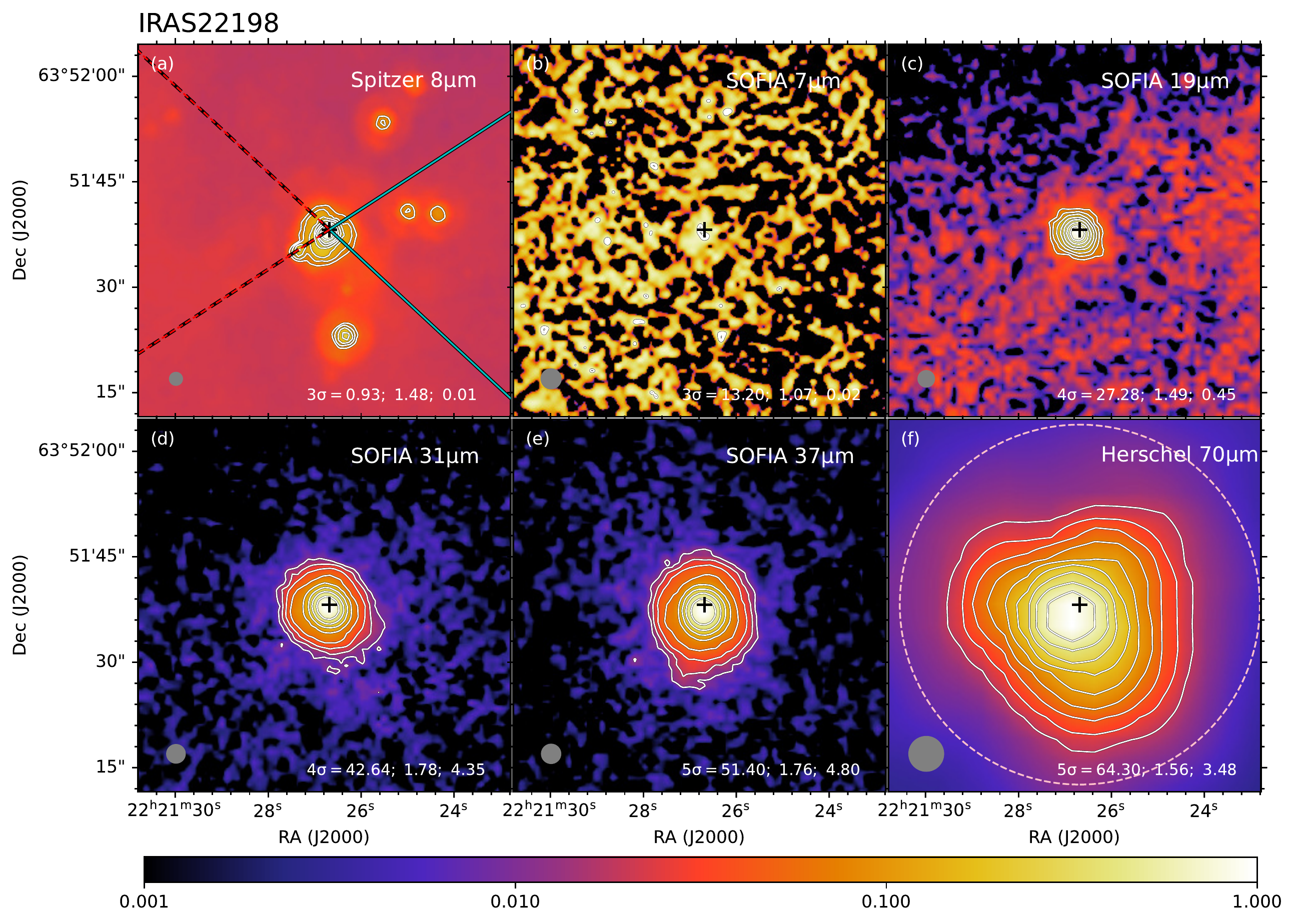}
\caption{
Multi-wavelength images of IRAS~22198+6336, following the format of
Figure 1. The black cross in all panels denotes the position of the
3.6\,cm source in S\'{a}nchez-Monge et al. (2008) at R.A.(J2000) =
22$^h$21$^m$26$\fs$68, Decl.(J2000) =
$+$63$\arcdeg$51$\arcmin$38$\farcs$2. The lines in panel (a) show the
orientation of outflow axes, with the solid spans tracing blue-shifted
direction and the dashed spans red-shifted direction. The outflow axis
angles are from the CO(1-0) outflow emission of S\'{a}nchez-Monge et
al. (2010). \label{fig:IRAS22198}}
\end{figure*}

IRAS 22198+6336 was previously considered to be a massive YSO (Palla
et al. 1991; Molinari et al. 1996; S\'{a}nchez-Monge et al. 2008)
until an accurate distance of 764 $\pm$ 27 pc was derived from the parallax measurements of 22\,GHz associated water masers
(Hirota et al. 2008). These authors, after reanalyzing the
protostellar SED, then proposed IRAS 22198+6336 is an intermediate-mass
deeply embedded YSO with spectral type of late-B, equivalent to a
Class 0 object in low-mass star formation. S\'{a}nchez-Monge et
al. (2010) detected a compact source at 3.5, 2.7, and 1.3 mm
coincident with the centimeter source reported by S\'{a}nchez-Monge et
al. (2008) and surrounded by a faint structure extended toward the
southwest. The high rotational
temperature (100-150 K) derived from CH$_3$CN and CH$_3$OH, together
with the chemically rich spectrum, is clear evidence that IRAS 22198
is an intermediate-mass hot core. The CO(1-0) emission in
S\'{a}nchez-Monge et al. (2010) reveals an outflow with a quadrupolar
morphology clearly centered on the position of the main dust
condensation. Observations of the high-velocity emission of different
outflow tracers HCO$^+$(1-0), HCN(1-0) and SiO(2-1) seem to favor the
superposition of two bipolar outflows. Higher angular resolution
observations at 1.3 mm by Palau et al. (2013) reveal a counterpart of
the cm source (MM2 in their nomenclature) and a faint extension to its south
(MM2-S). Palau et al. suggest that MM2 is likely driving the
southwest-northeast outflow, while an unresolved close companion of MM2
or MM2-S, which is only detected at 3.6$\mu$m, could be
the driving source of the northwest-southeast outflow. Periodic flares
of the 6.7-GHz methanol maser have been detected in IRAS 22198 and
their characteristics can be explained by a colliding-wind binary
model (Fujisawa et al. 2014).

Our \textit{SOFIA} images reveal the MIR counterpart of the centimeter/millimeter source. Extended emission is seen towards the blue-shifted outflow in the southwest at 19 and 31\,$\mu$m. In contrast, the extended emission at\,$\mu$m directly points to the south. Faint extended emission is also seen along the axes of the two outflows at 70\,$\mu$m.

\subsubsection{NGC~2071}
\begin{figure*}
\epsscale{1.2}
\plotone{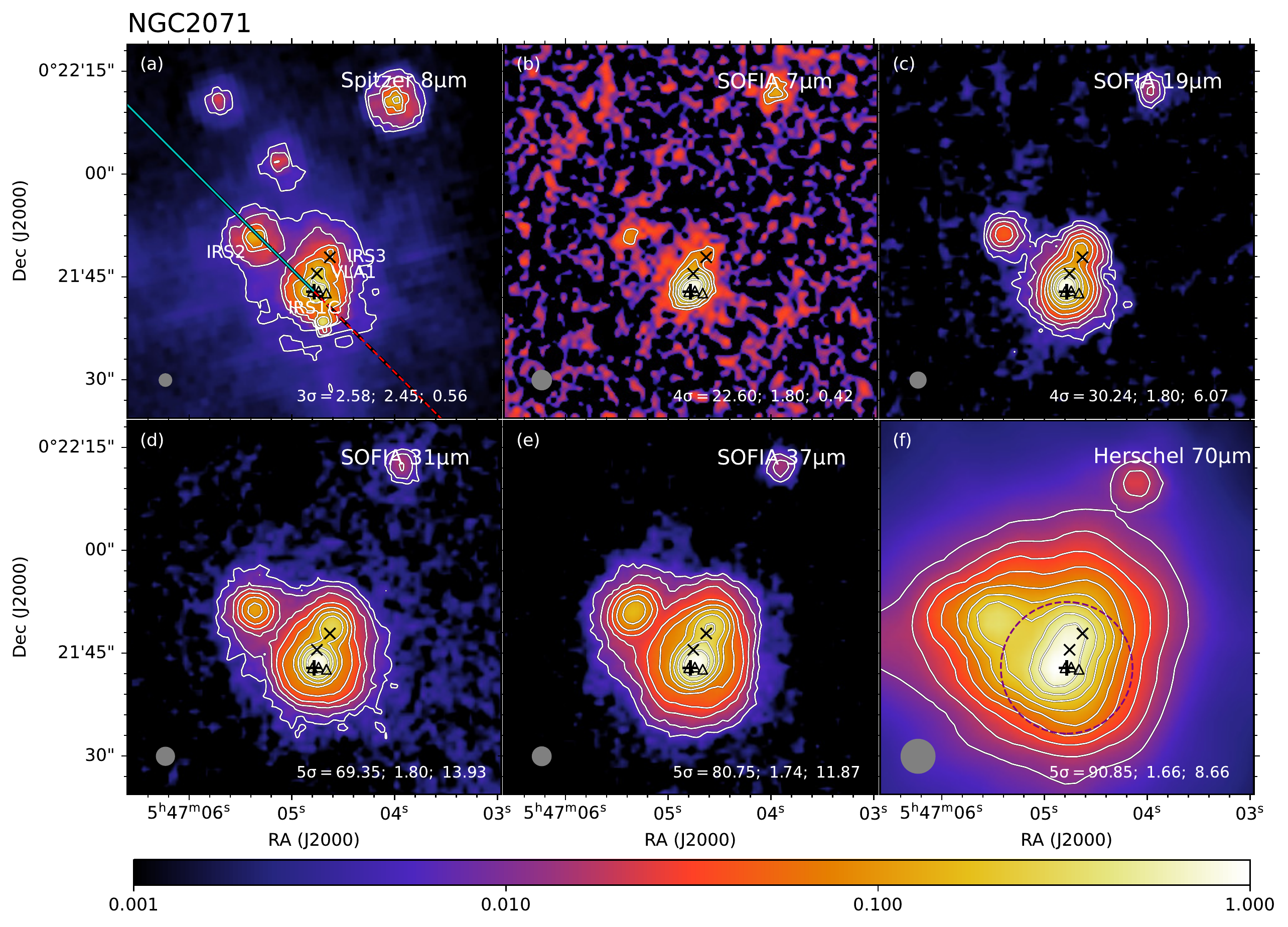}
\caption{
Multi-wavelength images of NGC~2071. The black cross in all panels
denotes the position of the 1.3\,cm source IRS 1C in Trinidad et
al. (2009) at R.A.(J2000) = 05$^h$47$^m$04$\fs$741, Decl.(J2000) =
$+$00$\arcdeg$21$\arcmin$42$\farcs$96. The $\times$ signs from north
to south mark the positions of the 1.3\,cm sources IRS3 and VLA1,
respectively. The triangle signs from east to west mark the positions
of the 1.3\,cm sources IRS1E, IRS1W, and IRS1Wb, respectively. The
lines in panel (a) show the orientation of the outflow axis (flow I),
with the solid span tracing the blue-shifted direction and the dashed
span the red-shifted direction. The outflow axis angle is from the
high-velocity CO(1-0) main outflow emission of Stojimirovi\'{c} et
al. (2008). Note that the center of the outflow has an uncertainty of
$\sim$5\arcsec\ and is not necessarily at IRS1C. \label{fig:NGC2071}}
\end{figure*}

NGC~2071 is a reflection nebula located at a distance of 390 pc in the
L1630 molecular cloud of Orion B (Anthony-Twarog 1982). The three
brightest members of the infrared cluster at 10 $\mu$m, IRS1, IRS2 and
IRS3, are each associated with compact radio sources at 5 GHz (Snell
\& Bally 1986). The radio continuum emission of IRS1 and IRS3 and the
water masers associated with them suggest that both sources are
associated with thermal jets (Smith \& Beck 1994; Torrelles et
al. 1998; Seth et al. 2002). Higher resolution VLA observations
(Trinidad et al. 2009) break IRS1 into three continuum peaks (IRS1E,
1C and 1W), aligned in the east-west direction. Both the morphology
and spectral index suggest that IRS1C is a thermal radio jet, while
IRS1E and IRS1W could be condensations ejected by IRS1C. An energetic
bipolar CO outflow has been observed toward NGC~2071, extending in the
northeast-southwest direction and reaching $\sim$15' in length (Bally
1982). In addition, shock-excited molecular hydrogen emission at 2.12
$\mu$m has also been reported showing a spatial extent similar to that
of the CO outflow and revealing several H$_2$ outflows in the field,
including one (flow II) perpendicular to the main outflow (flow I)
(Eisl\"{o}ffel 2000). Stojimirovi\'{c} et al. (2008) also detected
CO(1-0) emission in the direction of flow II. Trinidad et al. (2009)
tried to identify individual driving sources for each outflow based on
the observations of Eisl\"{o}ffel (2000) and the elongation of the
IRS3 jet. However, we note that higher resolution observations of the
outflows are needed to better distinguish the driving sources in this
region.

Based on radio continuum emission indicating presence of thermal jets
and water masers that are tracing disk-YSO-outflow systems, it has
been proposed that IRS1 and IRS3 are intermediate- and low-mass YSOs,
respectively (Smith \& Beck 1994; Torrelles et al. 1998; Seth et
al. 2002, Trinidad et al. 2009). In our \textit{SOFIA} images, the
three sources IRS1, IRS2 and IRS3 are revealed at all wavelengths
(see~Fig.~\ref{fig:NGC2071}). Here, we will focus on the SED of the
IRS1 source, but the aperture we adopt also includes IRS3.

\subsubsection{Cepheus E}
\begin{figure*}
\epsscale{1.2}
\plotone{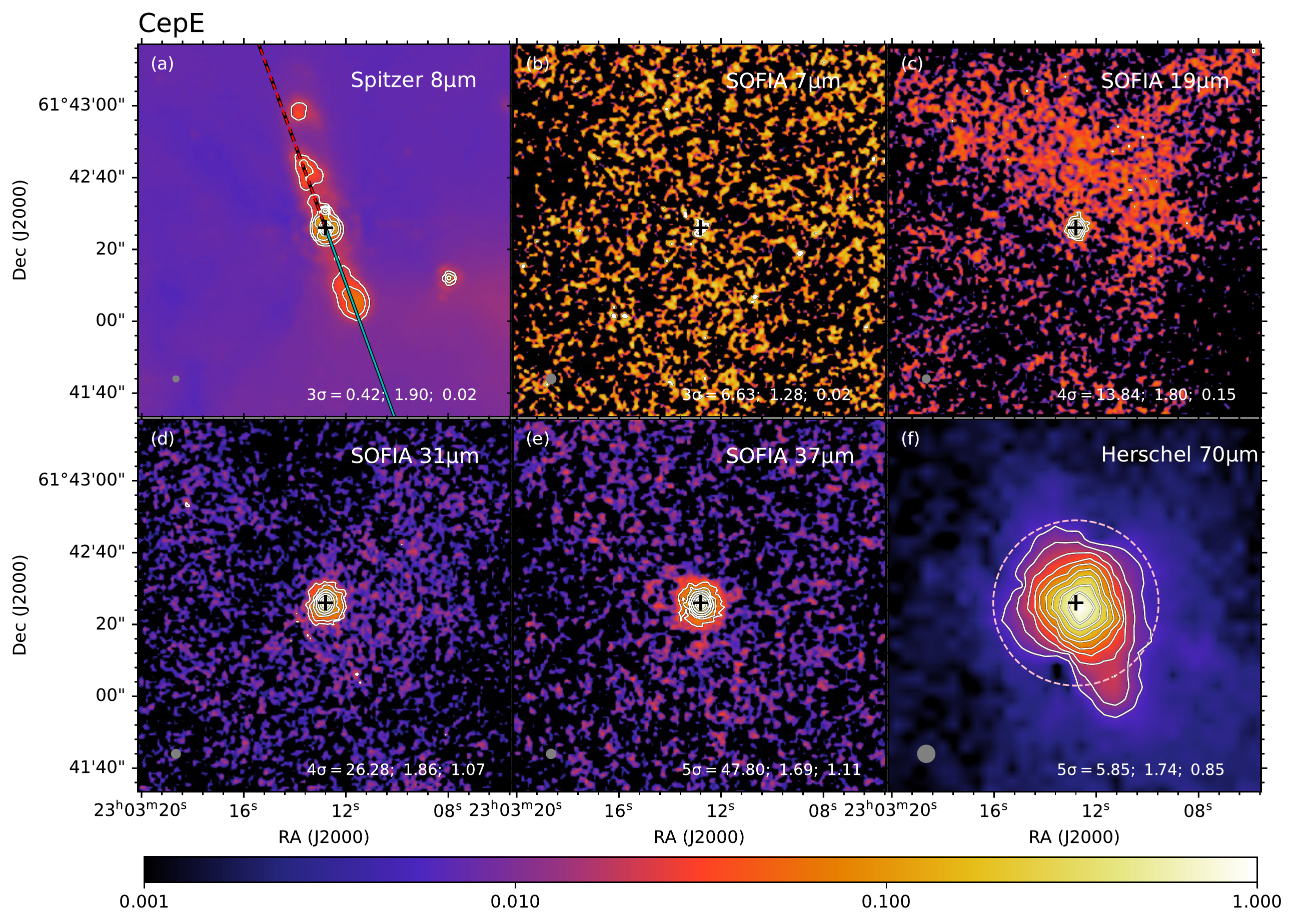}
\caption{
Multiwavelength images of Cep E. The black cross in all panels denotes
the position of the 1.3\,mm source CepE-A in Ospina-Zamudio et
al. (2018) at R.A.(J2000) = 23$^h$03$^m$12$\fs$8, Decl.(J2000) =
$+$61$\arcdeg$42$\arcmin$26$\arcsec$. The lines in panel (a) show the
orientation of the outflow axis, with the solid span tracing the
blue-shifted direction and the dashed span the red-shifted
direction. The outflow axis angle is defined by the CO(2-1) outflow
emission of Lefloch et al. (2015).\label{fig:CepE}}
\end{figure*}

The Cepheus E (Cep E) molecular cloud is located at a distance of
730~pc (Sargent 1977). Since its early discovery by Wouterloot \&
Walmsley (1986) and Palla et al. (1993), subsequent studies have
confirmed the central source Cep E-mm to be an isolated
intermediate-mass protostar in the Class 0 stage (Lefloch et al. 1996;
Moro-Mart\'{i}n et al. 2001). The source drives a very luminous molecular
outflow and jet (Lefloch et al. 2011, 2015), terminated by the bright
Herbig-Haro object HH377 in the south (Ayala et al. 2000). The
21\arcsec-long jet, the HH 377 terminal bow-shock, and the outflow
cavity are clearly revealed in multiple CO transitions and the [OI] 63
$\mu$m line (Gusdorf et al. 2017).  The observations are interpreted
by means of time-dependent magneto-hydrodynamics (MHD) shock models by
(Lefloch et al. 2015). Ospina-Zamudio et al. (2018) reveal Cep E-mm as
a binary protostellar system with \textit{NOEMA} observations. They
identified two components from a two-component fit to the
visibilities, Cep E-A and Cep E-B, which are separated by $\sim$
1.7\arcsec. Ospina-Zamudio et al. argued Cep E-A dominates the core
continuum emission and powers the well-known, high-velocity jet
associated with HH 377, while the lower flux source Cep E-B powers
another high-velocity molecular jet revealed in SiO(5-4) propagating
in a direction close to perpendicular with respect to the Cep E-A
jet.
%jct5 - I've commented this out in response to Maite's comment
The spectra of molecular lines observed by NOEMA show bright
emission of O- and N-bearing complex organic molecules (COMs) around
Cep E-A and no COM emission towards Cep E-B.

From our \textit{SOFIA}
images (Fig.~\ref{fig:CepE}), we are not able to resolve the potential
binary system, so our modeling will be an approximation of the
properties of Cep E-A, assuming it dominates the system. The IR
emission along the main jet is clearly seen in the {\it Spitzer}~8
$\mu$m image and also in the {\it Herschel} 70 $\mu$m image, since
these space-based observations are more sensitive to fainter emission
features.

\subsubsection{L1206}

\begin{figure*}
\epsscale{1.2}
\plotone{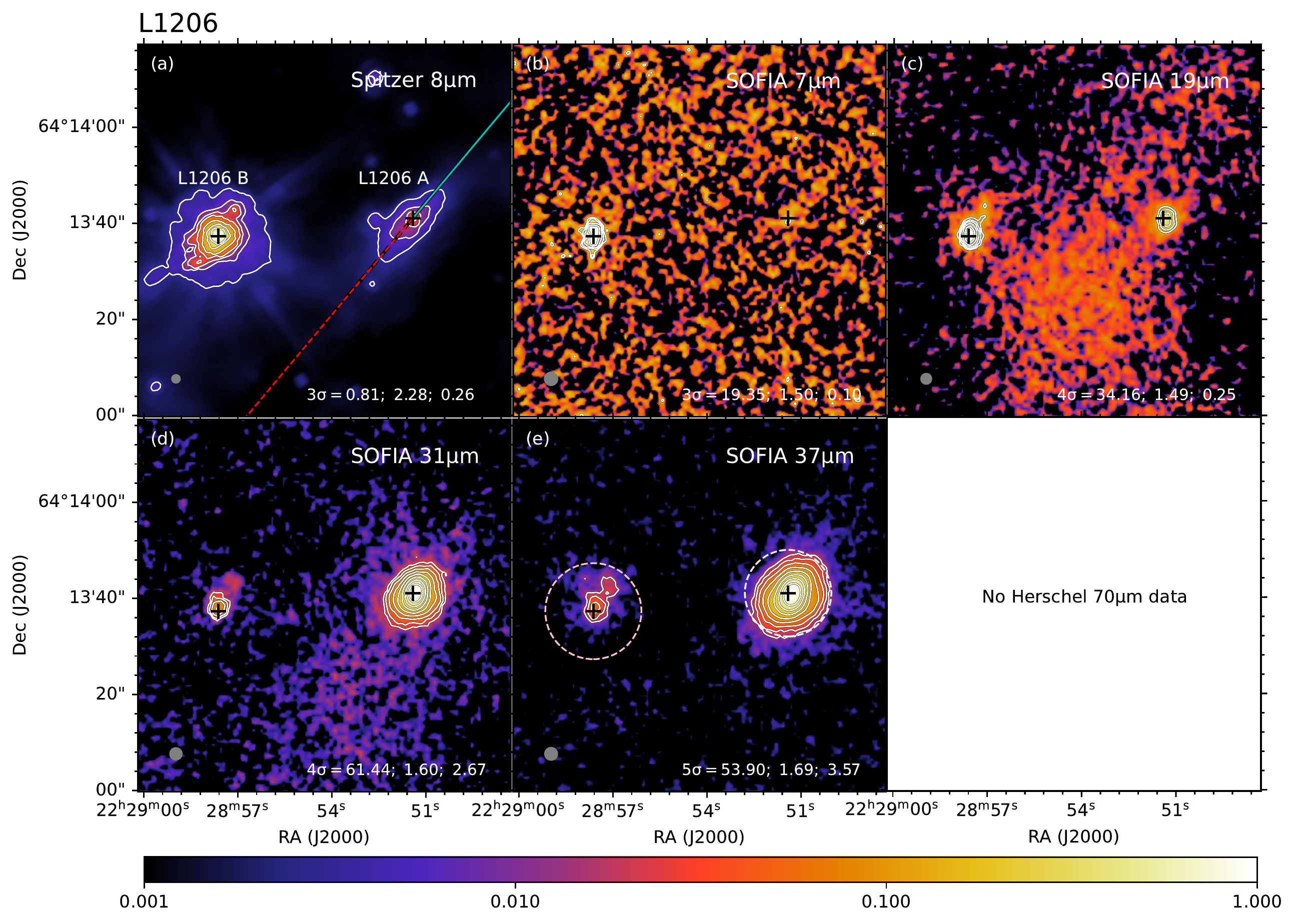}
\caption{
Multi-wavelength images of L1206. The black crosses in all panels from
east to west denote the position of the 8$\mu$m peak of L1206 B at
R.A.(J2000) = 22$^h$28$^m$57$\fs$626, Decl.(J2000) =
$+$64$\arcdeg$13$\arcmin$37$\farcs$348 and the position of L1206 A
coincident with that of the 2.7\,mm source OVRO 2 in Beltr\'{a}n et al. (2006)
at R.A.(J2000) = 22$^h$28$^m$51$\fs$41, Decl.(J2000) =
$+$64$\arcdeg$13$\arcmin$41$\farcs$1, respectively. The lines in panel
(a) show the orientation of the outflow axis from L1206 A, with the solid
span tracing blue-shifted direction and the dashed span red-shifted
direction. The outflow axis angle is given by the CO(1-0) outflow
emission of Beltr\'{a}n et al. (2006).\label{fig:L1206}}
\end{figure*}

L1206, also known as IRAS 22272+6358, is located at a distance of 776
pc from the trigonometric parallaxes of 6.7 GHz methanol masers (Rygl
et al. 2010). There are two MIR sources presented in our field of
view. The western source IRAS 22272+6358 A (hereafter referred to as
L1206 A) has no optical counterpart, and at near-infrared wavelengths,
it has only been seen in scattered light (Ressler \& Shure
1991). Given its extremely low 60/100 $\mu$m color temperature, L1206
A is believed to be very embedded, cold and young (Ressler \& Shure
1991, Beltr\'{a}n et al. 2006). It has been detected at 2.7 and 2 mm,
but not at 2 or 6 cm (Wilking et al. 1989; McCutcheon et al. 1991;
Sugitani et al. 2000; Beltr\'{a}n et al. 2006). The 2.7 mm continuum
observations by Beltr\'{a}n et al. (2006) revealed four sources, OVRO
1, OVRO 2, OVRO 3, and OVRO 4, in a 12\arcsec\ vicinity of L1206 A. The
strongest millimeter source OVRO 2 is most likely the YSO associated
with L1206 A, and is probably the driving source of the CO molecular
outflow detected in the region. The dust emission morphology and
properties of OVRO 2 suggest that this intermediate-mass protostar is
probably in transition between Class 0 and I.

The K, L, L' and M filter images of L1206 A reveal clearly lobes in a
bipolar system (Ressler \& Shure 1991). There is a distinct
3-4\arcsec\ gap between the two lobes at the K, L, L' bands. Since the
proposed illuminating source lies within this gap, it is suggested by
Ressler \& Shure (1991) that this gap is produced by the extreme
extinction of a thick, circumstellar disk. We also see such a gap in
the 3.6, 4.5, and 5.8 $\mu$m images. The CO(1-0) observations of
Beltr\'{a}n et al. (2006) reveal a very collimated outflow driven by
OVRO 2 with a very weak southeastern red lobe and a much stronger
northwestern blue lobe. The relative brightness of the red lobe also
decreases monotonically at K, L, L' bands (Ressler \& Shure
1991). Beltr\'{a}n et al. (2006) suggested a scenario in which
photodissociation produced by the ionization front coming from the
bright-rimmed diffuse H II region in the south could be responsible
for the weakness of the redshifted lobe and its overall morphology. 

The elongation along the outflow direction of L1206 A is clearly revealed at 8\,$mu$m. We
see a slight extension along the outflow direction in our
\textit{SOFIA} images, especially at 31$\mu$m and 37 $\mu$m (see
Fig.~\ref{fig:L1206}).

IRAS 22272 + 6358 B (hereafter referred to as L1206 B) is a bluer but
less luminous object, which lies approximately 40\arcsec\ to the east
of L1206 A. Since L1206 B is directly visible at NIR and is likely to
be a less obscured young stellar object, Ressler \& Shure (1991)
suggested that L1206 B is most likely a late Class I object or perhaps
an early Class II object, whose photospheric spectrum is heavily
extinguished by the parent cloud and is also affected by emission from
a circumstellar disk. 

From our \textit{SOFIA} images, it can be seen
that the emission of L1206 B becomes weaker as one goes to longer
wavelengths, which also indicates that L1206 B may be more evolved
than L1206 A.

\subsubsection{IRAS 22172+5549}

\begin{figure*}
\epsscale{1.2}
\plotone{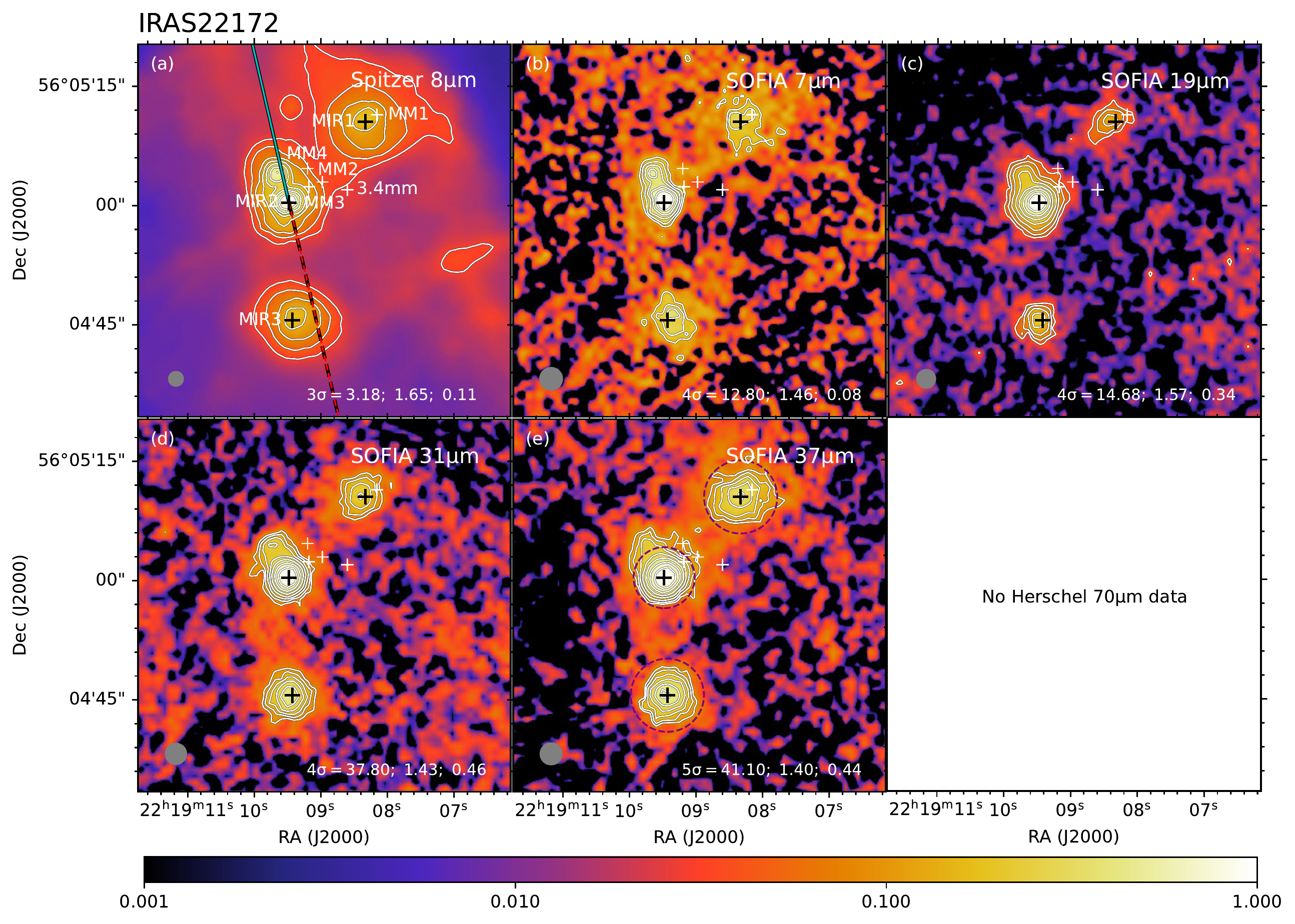}
\caption{
Multi-wavelength images of IRAS~22172. The black crosses in all panels
from north to south denote the positions of the MIR peaks at 37$\mu$m
MIR1 at R.A.(J2000) = 22$^h$19$^m$08$\fs$328, Decl.(J2000) =
$+$56$\arcdeg$05$\arcmin$10$\farcs$522, MIR2 at R.A.(J2000) =
22$^h$19$^m$09$\fs$478, Decl.(J2000) =
$+$56$\arcdeg$05$\arcmin$00$\farcs$370, and MIR3 at R.A.(J2000) =
22$^h$19$^m$09$\fs$430, Decl.(J2000) =
$+$56$\arcdeg$04$\arcmin$45$\farcs$581, respectively. The white
crosses from north to south mark the positions of the 1.3\,mm sources
MM1, MM4, MM2, MM3 in Palau et al. (2013) and the 3.4\,mm source in
Molinari et al. (2002) (also the mm core I22172-C in Fontani et
al. 2004), respectively. The lines in panel (a) show the orientation
of the outflow axis from MIR2, with the solid span tracing blue-shifted
direction and the dashed span red-shifted direction. The outflow axis
angle is from the CO(1-0) outflow emission of Fontani et
al. (2004). \label{fig:IRAS22172}}
\end{figure*}

IRAS 22172+5549 is located at a kinematic distance of 2.4 kpc
(Molinari et al. 2002). As a luminous IRAS source in the survey of
Molinari et al. (2002), IRAS 22172 shows the presence of a compact
dusty core without centimeter continuum emission, with prominent wings
in the HCO$^+$(1-0) line. Fontani et al. (2004) studied the 3\,mm
continuum and CO(1-0) emission in this region, finding a CO bipolar
outflow centered at MIR2 (IRS1 in their nomenclature), which is offset
by $\sim$ 7.5\arcsec\ from the 3.4\,mm peak. They suggested that the
dusty core might host a source in a very early evolutionary stage
prior to the formation of an outflow. From the outflow parameters,
they proposed that MIR2, as the driving source, must be relatively
massive. Palau et al. (2013) carried out higher angular resolution
1.3\,mm and CO(2-1) observations. They detected more mm
sources, including one confirmed protostar with no infrared
emission that is driving a small outflow (MM2), two protostellar
candidates detected only in the millimeter range (MM3 and MM4), and
one protostellar object detected in the mm and infrared, with
no outflow (MM1). MIR2 is still detected only in the infrared and is
driving the larger CO(1-0) outflow. No mm emission or
molecular outflows are detected towards MIR1 or MIR3. It is clear that
IRAS 22172 harbors a rich variety of YSOs at different evolutionary
stages. 

Our \textit{SOFIA} images (see Fig.~\ref{fig:IRAS22172})
reveal extended emission along the blue-shifted outflow from MIR2,
which could come from the outflow cavity.

\subsubsection{IRAS 21391+5802}

\begin{figure*}
\epsscale{1.2}
\plotone{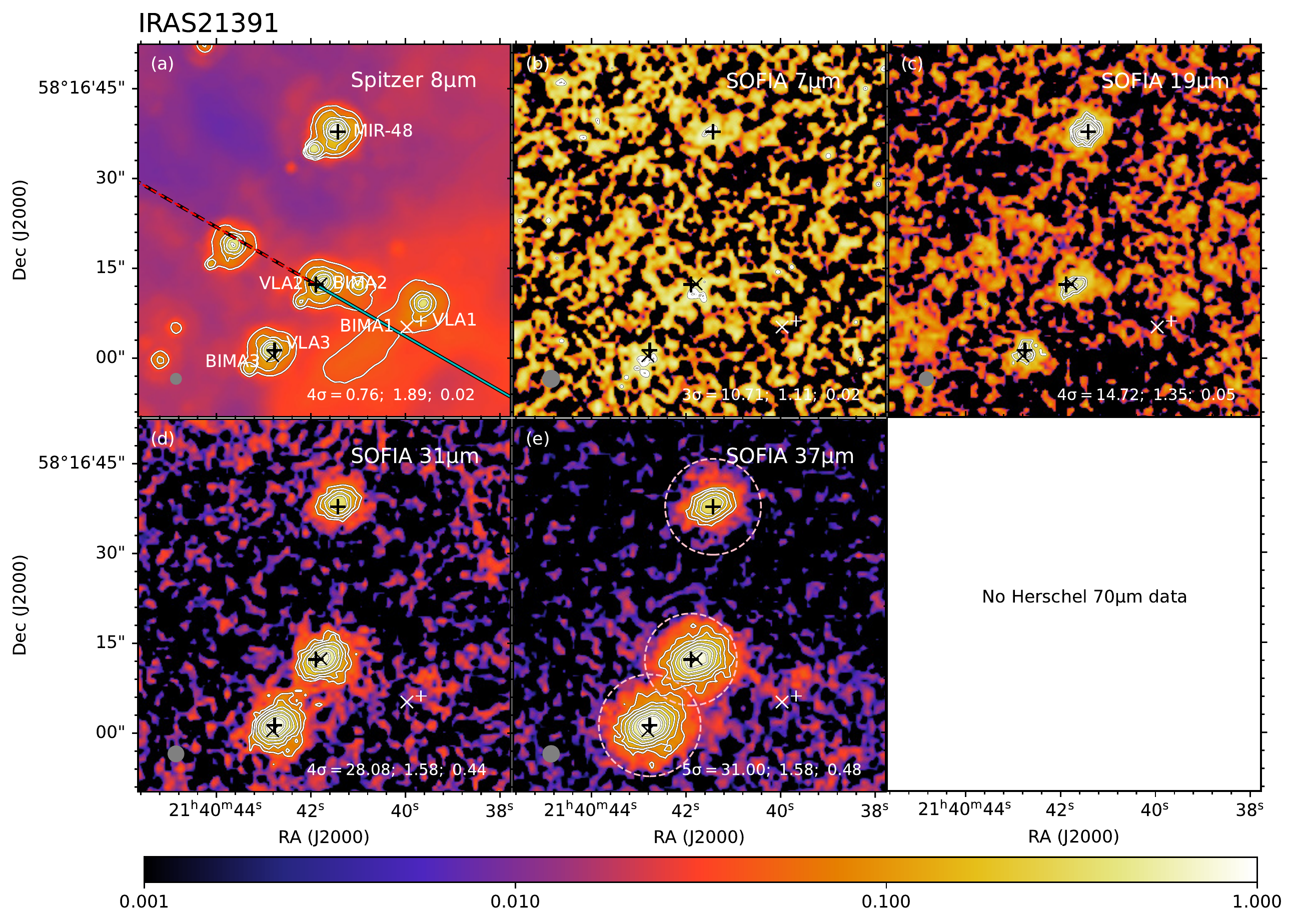}
\caption{
Multi-wavelength images of IRAS~21391. The black crosses in all panels
from north to south denote the positions of the MIR source MIR-48 at
R.A.(J2000) = 21$^h$40$^m$41$\fs$43, Decl.(J2000) =
$+$58$\arcdeg$16$\arcmin$37$\farcs$8 in Choudhury et al. (2010) and
3.6\,cm sources VLA2 at R.A.(J2000) = 21$^h$40$^m$41$\fs$90,
Decl.(J2000) = $+$58$\arcdeg$16$\arcmin$12$\farcs$3 and VLA3 at
R.A.(J2000) = 21$^h$40$^m$42$\fs$77, Decl.(J2000) =
$+$58$\arcdeg$16$\arcmin$01$\farcs$3 in Beltr\'{a}n et al. (2002). The
white cross sign marks the position of the 3.6\,cm source VLA1. The
$\times$ signs from east to west mark the positions of the 3.1\,mm
sources BIMA3, BIMA2 and BIMA1, respectively. The lines in panel (a)
show the orientation of the outflow axis from VLA2/BIMA2, with the
solid span tracing blue-shifted direction and the dashed span
red-shifted direction. The outflow axis angle is given by the
high-velocity CO(1-0) main outflow emission of Beltr\'{a}n et
al. (2002).\label{fig:IRAS21391}}
\end{figure*}

IRAS 21391+5802 is deeply embedded in the bright-rimmed globule IC
1396N located at a distance of 750 pc (Matthews 1979). This region
exhibits all of the signposts of an extremely young object, such as
strong sub-mm and mm dust continuum emission (Wilking et al. 1993;
Sugitani et al. 2000; Codella et al. 2001), line emission from
high-density gas tracers (Serabyn et al. 1993; Cesaroni et al. 1999;
Codella et al. 2001), and water maser emission (Felli et al. 1992;
Tofani et al. 1995; Patel et al. 2000; Valdettaro et
al. 2005). Sugitani et al. (1989) discovered an extended CO bipolar
outflow, which was also mapped later by Codella et al. (2001). NIR
images of the region have revealed a collimated 2.12$\mu$m H$_2$ jet
driven by IRAS 21391 (Nisini et al. 2001, Beltr\'{a}n et
al. 2009). Based on mm observations, Beltr\'{a}n et al. (2002)
resolved IRAS 21391 into an intermediate-mass source named BIMA 2,
surrounded by two less massive and smaller objects, BIMA 1 and BIMA
3. Choudhury et al. (2010) identified MIR-50 and 54 as the
mid-infrared counterparts of BIMA 2 and BIMA 3 and did not detect any
source associated with BIMA 1. The source located $\sim$ 25\arcsec
\ to the north of BIMA 2 was identified as MIR-48. BIMA 1, BIMA 2 and
BIMA 3 are all associated with 3.6 cm continuum emission (Beltr\'{a}n
et al. 2002). Figure \ref{fig:IRAS21391} shows the region as seen by
{\it Spitzer} at 8~$\rm \mu m$ and by {\it SOFIA}-FORCAST. Our
analysis focusses on the MIR-48, BIMA 2 and BIMA3 sources.

A strong CO(1-0) outflow along the east-west direction is centered at
the position of BIMA 2, and another collimated, weaker, and
smaller bipolar outflows elongated along the north-south direction are
associated with BIMA 1, which is only
detected at low velocities (see Figure 4 in Beltr\'{a}n et al. 2002). At the position of MIR-48, we see weak, overlapping blue- and red-shifted CO(1-0) emission, which is also only
detected at low velocities. There is no molecular emission detected
towards BIMA 3. The east-west outflow driven by BIMA 2 is highly
collimated, and the collimation remains even at low outflow
velocities. Beltr\'{a}n et al. (2002) interpreted the complex
morphology of the outflows as being the result of the interaction of
the high velocity gas with dense clumps surrounding the
protostar. They also suggested that BIMA 2 fits very well correlations
between source and outflow properties for low-mass Class 0 objects
given by Bontemps et al. (1996).

Neri et al. (2007) used still higher angular resolution millimeter
interferometric observations to reveal that BIMA 2 is a cluster of
multiple compact sources with the primary source named IRAM 2A. The
detection of warm CH$_3$CN in IRAM 2A implies that this is the most
massive protostar and could be the driving source of this energetic
outflow. This interpretation is also supported by the morphology of
the 1.2 mm and 3.1 mm continuum emission, which are extended along the
outflow axis tracing the warm walls of the biconical cavity (Fuente et
al. 2009). The CH$_3$CN abundance towards IRAM 2A is similar to that
found in low-mass hot corinos and lower than that expected towards IM
and high mass hot cores. Based on the low CH$_3$CN abundance, Fuente
et al. (2009) suggested that IRAM 2A is a low-mass or a Herbig Ae star
instead of the precursor of a massive Be star, or alternatively, IRAM
2A is a Class 0/I transition object that has already formed a small
photodissociation region (PDR).

For BIMA 1 and BIMA 3, Beltr\'{a}n et al. (2002) suggested they are
more evolved low-mass objects given their small dust emissivity index
and the more compact appearance of their dust emission.

While extended morphologies of the three sources are revealed in our \textit{SOFIA} images (see Fig.~\ref{fig:IRAS21391}), the extension of BIMA 2 does not follow the northeast-southwest direction of the major outflow or the north-south direction of the weak, low-velocity outflow. 

\subsection{General Results from the SOFIA Imaging}

\begin{figure*}
\epsscale{1.17}
\plotone{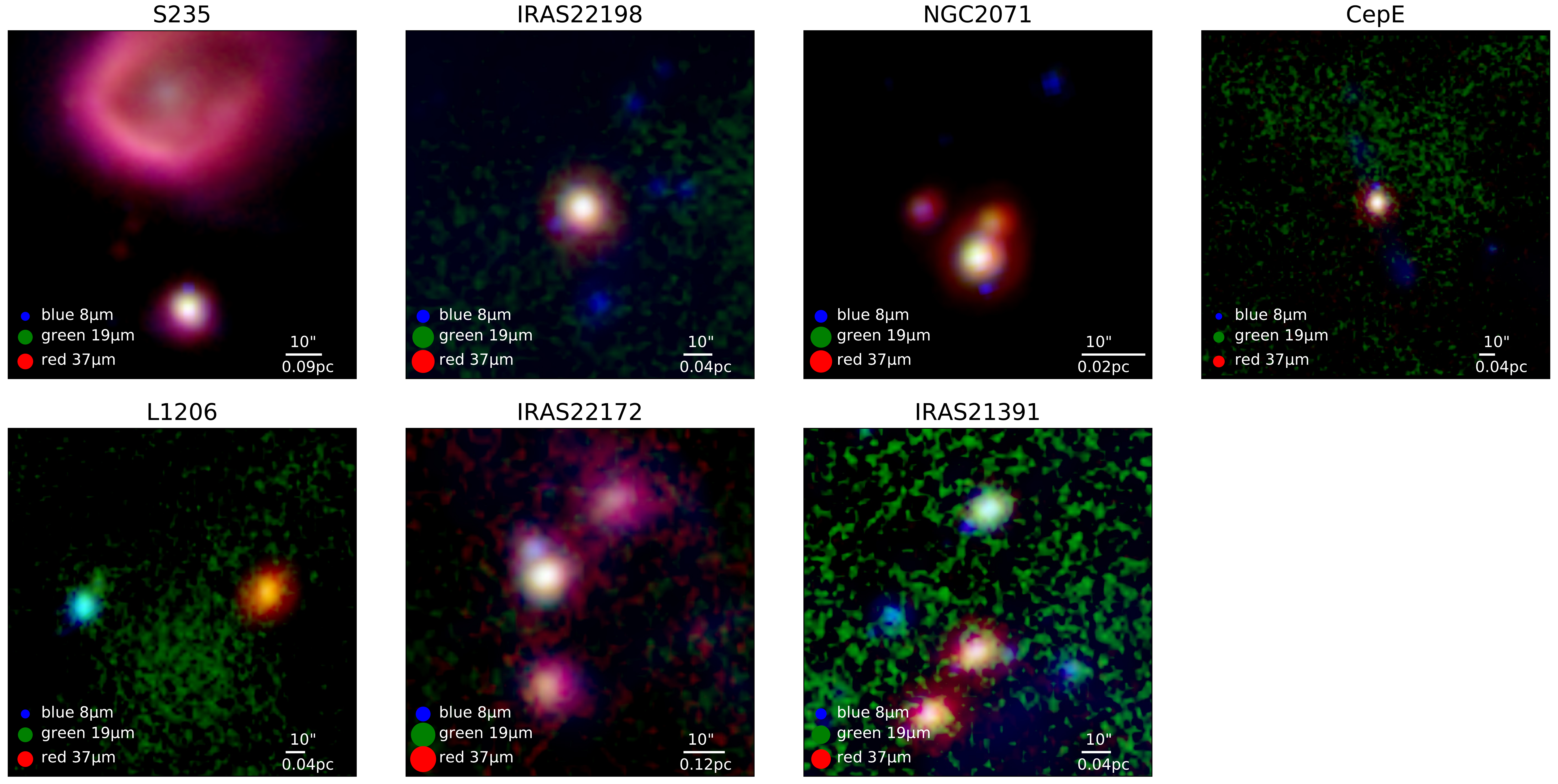}
\caption{
Gallery of RGB images of the seven new regions analyzed in this paper,
as labeled. The color intensity scales are stretched as arcsinh and
show a dynamic range of 100 from the peak emission at each wavelength.
%except for the 19\,$\mu$m image of G49.27, where only a dynamic range of 10 is shown due to its relatively low signal to noise ratio. 
The legend shows the wavelengths used and the beam sizes at these
wavelengths. {\it SOFIA}-FORCAST 37\,$\mu$m is shown in red. {\it
  SOFIA}-FORCAST 19\,$\mu$m is shown in green. {\it Spitzer} 8\,$\mu$m is shown in blue. \label{fig:rgb}}
\end{figure*}

Most of the sources presented in this paper are associated with
outflows. In a few cases, such as IRAS~22198, L1206 A and IRAS~22172 MIR2, the
SOFIA 20 to 40 $\mu$m images show modest extensions in the directions
of the outflow axes, which was a common feature of the high-mass
protostars in Papers I and II. However, the appearance of most of the
IM protostars in the {\it SOFIA} images is quite compact, i.e., only
a few beams across, and relatively round. In some of these
cases, such as IRAS 22198, Cep E and IRAS~21391 (BIMA 2) Spitzer
$8\:{\rm \mu m}$ images, which are sensitive to lower levels of
diffuse emission, do reveal outflow axis elongation, which the {\it
  SOFIA} images are not able to detect.
%and the 20 to 40 $\mu$m images of these IM protostars appear
%relatively round. This is in contrast to the high-mass protostars in
%Papers I and II, whose 20 to 40 $\mu$m images clearly show extension
%along the outflow axes.
One contributing factor here is likely to be that the IM protostars
are intrinsically less luminous than high-mass protostars and so
produce less extended MIR emission. Another factor may be that the
mass surface densities of their clump environments are lower than
those of high-mass protostars (this is revealed in the derived values
of $\Sigma_{\rm cl}$ from the SED fitting; see Section~\ref{S:fitting
  results}) and thus their MIR to FIR emission can appear more
compact and more apparently symmetric. Three-color images of all the
sources are presented together in Figure~\ref{fig:rgb}.

We notice that three of our sources are resolved into at least two
components by higher angular resolution mm observations (within $\sim$
0.01pc) including IRAS 22198, Cep E, IRAS 21391 BIMA2. A few mm
sources are detected close to the main MIR source in IRAS 22172
located 3\arcsec - 8\arcsec (0.03 - 0.09 pc) away and a few mm sources
are detected close to L1206 A located $\sim$ 12\arcsec (0.04 pc)
away. Several jet-like condensations are revealed by radio
observations in NGC 2071 IRS1 (within $\sim$ 0.01pc). This indicates
that at least some of the
%intermediate-mass or low-mass
protostars in our sample may have nearby companions.

\begin{figure*}
\epsscale{1.17}
\plotone{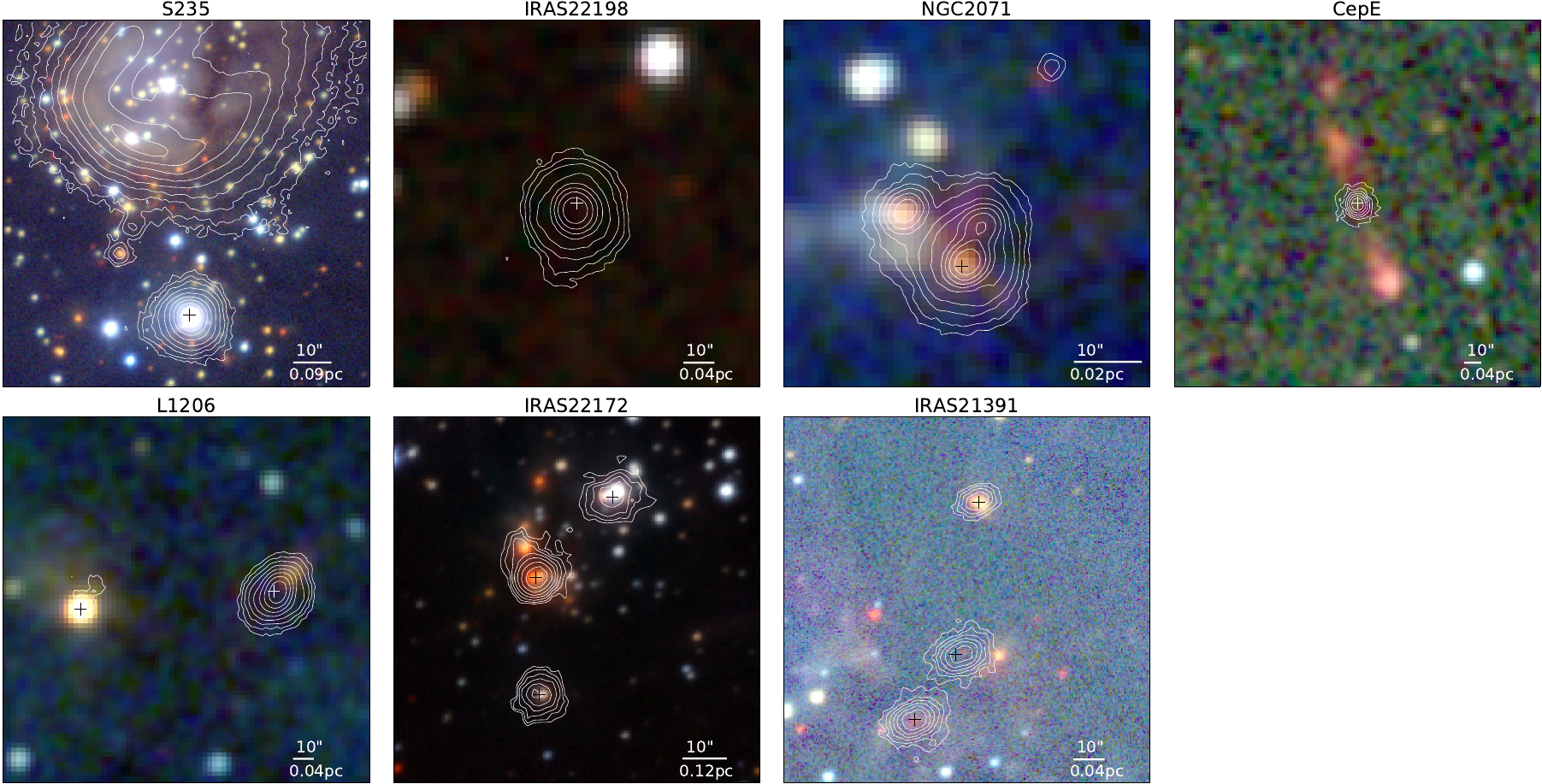}
\caption{
NIR RGB images of the seven new regions analyzed in this paper, as
labeled. The data of S235, IRAS~22172 and IRAS~21391 are from the
UKIDSS survey (Lawrence et al. 2007). The data of IRAS~22198,
NGC~2071, Cep E and L1206 are from the 2MASS survey (Skrutskie et
al. 2006). K band data are shown in red, H band data in green and J
band data in blue. The white contours are the {\it SOFIA} 37$\mu$m
emission, with the same levels as displayed in the previous individual
figures for each source. The crosses in each panel are the same as
those in the previous individual figures. The scale bar is shown in
the right corner of each panel.\label{fig:nir}}
\end{figure*}

From Figure~\ref{fig:nir}, we see that three of the sources have
high-resolution UKIDSS NIR imaging: S235, IRAS 22172 and IRAS
21391. These images show the presence of a number of NIR sources in
the vicinities of the protostars, especially for S235 and IRAS 22172,
which may be associated clusters of YSOs. On the other hand, IRAS
22198, NGC 2071, Cep E and L1206 appear more isolated in their NIR
images, although is must be noted that these images have lower
resolution and higher noise levels.
%seem to be relatively isolated, but their  and that makes their
%environments ambiguous especially like NGC 2071.
We also note that S235 B is located (in projection) near the center of
its cluster, while IRAS22172 MIR2 is closer to the eastern edge of its
cluster.

%jct5 - these figures switched in order to correspond to order they are introduced in text

\subsection{Results of SED Model Fitting}

\subsubsection{The SEDs}\label{S:SED results}

\begin{figure*}
\epsscale{1.2}
\plotone{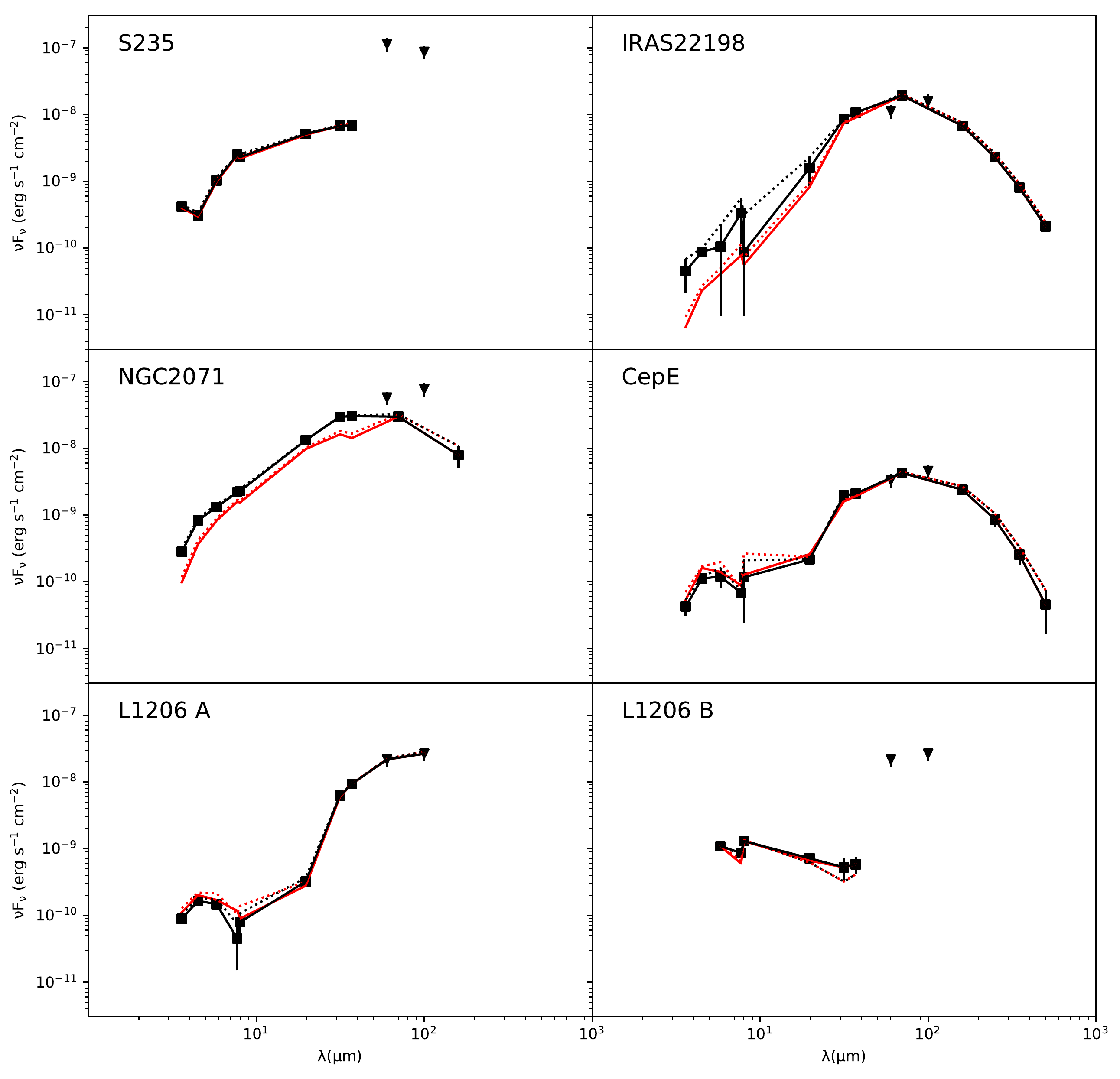}
%\vspace{-0.5in}
\caption{
SEDs of the 14 presented sources. Total fluxes with no background
subtraction applied are shown with dotted lines. The fixed aperture case
is black dotted; the variable aperture (at $<70\:{\rm \mu m}$) case is
red dotted. The background subtracted SEDs are shown with solid lines:
black for fixed aperture (the fiducial case); red for variable
aperture. Black solid squares indicate the actual measured values that
sample the fiducial SED. Black triangles denote the flux densities
measured with \textit{IRAS}. The down arrows in G305 A and IRAS16562 N denote that those data points are fluxes with no background subtraction and are treated as upper limits.
\label{fig:SEDs}}
\end{figure*}

\begin{figure*}[!htb]
    \ContinuedFloat
    \centering
    \captionsetup{list=off,format=cont, labelsep=space}
    \includegraphics[width=18cm]{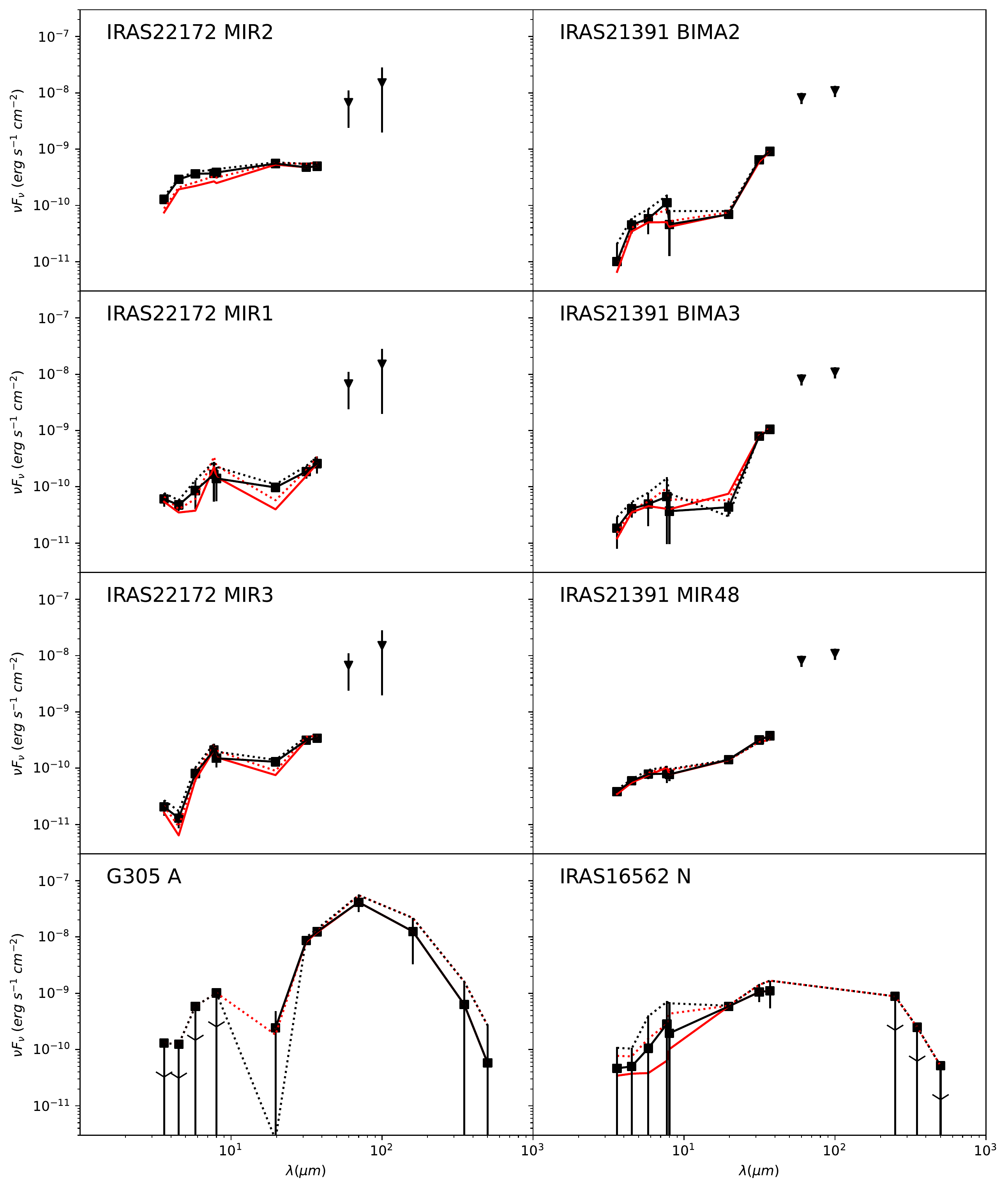}
    \caption{(cont.)}
\end{figure*}

Figure~\ref{fig:SEDs} shows the SEDs of the 14 sources presented in
this paper. There are 10 sources that lack {\it Herschel} 70 and 160
$\mu$m observations, which makes it difficult to determine the
location of the peak of their SEDs. For the remaining 4 sources,
NGC~2071 has a SED that peaks between 37 and 70~$\mu$m, while IRAS
22198, Cep E and G305 A have their peaks around 70 $\mu$m. It is
noticeable that L1206 B, IRAS22172 MIR2, IRAS22172 MIR1, IRAS21391
MIR48 and IRAS16562 N have very flat MIR SEDs, especially L1206 B
%and IRAS22172 MIR2
even shows decreasing flux densities as the wavelength increases.

\subsubsection{ZT Model Fitting Results}\label{S:fitting results}

We now consider the results of fitting the ZT protostellar radiative transfer models to the SEDs.
Note that a general comparison of differences in results when using the Robitaille et al. (2007) radiative transfer models was carried out in Paper I, with some of the main results being that the Robitaille et al. models often give solutions with very low accretion rates, which are not allowed in the context of the ZT models. As discussed in Paper I, our preference is to use the ZT models for analysis of the SOMA sources, since these models have been developed specifically for massive star formation under a physically self-consistent scenario, including full protostellar evolution, and with relatively few free parameters. Figure~\ref{fig:SEDsZT} shows the results of fitting the ZT protostellar radiative transfer models to the fixed aperture,
background-subtracted SEDs, which is the fiducial analysis method
presented in Papers I and II. In general, reasonable fits can be found
to the observed SEDs, i.e., with relatively low values of reduced
$\chi^2$.
%jct2 - we need to be careful about the definition of \chi^2. I think we agreed it is defined already as the reduced chi^2, i.e., already divided by N. So I have removed the N here.
%Then the header of column 2 in Table 4 should be just $\chi^2$. The scale bar label of Fig. 12 is chi^2, as it is.
%jct2 - note, there is then a mistake in the table header of column 2 in Paper II. Let's be careful!!!

A summary of fitted parameter results in the $\Sigma_{\rm cl}$ - $M_c$
- $m_*$ parameter space is shown for each source in
Figure~\ref{fig:primary}. Note that the clump environment mass surface
density, $\Sigma_{\rm cl}$ (ranging from 0.1 to 3~$\rm g\:cm^{-2}$),
and initial core mass, $M_c$ (ranging from 10 to 480~$M_\odot$), are
the primary physical parameters of the initial conditions of the ZT
models, while the current protostellar mass, $m_*$ (ranging from
0.5~$M_\odot$ up to about 50\% of $M_c$, with this efficiency set by
disk wind driven outflow feedback), describes the evolutionary state
of stars forming from such cores. The two other independent parameters
of the models are the angle of the line of sight to the outflow axis,
$\theta_{\rm view}$, and the amount of foreground extinction, $A_V$,
with all other model parameters being completely specified by
$\Sigma_{\rm cl}$, $M_c$, and $m_*$. Note that $L_{\rm bol,iso}$ represents the isotropic bolometric luminosity, i.e., without correction for the inclination, and $L_{\rm bol}$ represents the intrinsic bolometric luminosity. The best five model fits for each source are listed in Table~\ref{tab:models}. Note that $\chi^2$ listed
in this table is the reduced $\chi^2$, i.e., already normalized by the
number of data points used in the fitting. Note, also that Table~4 of
Paper II listed, incorrectly, this as quantity as $\chi^2/N$, rather
than as $\chi^2$ used here and in Paper I.

The best-fit models indicate that S235 and G305 A are more likely to
be high-mass protostars, with most of the models (except the best
model for S235) returning protostellar masses $m_{*} \geq 12 \:
M_{\odot}$, accretion rates $\dot{m}_{*} \sim 10^{-5} - \rm a \ few
\ \times 10^{-4} \: M_{\odot} \: {\rm yr}^{-1}$, initial core masses
$M_{c} \sim 50-400 \: M_{\odot}$, clump mass surface densities
$\Sigma_{\rm cl} \sim 0.1-1 \: \rm g \ cm^{-2}$, and isotropic
luminosities $L_{\rm bol,iso} \sim 10^{3} - \rm a \ few \ \times 10^{4} \:
L_{\odot}$.

We find that IRAS 22198, NGC 2071, L1206 A, L1206 B, IRAS22172 MIR2,
IRAS22172 MIR3, IRAS21391 MIR48, and IRAS16562 N are likely to {\it
  currently} be intermediate-mass protostars, with most models
returning protostellar masses $m_{*} \sim 2 - 8 \ M_{\odot}$,
%jct3 - this range should be up to 8 Msun for intermediate mass... we say ``most'' models.
accretion rates $\dot{m}_{*} \sim 10^{-5} - 10^{-4} \: M_{\odot} \:
{\rm yr}^{-1}$, initial core masses $M_{c}$ ranging from 10 to 480
$M_{\odot}$, clump mass surface densities $\Sigma_{\rm cl}$ ranging
from 0.1 to 3.2 g cm$^{-2}$, and isotropic luminosities $L_{\rm bol,iso}
\sim 10 - \rm a \ few \ \times 10^{2} \: L_{\odot}$. However, given
the estimated remaining envelope masses around these protostars, for
many models the final outcome would be a massive star, since star
formation efficiencies are typically $\sim50\%$ in the models (see
also Tanaka et al. 2017; Staff et al. 2019).

Considering the remaining sources, we see that Cep E, IRAS22172 MIR1,
IRAS21391 BIMA2, IRAS21391 BIMA3 are likely to {\it currently} be
low-mass protostars, with most models returning protostellar masses
$m_{*} \sim 0.5 - 2 \ M_{\odot}$, accretion rates $\dot{m}_{*} \sim
10^{-5} - 10^{-4} \: M_{\odot} \: {\rm yr}^{-1}$, initial core masses
$M_{c}$ ranging from 10 to 160 $M_{\odot}$, the clump mass surface
densities $\Sigma_{\rm cl}$ ranging from 0.1 to 0.3 g cm$^{-2}$, and
isotropic luminosities $L_{\rm bol,iso} \sim 10^{2} \: L_{\odot}$. Given
that the models used for the fitting all have initial core masses of
$10\:M_\odot$ or greater, then the outcome of the evolution would
always be formation of at least intermediate-mass stars. However,
within the degeneracies of the model fits, there are some solutions
that would imply we are catching a massive star in the very earliest
stages of its formation.

Below, we describe the fitting results of each individual source and
compare then with previous estimates from the literature.

\begin{figure*}
\epsscale{1.10}
\plotone{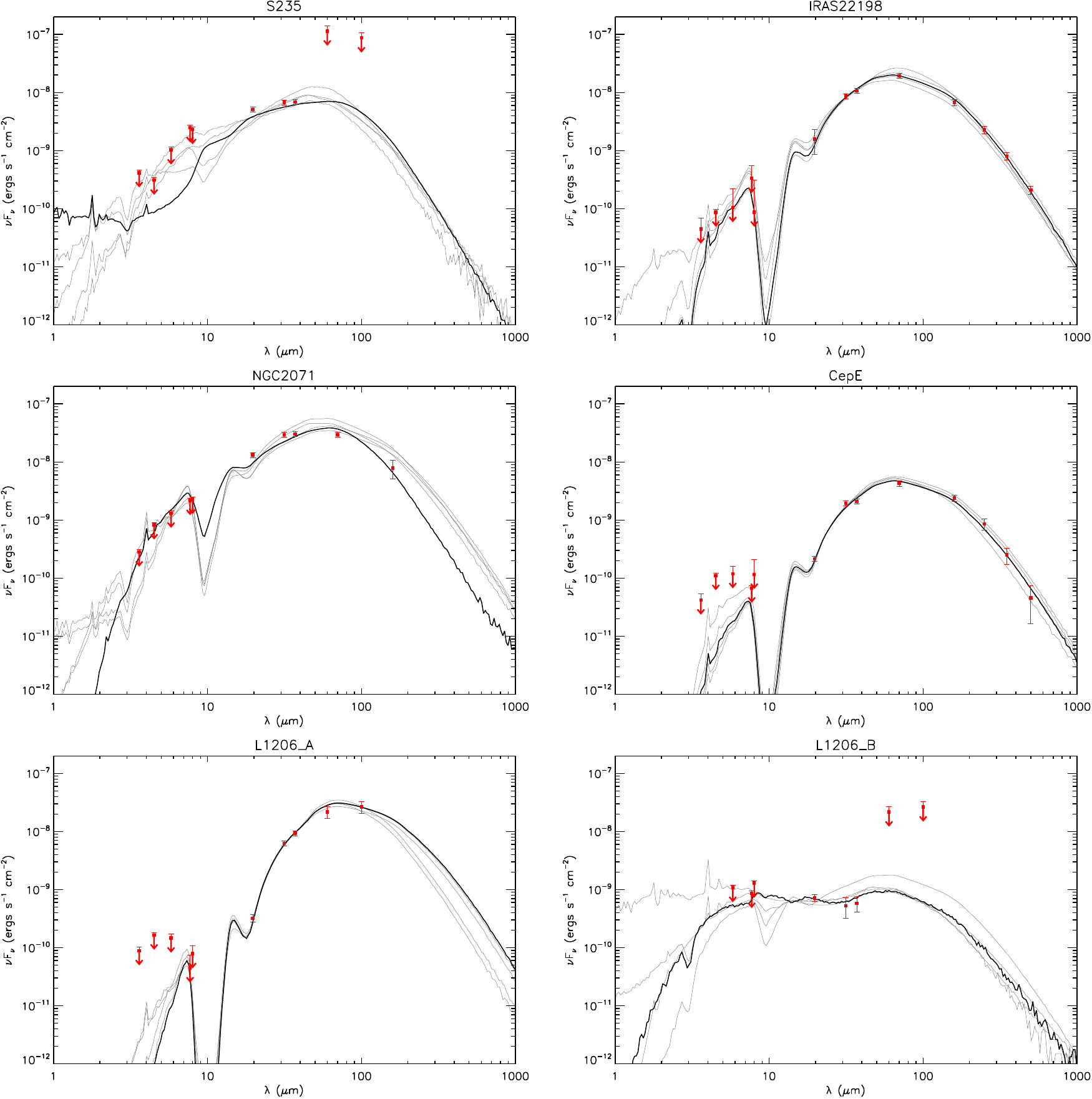}
%\vspace{-3cm}
\caption{
Protostar model fitting to the fixed aperture, background-subtracted
SED data using the ZT model grid. For each source, the best fit model
is shown with a solid black line and the next four best models are
shown with solid gray lines. Flux values are those from
Table~\ref{tab:flux}. Note that the data at $\lesssim8\:{\rm \mu m}$
are treated as upper limits (see text). The resulting model parameter
results are listed in Table~\ref{tab:models}.\label{fig:SEDsZT}}
\end{figure*}

\begin{figure*}[!htb]
    \ContinuedFloat
    \centering
    \captionsetup{list=off,format=cont, labelsep=space}
    \includegraphics[width=18cm]{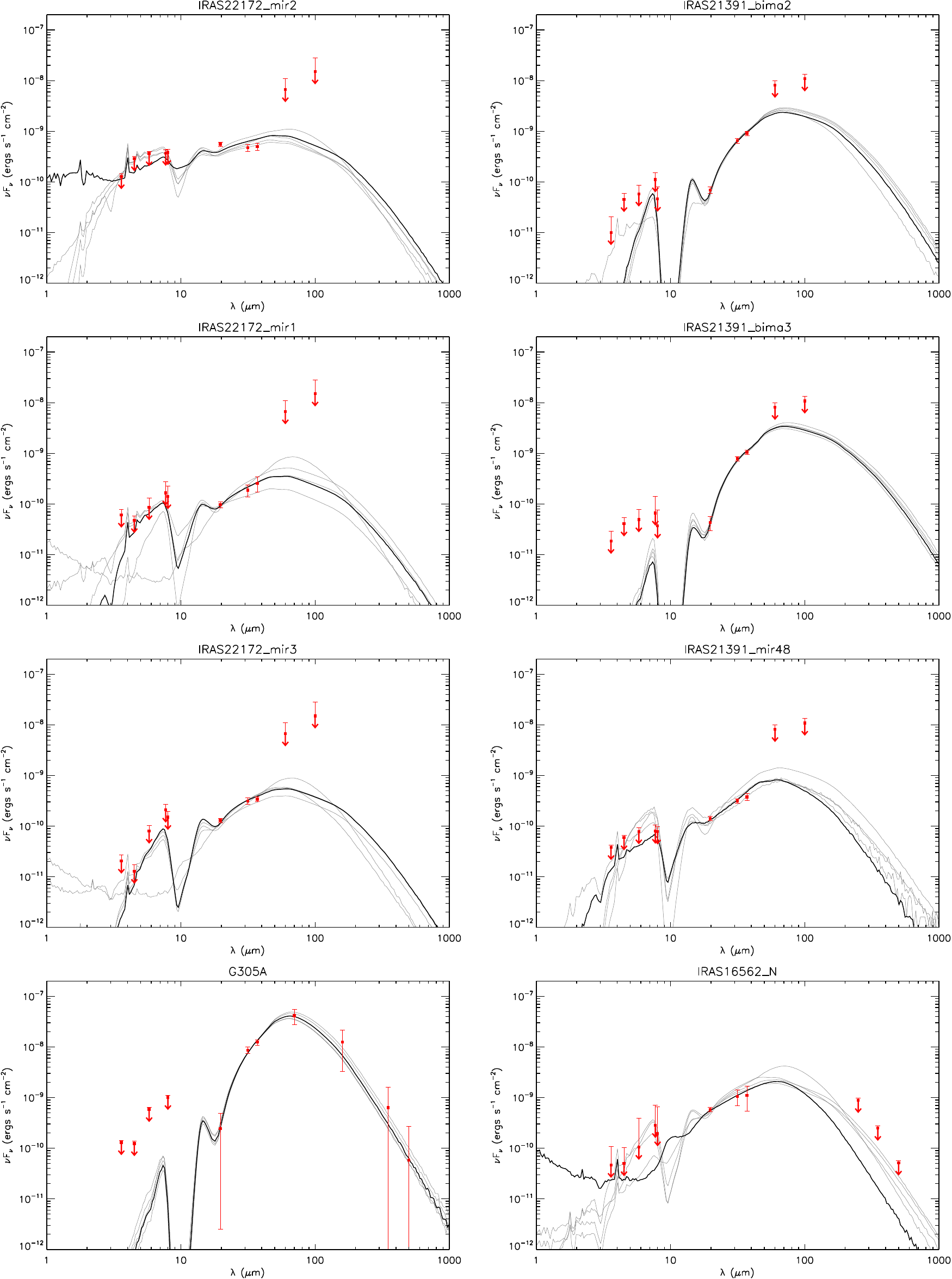}
    \caption{(cont.)}
\end{figure*}

\textbf{S235:} From the best five model fits, this source has an
estimated isotropic bolometric luminosity of $\sim1$ to
$2\times10^3\:L_\odot$. However, the intrinsic bolometric luminosity
of these models spans a much wider range from $3\times 10^3$ to
$2\times 10^5\:L_\odot$. We note that for this source there are
effectively only three measurements of the SED, all from the {\it
  SOFIA} FORCAST data, with observations at other wavelengths being
used as upper limits. The large intrinsic luminosities for this source
are possible because of the ``flashlight effect'', i.e., most of the
flux is not directed towards us due to high local extinction in the
core.  This range of intrinsic luminosities means that there is a wide
range of protostellar properties that are consistent with the observed
SED, i.e., there are significant degeneracies in the derived
protostellar parameters (see Fig.~\ref{fig:primary}). In particular,
while the best fit model has a low initial core mass ($10\:M_\odot$)
and current protostellar mass ($2\:M_\odot$) forming from a high
$\Sigma_{\rm cl}$ environment ($3\:{\rm g\:cm}^{-2}$) that is viewed
at a relatively small angle to the outflow axis, the next four best
models are all with larger core and protostellar masses in lower
density environments viewed at angles nearly orthogonal to the outflow
axis, i.e., close to the equatorial plane where there would be the
most line of sight extinction.
%first appears to be an intermediate-mass source given the low initial
%core mass ranging from 10 to 80 $M_{\odot}$ and the low luminosity
%$\sim 10^{3} L_{\odot}$. However,
%the stellar mass and the clump surface density vary a lot between the
%best five models.

Among previous studies of S235, Felli et al. (2006) used JHK band
images and MSX fluxes and derived a luminosity of $410 L_{\odot}$,
which they claimed must be considered to be a lower limit because the
FIR part of the spectrum is not taken into account in their
calculation. Dewangan \& Anandarao (2011) used JHK band images
and 2MASS and IRAC fluxes to do SED fitting with models from
Robitaille et al. (2006, 2007). They derived $m_{*} \sim 6.5
M_{\odot}$, $L_{\rm bol} \sim 575 L_{\odot}$ and $M_{\rm env} \sim 9
M_{\odot}$.
%Four of our best five models return a stellar mass higher than 12
%$M_{\odot}$, which indicates a massive protostar.
The stellar source itself has been classified as a B1V star by Boley
et al. (2009), with emission-line profiles indicative of an accretion
disk.
%In the model grid there are only $\sim$ 30 models returning a surface
%temperature consistent with a B1V star. In the 10 best models of S235,
%only the 8th best model has a $T_{*} \ \sim$ 25000 K.
Based on the intensity of the reflected component, it was concluded
that the accretion disk must be viewed nearly edge-on, which agrees
with four of our best models and explains the discrepancy between
$L_{\rm bol,iso}$ and $L_{\rm bol}$.
%Felli et al. (1997) derived a high visual extinction of $A_{v} \approx$ 8-12 mag. It can also be due
%to an edge-on disk.
Boley et al. (2009) estimated a mass accretion rate of $2-6 \times
10^{-6} M_{\odot} \rm yr^{-1}$ for a B1V star with a mass of 13
$M_{\odot}$ using the Br$\gamma$ luminosity, which is comparable with
the mass-loss rate of $4 \times 10^{-6} M_{\odot} \rm yr^{-1}$ derived
by Felli et al. (2006) from the radio flux density. However, our best
models have disk accretion rates more than ten times higher. It
should be noted that the accretion rate is not a free parameter in the
ZT models and that the range of accretion rates is generally
relatively high, being set by the properties of the initial cores and
the mass surface density of their clump environments.

%There are only 3 models that return a disk accretion rate lower than
%$10^{-5} M_{\odot} \rm yr^{-1}$ in the whole model grid and they all
%have very large $\chi^{2}$ for S235.

\textbf{IRAS~22198:} The best models are those with a protostar with
current mass of 2 - 4 $M_{\odot}$, forming in a low mass surface
density clump (0.1 - 0.3 $\rm g \ cm^{-2}$). Our estimate of the
isotropic luminosity is about 600~$L_\odot$, with the intrinsic
luminosity being about $800\:L_\odot$. S\'{a}nchez-Monge et al. (2010)
fit the SED of IRAS 22198 from NIR to centimeter wavelengths with a
modified blackbody plus a thermal ionized wind and derived a
bolometric luminosity of $\sim$370 $L_{\odot}$ and an envelope mass of
$\sim$5 $M_{\odot}$, remarking that the SED of IRAS 22198 resembles
that of Class 0 objects (Andre et al. 1993). Our derived isotropic
luminosity is slightly higher, while our envelope mass is much higher,
$\sim50\:M_\odot$, than their results. However, their $M_{\rm env}$
was derived from interferometric flux measurements and thus should be
treated as a lower limit. The single-dish measurement at mm
wavelengths of the dense core mass is 17 $M_{\odot}$ within a radius
of 2,650~au (3.5\arcsec) (Palau et al. 2013). Thus the reason for our
larger mass estimate is likely due to our analysis applying to a much
larger scale, i.e., within a radius of 0.089~pc (26\arcsec).
%, again much smaller than our envelope mass. In the 432 models there
%are only 9 models with $M_{\rm env}$ lower than 3 $M_{\odot}$. While
%their $M_c$ ranges from 10 to 100 $M_{\odot}$ and their $\Sigma_{\rm
%  cl}$ ranges from 0.1 to 3.2 $\rm g \ cm^{-2}$, which appear similar
%to the range of the best models, those models typically have
%relatively large $m_{*}$ ranging from 4 to 48 $M_{\odot}$.
%This is expected since high stellar mass leads to low remaining mass
%in the envelope. However, as indicated from
%Figure~\ref{fig:m_sigma_ms}, models with high $m_{*}$ have very large
%$\chi^{2}$.
%$M_c$ sampled by ZT models start from 10 $M_{\odot}$ so there are few
%models with both low $m_{*}$ and low $M_{\rm env}$. Thus smaller $M_c$
%may be favored in this case.

\textbf{NGC~2071:} The best models suggest a currently
intermediate-mass protostar with a mass of 2 - 4 $M_{\odot}$ forming
within a core with initial mass of 10 - 60 $M_{\odot}$. Trinidad et
al. (2009) estimated a central mass of $\sim 5 \pm 3 M_{\odot}$ for
IRS1 and $\sim 1.2 \pm 0.4 M_{\odot}$ for IRS3 based on the observed
velocity gradient of the water masers, which is consistent with our
estimate.
%seems consistent with the luminosity of 520 $L_{\odot}$ for the
%$\sim$30$\arcsec$ diameter infrared cluster NGC2071 (Butner et
%al. 1990), and also consistent with our results.
The single-dish measurement at mm wavelength of the dense core mass is
39 $M_{\odot}$ within a radius of 4,700~AU (11\arcsec) (Palau et
al. 2013), which is similar to the $M_{\rm env}$ returned by most of
our best fit models inside 10\arcsec.

\textbf{Cep E:} The best 5 models all return a $\Sigma_{\rm cl}$ of
0.1 $\rm g \ cm^{-2}$ and most models have $m_{*}$ as low as 1 - 2
$M_{\odot}$. Crimier et al. (2010) modeled the MIR to mm SED with the
1D radiative transfer code DUSTY and derived a luminosity of $\sim$100
$L_{\odot}$ and an envelope mass of 35 $M_{\odot}$, which are similar
to our results.

\textbf{L1206:} The best models of L1206 A involve a protostar forming
inside a relatively massive initial core (40 - 480 $M_{\odot}$) with
low clump mass surface density (0.1 - 0.3 $\rm g \ cm^{-2}$). All the
best 5 models give a value of $m_{*}=4\:M_{\odot}$. Ressler \& Shure
(1991) found a total luminosity of 1100 $L_{\odot}$ by fitting four
IRAS fluxes plus the 2.7 mm data of Wilking et al. (1989) with a
single-temperature dust spectrum at 1~kpc, which is similar to our
result. Beltr\'{a}n et al. (2006) estimated the core mass of OVRO 2 to
be 14.2 $M_{\odot}$ from the 2.7~mm dust continuum emission at a
distance of 910 pc. This core mass estimate is derived from
interferometric observations that may be missing flux, and indeed three of
our best-fit models give a much higher value of $M_{\rm env}$.
%There are $\sim$ 50 models with $M_{\rm env} < 15 M_{\odot}$, and most
%of them have $M_c \la 50 \ M_{\odot}$ and $m_{*} \ga 8
%\ M_{\odot}$.
Ressler \& Shure (1991) suggested that L1206 A is seen only in scattered
light because of heavy obscuration by an almost edge-on circumstellar
disk. Four of the best five models return a nearly edge-on line of
sight.

L1206 B has a very flat and slightly decreasing SED at short
wavelengths. A circumstellar disk could explain the infrared excess,
as suggested by Ressler \& Shure (1991), and the protostar may have
already cleared a significant portion of its envelope, thus explaining
the decreasing spectrum between 10 and 30 $\mu$m. The favored ZT
models have a wide range of stellar mass $m_{*} \sim 0.5-12
\ M_{\odot}$, but low initial core mass $M_{\rm c} \sim 10-40
\ M_{\odot}$, low current envelope mass of 1 to 9~$M_\odot$ and low
mass surface density $\Sigma_{\rm cl} \sim 0.1-0.3\: \rm g \ cm^{-2}$
of the clump environment.

\textbf{IRAS~22172:} The models for the three MIR sources all involve
protostars with masses $\sim$ 1 - 4 $M_{\odot}$ forming in relatively
low-mass initial cores of 10 - 40 $M_{\odot}$. Fontani et al. (2004)
divided the SED between the NIR cluster and the cold 3.4 mm core
(their I22172-C) and performed two grey-body fits to the SED. The
grey-body fit to the MSX and IRAS data with $\lambda \leqslant 25
\mu$m, which represent the emission due to the cluster of stars
surrounding the mm core I22172-C, yields a luminosity of $2.2 \times
10^2\:L_{\odot}$. Based on the beam size and the MSX 21$\mu$m
emission, their photometry should cover the whole field, i.e., all the
three MIR sources. However, in our analysis we derive a much higher
combined luminosity from the region, with contributions from the three
MIR sources analyzed.
%their luminosity estimate is only comparable to one source estimated here.
%For the mm core, they use data after 25$\mu$m from IRAS, SCUBA, OVRO, and VLA have derived the following best fit parameters: dust temperature = 27 K, angular diameter of the source = 9\arcsec, envelope mass = $83 M_{\odot}$, luminosity = $2.2 \times 10^3 L_{\odot}$. From their Figure 8, the second half of the SED should cover an area with $\sim$ 10\arcsec diameter, which overlaps partly with MIR2.
The single-dish measurement at mm wavelengths of the dense core mass
of MIR2 is 150 $M_{\odot}$ (Palau et al. 2013), much higher than the
$M_{\rm env}$ given by our models. However, their core radius,
represented by the deconvolved FWHM/2, is about 10\arcsec, while our
mass estimate is based on an aperture radius of 4\arcsec.

%which means they measured a larger area than the envelope of MIR2
%alone, but still even the initial core mass favored by our models is
%much smaller than 150 $M_{\odot}$.

%jct3 - this footnote is not appearing on the same page. - try and fix.... I tried moving the footnote
\textbf{IRAS~21391:} Previous SED fitting with low-resolution data
estimated the bolometric luminosity of IRAS~21391 to range from 235
$L_{\odot}$ (Saraceno et al. 1996) to 440 $L_{\odot}$ (Sugitani et
al. 2000). Our fitting results for the three sources BIMA 2, BIMA 3
and MIR 48\footnote{Note that we follow the nomenclature in
  Beltr\'{a}n et al. (2002), but the photometry centers of IRAS 21391
  BIMA2 and IRAS 21391 BIMA3 are VLA2 and VLA3, respectively.} all
return isotropic luminosities $\la$ 100 $L_{\odot}$. By using the
relationship between the momentum rate and the bolometric luminosity
(Cabrit \& Bertout 1992), Beltr\'{a}n et al. (2002) inferred a
bolometric luminosity of 150 $L_{\odot}$ for BIMA 2.

Choudhury et al. (2010) fit the 1 - 24 $\mu$m SED derived from optical
BVRI, Spitzer IRAC and MIPS observations with Robitaille et al. (2007)
models and derived a luminosity of 197 $L_{\odot}$ and a stellar mass
of 6 $M_{\odot}$ for BIMA 2 (their MIR-50), which are both higher than
our results. As indicated by Figure~\ref{fig:m_sigma_ms}, ZT models
with $m_{*}$ higher than 5 $M_{\odot}$ have a very large
$\chi^{2}$. The envelope mass of Choudhury et al. (2010) of 41
$M_{\odot}$ is also slightly higher than the $M_c$ and $M_{\rm env}$
in our first 3 best models. However, their disk accretion rate is
about 1000 times lower than that in our best models, which is a known
issue when comparing Robitaille et al. (2007) and ZT models (see
discussion in De Buizer et al. 2017). Beltr\'{a}n et al. (2002)
estimated the circumstellar mass to be 5.1 $M_{\odot}$ based on BIMA
3.1 mm continuum observations, which should be treated as a lower
limit of $M_{\rm env}$ given that it is an interferometric measurement
subject to missing flux. Beltr\'{a}n et al. (2002) suggested that the
axis of the outflow should be close to the plane of the sky, given the
morphology of the CO(1-0) outflows at low velocities with blue-shifted
and redshifted gas in both lobes. However, in our best 5 models, only
the third model has a more edge-on inclination.

Our best models for IRAS 21391 BIMA3 involve a protostar with a current
stellar mass of 0.5 $M_{\odot}$ with a bolometric luminosity $\sim$
100 $L_{\odot}$.
%It is not suggestive from our results that BIMA 3 is
%less massive than BIMA 2.
The best-fit model in Choudhury et al. (2010) for BIMA 3 (their
MIR-54) yields a luminosity of 33.4 $L_{\odot}$ and a stellar mass of
1.5 $M_{\odot}$. Beltr\'{a}n et al. (2002) derived a circumstellar
mass of 0.07 $M_{\odot}$ for BIMA 3, which is much lower than the
predicted $M_{\rm env}$ by our best models.

Our best models for IRAS 21391 MIR48 involve a protostar with a mass
ranging from 1 to 12 $M_{\odot}$. The best-fit model in Choudhury et
al. (2010) for MIR-48 yields a luminosity of 280 $L_{\odot}$ and a
stellar mass of 5 $M_{\odot}$, which is similar to the isotropic
luminosity and the stellar mass in our best two models.

%jct2 - I don't think a clearpage command is good if we want to make a nice PDF for arXiv.
%jct2 - I changed chi^2/N -> chi^2
%\clearpage
\renewcommand{\arraystretch}{0.9}
\startlongtable
\begin{deluxetable*}{ccccccccccccc}
\tabletypesize{\scriptsize}
\tablecaption{Parameters of the Best Five Fitted Models \label{tab:models}} 
\tablewidth{18pt}
\tablehead{
\colhead{Source} &\colhead{$\chi^{2}$} & \colhead{$M_{\rm c}$} & \colhead{$\Sigma_{\rm cl}$} & \colhead{$R_{\rm core}$}  &\colhead{$m_{*}$} & \colhead{$\theta_{\rm view}$} &\colhead{$A_{V}$} & \colhead{$M_{\rm env}$} &\colhead{$\theta_{w,\rm esc}$} & \colhead{$\dot {M}_{\rm disk}$} & \colhead{$L_{\rm bol, iso}$} & \colhead{$L_{\rm bol}$} \\
\colhead{} & \colhead{} & \colhead{($M_\odot$)} & \colhead{(g $\rm cm^{-2}$)} & \colhead{(pc) ($\arcsec$)} & \colhead{($M_{\odot}$)} & \colhead{(\arcdeg)} & \colhead{(mag)} & \colhead{($M_{\odot}$)} & \colhead{(deg)} &\colhead{($M_{\odot}$/yr)} & \colhead{($L_{\odot}$)} & \colhead{($L_{\odot}$)} \\
 \vspace{-0.4cm}
}
\startdata
S235
& 1.26 & 10 & 3.2 & 0.013 ( 2 ) & 2.0 & 39 & 0.0 & 6 & 35 & 1.8$\times 10^{-4}$ & 1.4$\times 10^3$ & 2.6$\times 10^3$ \\
$d$ = 1.8 kpc
& 2.55 & 60 & 1.0 & 0.057 ( 7 ) & 24.0 & 89 & 11.1 & 5 & 71 & 1.9$\times 10^{-4}$ & 2.1$\times 10^3$ & 9.3$\times 10^4$ \\
$R_{ap}$ = 12 \arcsec
& 2.74 & 50 & 0.1 & 0.165 ( 19 ) & 12.0 & 89 & 4.0 & 15 & 59 & 3.4$\times 10^{-5}$ & 1.4$\times 10^3$ & 1.4$\times 10^4$ \\
= 0.10 pc
& 3.00 & 80 & 1.0 & 0.066 ( 8 ) & 32.0 & 89 & 15.2 & 3 & 79 & 1.4$\times 10^{-4}$ & 1.6$\times 10^3$ & 1.6$\times 10^5$ \\
& 3.02 & 50 & 0.3 & 0.093 ( 11 ) & 16.0 & 80 & 0.0 & 8 & 68 & 7.1$\times 10^{-5}$ & 1.4$\times 10^3$ & 3.1$\times 10^4$ \\
\hline\noalign{\smallskip}
IRAS22198
& 0.18 & 80 & 0.1 & 0.208 ( 56 ) & 4.0 & 89 & 29.3 & 71 & 18 & 3.7$\times 10^{-5}$ & 6.0$\times 10^2$ & 8.5$\times 10^2$ \\
$d$ = 0.8 kpc
& 0.27 & 60 & 0.1 & 0.180 ( 49 ) & 4.0 & 62 & 41.4 & 51 & 21 & 3.4$\times 10^{-5}$ & 6.1$\times 10^2$ & 8.9$\times 10^2$ \\
$R_{ap}$ = 26 \arcsec
& 1.08 & 100 & 0.1 & 0.233 ( 63 ) & 4.0 & 89 & 35.4 & 91 & 15 & 4.0$\times 10^{-5}$ & 6.5$\times 10^2$ & 8.8$\times 10^2$ \\
= 0.09 pc
& 1.47 & 40 & 0.3 & 0.083 ( 22 ) & 2.0 & 22 & 9.1 & 35 & 17 & 5.3$\times 10^{-5}$ & 6.5$\times 10^2$ & 7.5$\times 10^2$ \\
& 1.78 & 50 & 0.1 & 0.165 ( 44 ) & 4.0 & 62 & 25.3 & 41 & 24 & 3.2$\times 10^{-5}$ & 5.1$\times 10^2$ & 7.9$\times 10^2$ \\
\hline\noalign{\smallskip}
NGC2071
& 3.14 & 10 & 3.2 & 0.013 ( 7 ) & 4.0 & 58 & 57.6 & 2 & 56 & 1.9$\times 10^{-4}$ & 5.0$\times 10^2$ & 1.9$\times 10^3$ \\
$d$ = 0.4 kpc
& 3.59 & 30 & 0.1 & 0.127 ( 67 ) & 4.0 & 65 & 12.1 & 21 & 33 & 2.7$\times 10^{-5}$ & 3.6$\times 10^2$ & 7.7$\times 10^2$ \\
$R_{ap}$ = 10 \arcsec
& 5.79 & 40 & 0.1 & 0.147 ( 78 ) & 4.0 & 62 & 11.1 & 30 & 27 & 3.0$\times 10^{-5}$ & 4.4$\times 10^2$ & 7.5$\times 10^2$ \\
= 0.02 pc
& 7.06 & 60 & 0.1 & 0.180 ( 95 ) & 2.0 & 29 & 0.0 & 55 & 15 & 2.5$\times 10^{-5}$ & 3.2$\times 10^2$ & 3.5$\times 10^2$ \\
& 7.57 & 50 & 0.1 & 0.165 ( 87 ) & 2.0 & 29 & 0.0 & 46 & 16 & 2.4$\times 10^{-5}$ & 2.8$\times 10^2$ & 3.1$\times 10^2$ \\
\hline\noalign{\smallskip}
CepE
& 0.63 & 30 & 0.1 & 0.127 ( 36 ) & 1.0 & 83 & 29.3 & 27 & 15 & 1.5$\times 10^{-5}$ & 1.3$\times 10^2$ & 1.7$\times 10^2$ \\
$d$ = 0.7 kpc
& 0.70 & 30 & 0.1 & 0.127 ( 36 ) & 2.0 & 65 & 60.6 & 25 & 23 & 2.0$\times 10^{-5}$ & 1.5$\times 10^2$ & 2.4$\times 10^2$ \\
$R_{ap}$ = 23 \arcsec
& 0.80 & 40 & 0.1 & 0.147 ( 42 ) & 1.0 & 89 & 21.2 & 38 & 12 & 1.6$\times 10^{-5}$ & 1.3$\times 10^2$ & 1.7$\times 10^2$ \\
= 0.08 pc
& 1.40 & 50 & 0.1 & 0.165 ( 46 ) & 1.0 & 89 & 19.2 & 48 & 11 & 1.7$\times 10^{-5}$ & 1.4$\times 10^2$ & 1.7$\times 10^2$ \\
& 1.67 & 20 & 0.1 & 0.104 ( 29 ) & 4.0 & 71 & 100.0 & 10 & 43 & 2.1$\times 10^{-5}$ & 1.9$\times 10^2$ & 6.8$\times 10^2$ \\
\hline\noalign{\smallskip}
L1206 A
& 0.08 & 480 & 0.1 & 0.510 ( 136 ) & 4.0 & 89 & 45.5 & 474 & 6 & 6.1$\times 10^{-5}$ & 9.2$\times 10^2$ & 1.0$\times 10^3$ \\
$d$ = 0.8 kpc
& 0.09 & 400 & 0.1 & 0.465 ( 124 ) & 4.0 & 83 & 56.6 & 390 & 7 & 5.8$\times 10^{-5}$ & 9.4$\times 10^2$ & 1.0$\times 10^3$ \\
$R_{ap}$ = 9 \arcsec
& 0.17 & 50 & 0.3 & 0.093 ( 25 ) & 4.0 & 55 & 41.4 & 41 & 22 & 7.7$\times 10^{-5}$ & 8.8$\times 10^2$ & 1.4$\times 10^3$ \\
= 0.03 pc
& 0.21 & 40 & 0.3 & 0.083 ( 22 ) & 4.0 & 89 & 28.3 & 31 & 25 & 7.2$\times 10^{-5}$ & 7.3$\times 10^2$ & 1.4$\times 10^3$ \\
& 0.23 & 240 & 0.1 & 0.360 ( 96 ) & 4.0 & 89 & 74.7 & 229 & 9 & 5.1$\times 10^{-5}$ & 9.0$\times 10^2$ & 1.0$\times 10^3$ \\
\hline\noalign{\smallskip}
L1206 B
& 0.13 & 40 & 0.1 & 0.147 ( 39 ) & 12.0 & 89 & 8.1 & 2 & 82 & 9.5$\times 10^{-6}$ & 5.7$\times 10^1$ & 1.1$\times 10^4$ \\
$d$ = 0.8 kpc
& 0.45 & 30 & 0.3 & 0.072 ( 19 ) & 12.0 & 89 & 30.3 & 1 & 81 & 2.2$\times 10^{-5}$ & 7.0$\times 10^1$ & 1.2$\times 10^4$ \\
$R_{ap}$ = 10 \arcsec
& 0.55 & 10 & 0.3 & 0.041 ( 11 ) & 4.0 & 77 & 0.0 & 1 & 68 & 2.4$\times 10^{-5}$ & 4.9$\times 10^1$ & 6.7$\times 10^2$ \\
= 0.04 pc
& 0.71 & 10 & 0.1 & 0.074 ( 20 ) & 2.0 & 51 & 0.0 & 4 & 50 & 1.1$\times 10^{-5}$ & 8.1$\times 10^1$ & 1.3$\times 10^2$ \\
& 2.26 & 10 & 0.1 & 0.074 ( 20 ) & 0.5 & 22 & 34.3 & 9 & 20 & 7.8$\times 10^{-6}$ & 1.5$\times 10^2$ & 7.5$\times 10^1$ \\
\hline\noalign{\smallskip}
IRAS22172 MIR2
& 1.67 & 40 & 0.1 & 0.147 ( 13 ) & 2.0 & 22 & 0.0 & 36 & 19 & 2.2$\times 10^{-5}$ & 3.9$\times 10^2$ & 2.7$\times 10^2$ \\
$d$ = 2.4 kpc
& 2.27 & 30 & 0.1 & 0.127 ( 11 ) & 2.0 & 22 & 32.3 & 25 & 23 & 2.0$\times 10^{-5}$ & 8.0$\times 10^2$ & 2.4$\times 10^2$ \\
$R_{ap}$ = 4 \arcsec
& 2.39 & 20 & 0.1 & 0.104 ( 9 ) & 4.0 & 48 & 6.1 & 10 & 43 & 2.1$\times 10^{-5}$ & 3.4$\times 10^2$ & 6.8$\times 10^2$ \\
= 0.04 pc
& 2.51 & 30 & 0.1 & 0.127 ( 11 ) & 1.0 & 13 & 37.4 & 27 & 15 & 1.5$\times 10^{-5}$ & 8.7$\times 10^2$ & 1.7$\times 10^2$ \\
& 2.81 & 10 & 1.0 & 0.023 ( 2 ) & 2.0 & 39 & 50.5 & 5 & 39 & 7.5$\times 10^{-5}$ & 1.0$\times 10^3$ & 7.6$\times 10^2$ \\
\hline\noalign{\smallskip}
IRAS22172 MIR1
& 0.04 & 20 & 0.1 & 0.104 ( 9 ) & 2.0 & 34 & 25.3 & 15 & 30 & 1.7$\times 10^{-5}$ & 1.4$\times 10^2$ & 1.9$\times 10^2$ \\
$d$ = 2.4 kpc
& 0.04 & 20 & 0.1 & 0.104 ( 9 ) & 1.0 & 22 & 50.5 & 17 & 20 & 1.3$\times 10^{-5}$ & 2.7$\times 10^2$ & 1.5$\times 10^2$ \\
$R_{ap}$ = 5 \arcsec
& 0.20 & 10 & 3.2 & 0.013 ( 1 ) & 4.0 & 71 & 0.0 & 2 & 56 & 1.9$\times 10^{-4}$ & 1.9$\times 10^2$ & 1.9$\times 10^3$ \\
= 0.05 pc
& 0.23 & 10 & 0.1 & 0.074 ( 6 ) & 1.0 & 34 & 1.0 & 7 & 31 & 1.0$\times 10^{-5}$ & 8.1$\times 10^1$ & 1.1$\times 10^2$ \\
& 0.40 & 30 & 0.1 & 0.127 ( 11 ) & 1.0 & 22 & 16.2 & 27 & 15 & 1.5$\times 10^{-5}$ & 1.7$\times 10^2$ & 1.7$\times 10^2$ \\
\hline\noalign{\smallskip}
IRAS22172 MIR3
& 0.19 & 30 & 0.1 & 0.127 ( 11 ) & 1.0 & 22 & 0.0 & 27 & 15 & 1.5$\times 10^{-5}$ & 1.7$\times 10^2$ & 1.7$\times 10^2$ \\
$d$ = 2.4 kpc
& 0.39 & 30 & 0.1 & 0.127 ( 11 ) & 2.0 & 34 & 13.1 & 25 & 23 & 2.0$\times 10^{-5}$ & 1.9$\times 10^2$ & 2.4$\times 10^2$ \\
$R_{ap}$ = 5 \arcsec
& 0.45 & 10 & 3.2 & 0.013 ( 1 ) & 4.0 & 68 & 0.0 & 2 & 56 & 1.9$\times 10^{-4}$ & 2.1$\times 10^2$ & 1.9$\times 10^3$ \\
= 0.05 pc
& 0.61 & 10 & 1.0 & 0.023 ( 2 ) & 4.0 & 68 & 0.0 & 1 & 59 & 7.7$\times 10^{-5}$ & 1.5$\times 10^2$ & 1.1$\times 10^3$ \\
& 0.97 & 20 & 0.1 & 0.104 ( 9 ) & 1.0 & 29 & 0.0 & 17 & 20 & 1.3$\times 10^{-5}$ & 1.2$\times 10^2$ & 1.5$\times 10^2$ \\
\hline\noalign{\smallskip}
IRAS21391 BIMA2
& 0.04 & 20 & 0.1 & 0.104 ( 29 ) & 0.5 & 34 & 74.7 & 19 & 13 & 9.6$\times 10^{-6}$ & 8.0$\times 10^1$ & 9.0$\times 10^1$ \\
$d$ = 0.8 kpc
& 0.07 & 30 & 0.1 & 0.127 ( 35 ) & 0.5 & 22 & 74.7 & 29 & 10 & 1.1$\times 10^{-5}$ & 8.8$\times 10^1$ & 9.0$\times 10^1$ \\
$R_{ap}$ = 8 \arcsec
& 0.08 & 10 & 0.3 & 0.041 ( 11 ) & 2.0 & 71 & 19.2 & 5 & 43 & 3.0$\times 10^{-5}$ & 6.2$\times 10^1$ & 2.8$\times 10^2$ \\
= 0.03 pc
& 0.14 & 40 & 0.1 & 0.147 ( 40 ) & 0.5 & 22 & 59.6 & 39 & 8 & 1.1$\times 10^{-5}$ & 8.7$\times 10^1$ & 8.8$\times 10^1$ \\
& 0.18 & 50 & 0.1 & 0.165 ( 45 ) & 0.5 & 22 & 48.5 & 49 & 7 & 1.2$\times 10^{-5}$ & 8.7$\times 10^1$ & 8.7$\times 10^1$ \\
\hline\noalign{\smallskip}
IRAS21391 BIMA3
& 0.18 & 80 & 0.1 & 0.208 ( 57 ) & 0.5 & 86 & 2.0 & 79 & 5 & 1.4$\times 10^{-5}$ & 8.6$\times 10^1$ & 9.2$\times 10^1$ \\
$d$ = 0.8 kpc
& 0.20 & 100 & 0.1 & 0.233 ( 64 ) & 0.5 & 55 & 0.0 & 99 & 4 & 1.5$\times 10^{-5}$ & 8.9$\times 10^1$ & 9.1$\times 10^1$ \\
$R_{ap}$ = 8 \arcsec
& 0.23 & 60 & 0.1 & 0.180 ( 50 ) & 0.5 & 83 & 9.1 & 59 & 6 & 1.3$\times 10^{-5}$ & 8.0$\times 10^1$ & 8.7$\times 10^1$ \\
= 0.03 pc
& 0.24 & 120 & 0.1 & 0.255 ( 70 ) & 0.5 & 22 & 0.0 & 118 & 4 & 1.5$\times 10^{-5}$ & 9.0$\times 10^1$ & 8.8$\times 10^1$ \\
& 0.26 & 160 & 0.1 & 0.294 ( 81 ) & 0.5 & 22 & 0.0 & 158 & 3 & 1.6$\times 10^{-5}$ & 1.0$\times 10^2$ & 9.8$\times 10^1$ \\
\hline\noalign{\smallskip}
IRAS21391 MIR48
& 0.33 & 10 & 0.3 & 0.041 ( 11 ) & 4.0 & 89 & 43.4 & 1 & 68 & 2.4$\times 10^{-5}$ & 2.9$\times 10^1$ & 6.7$\times 10^2$ \\
$d$ = 0.8 kpc
& 0.58 & 10 & 0.1 & 0.074 ( 20 ) & 2.0 & 68 & 13.1 & 4 & 50 & 1.1$\times 10^{-5}$ & 2.5$\times 10^1$ & 1.3$\times 10^2$ \\
$R_{ap}$ = 8 \arcsec
& 2.70 & 40 & 0.1 & 0.147 ( 40 ) & 12.0 & 89 & 98.0 & 2 & 82 & 9.5$\times 10^{-6}$ & 5.7$\times 10^1$ & 1.1$\times 10^4$ \\
= 0.03 pc
& 3.75 & 30 & 0.3 & 0.072 ( 20 ) & 12.0 & 89 & 100.0 & 1 & 81 & 2.2$\times 10^{-5}$ & 7.0$\times 10^1$ & 1.2$\times 10^4$ \\
& 5.51 & 10 & 0.1 & 0.074 ( 20 ) & 1.0 & 39 & 92.9 & 7 & 31 & 1.0$\times 10^{-5}$ & 6.4$\times 10^1$ & 1.1$\times 10^2$ \\
\hline\noalign{\smallskip}
G305 A
& 0.16 & 240 & 0.3 & 0.203 ( 10 ) & 12.0 & 83 & 85.9 & 216 & 15 & 2.0$\times 10^{-4}$ & 3.1$\times 10^4$ & 4.1$\times 10^4$ \\
$d$ = 4.1 kpc
& 0.17 & 320 & 0.3 & 0.234 ( 12 ) & 12.0 & 71 & 79.8 & 293 & 13 & 2.2$\times 10^{-4}$ & 3.3$\times 10^4$ & 4.0$\times 10^4$ \\
$R_{ap}$ = 12 \arcsec
& 0.19 & 200 & 0.3 & 0.185 ( 9 ) & 12.0 & 80 & 81.8 & 173 & 17 & 1.9$\times 10^{-4}$ & 2.8$\times 10^4$ & 4.0$\times 10^4$ \\
= 0.24 pc
& 0.20 & 200 & 0.3 & 0.185 ( 9 ) & 16.0 & 83 & 97.0 & 162 & 22 & 2.2$\times 10^{-4}$ & 3.0$\times 10^4$ & 5.3$\times 10^4$ \\
& 0.20 & 400 & 0.3 & 0.262 ( 13 ) & 12.0 & 22 & 90.9 & 373 & 11 & 2.3$\times 10^{-4}$ & 3.7$\times 10^4$ & 4.0$\times 10^4$ \\
\hline\noalign{\smallskip}
IRAS16562 N
& 0.05 & 10 & 3.2 & 0.013 ( 2 ) & 4.0 & 62 & 0.0 & 2 & 56 & 1.9$\times 10^{-4}$ & 2.9$\times 10^2$ & 1.9$\times 10^3$ \\
$d$ = 1.7 kpc
& 0.14 & 50 & 0.1 & 0.165 ( 20 ) & 2.0 & 22 & 0.0 & 46 & 16 & 2.4$\times 10^{-5}$ & 3.1$\times 10^2$ & 3.1$\times 10^2$ \\
$R_{ap}$ = 8 \arcsec
& 0.28 & 10 & 1.0 & 0.023 ( 3 ) & 1.0 & 29 & 17.2 & 8 & 25 & 6.0$\times 10^{-5}$ & 5.6$\times 10^2$ & 7.7$\times 10^2$ \\
= 0.06 pc
& 0.37 & 60 & 0.1 & 0.180 ( 22 ) & 2.0 & 22 & 0.0 & 55 & 15 & 2.5$\times 10^{-5}$ & 3.5$\times 10^2$ & 3.5$\times 10^2$ \\
& 0.38 & 30 & 0.1 & 0.127 ( 15 ) & 4.0 & 62 & 7.1 & 21 & 33 & 2.7$\times 10^{-5}$ & 3.8$\times 10^2$ & 7.7$\times 10^2$ \\
\enddata
\end{deluxetable*}

\textbf{G305 A:} The best models are those with a high-mass protostar with a
current mass of 12 - 16 $M_{\odot}$ forming from a core with initial
mass of 200 - 400 $M_{\odot}$ and initial clump mass surface density
of 0.3 g cm$^{-2}$. In Paper II we mentioned G305A is likely to be
much younger and more embedded than G305B and in a hot core phase,
prior to the onset of an UC H II region.

\textbf{IRAS16562 N:} The best models involve a low-mass protostar with
current mass of 1 - 4 $M_{\odot}$ forming from a core with initial
mass of 10 - 60 $M_{\odot}$. $\Sigma_{\rm cl}$ is not well
constrained, varying from 0.1 to 3.2 g cm$^{-2}$.

\begin{figure*}
%\epsscale{0.8}
\centering
\includegraphics[width=15cm]{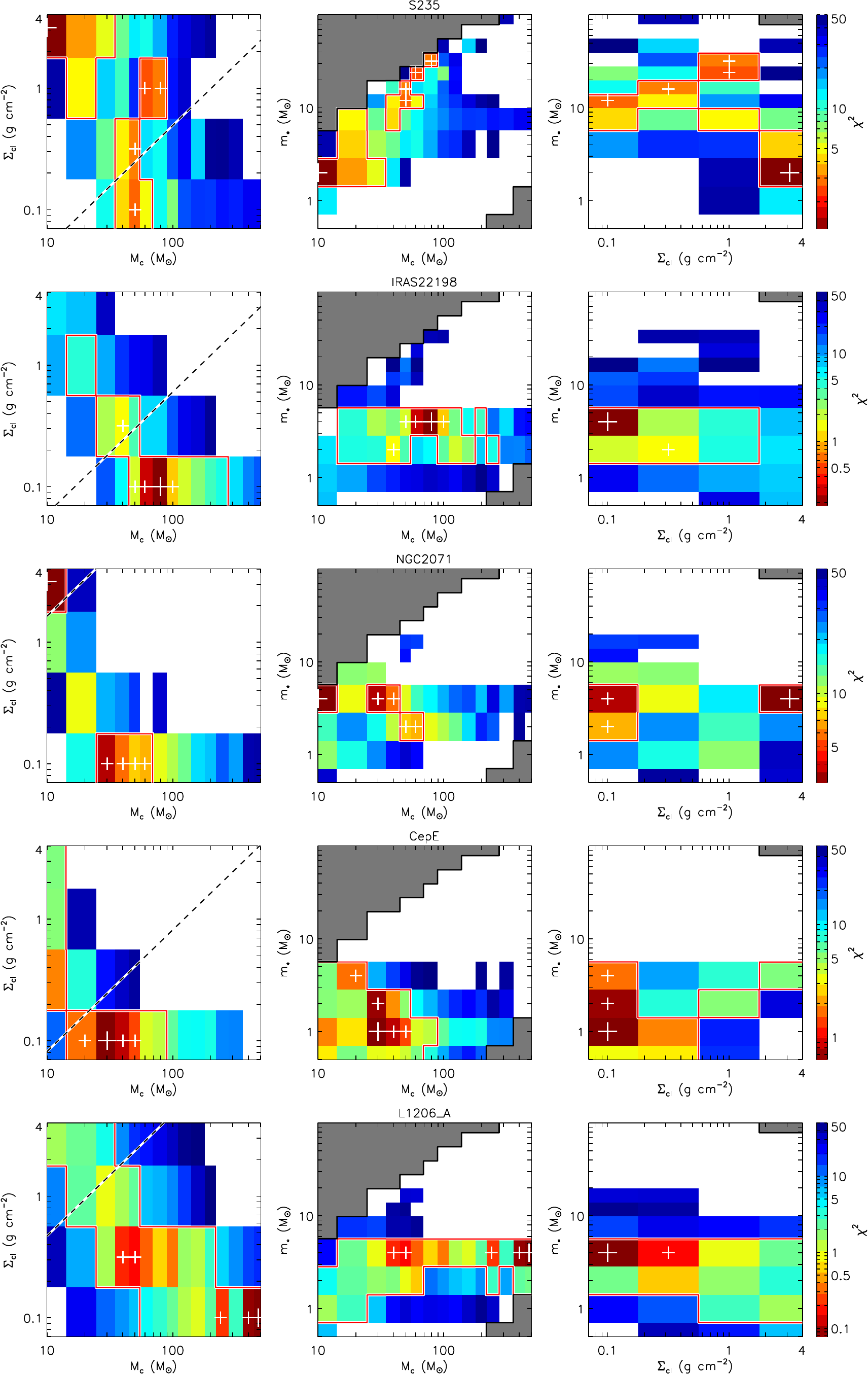}
\caption{
Diagrams of $\chi^{2}$ distribution in $\Sigma_{\rm cl}$ - $M_{c}$
space, $m_{*}$ - $M_{\rm c}$ space and $m_{*}$ - $\Sigma_{\rm
  cl}$ space. The white crosses mark the locations of the five best
models, and the large cross is the best model. The grey regions are
not covered by the model grid, and the white regions are where the
$\chi^{2}$ is larger than 50. The red contours are at the level of
$\chi^{2}$ = $\chi^{2}_{min}$ + 5. The dashed line denotes when $R_{c}
= R_{\rm ap}$.}\label{fig:primary}
\end{figure*}

\begin{figure*}[!htb]
    \ContinuedFloat
    \centering
    \captionsetup{list=off,format=cont, labelsep=space}
    \includegraphics[width=15cm]{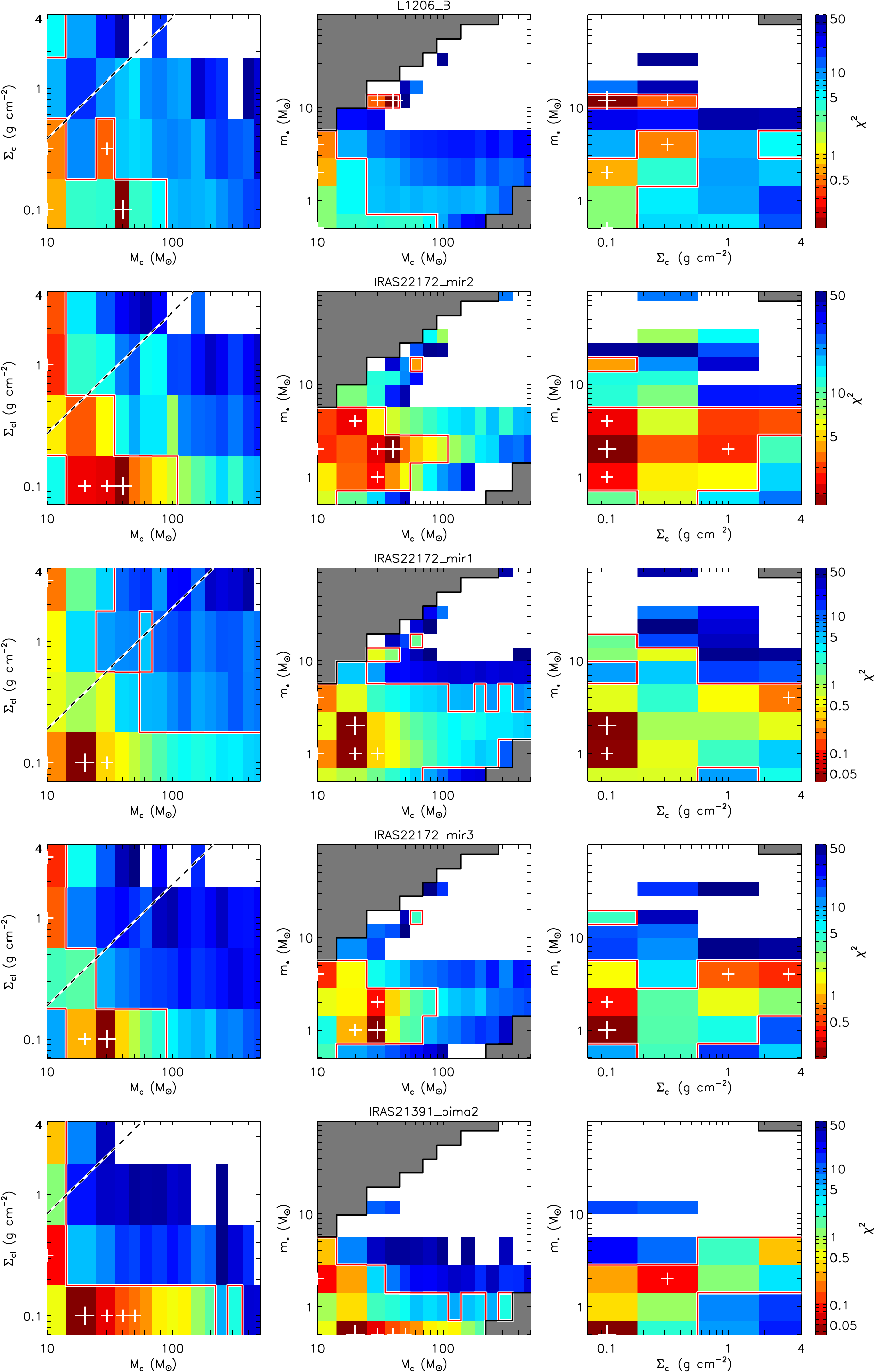}
    \caption{(cont.)}
\end{figure*}

\begin{figure*}[!htb]
    \ContinuedFloat
    \centering
    \captionsetup{list=off,format=cont, labelsep=space}
    \includegraphics[width=15cm]{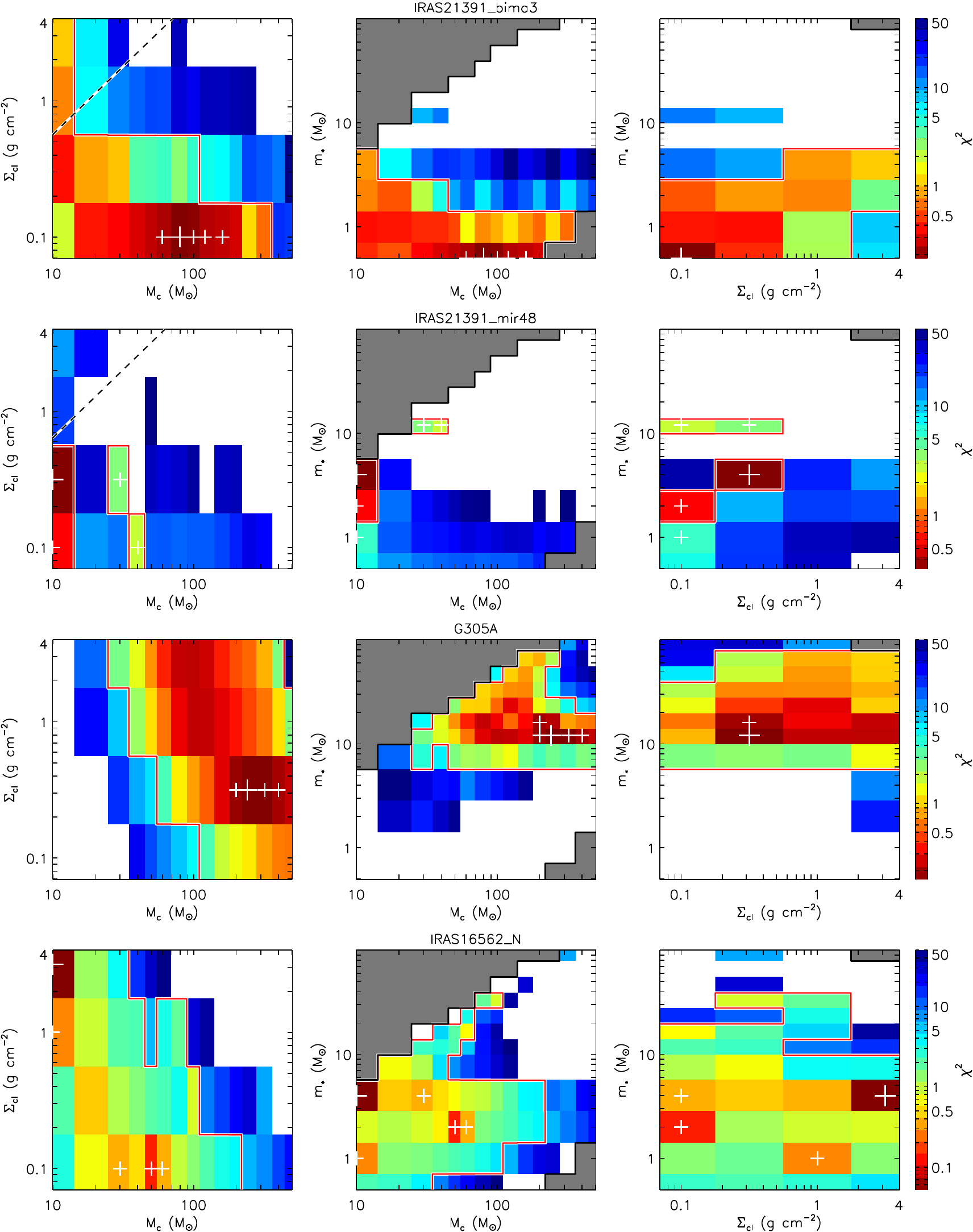}
    \caption{(cont.)}
\end{figure*}

Figure~\ref{fig:primary} shows the $\chi^{2}$ distribution in
$\Sigma_{\rm cl}$ - $M_{c}$ space, $m_{*}$ - $M_{c}$ space and $m_{*}$
- $\Sigma_{\rm cl}$ space for the 14 sources. As also discussed in Paper II,
these diagrams illustrate the full constraints in the primary
parameter space derived by fitting the SED data, and the possible
degeneracies. In general, all the three parameters span a larger range
compared with the sources of Papers I and II.
%where $m_{*}$ seems to be best constrained among the three.

Follow-up observations and analysis of SOMA sources can be helpful in
breaking degeneracies that arise from simple SED fitting. One example
of such follow-up work is that of Rosero et al. (2019), who examined
cm radio continuum data of the SOMA sources presented in Paper
I. Radio free-free emission from photoionized gas, first expected to
be present in the outflow cavity, is particularly useful for
contraining the mass of the protostar once it reaches $\gtrsim
10\:M_\odot$ and begins to contract to the zero age main
sequence. However, at lower masses most of the ionization associated
with the source is expected to be due to shock ionization, e.g., due
to internal shocks in the outflow (see also Fedriani et
al. 2019). Quantitative models for the amount of shock ionization and
associated radio emission have not yet been developed for the ZT
protostellar models. For the mainly intermediate-mass sources
presented in this paper, we anticipate that cm radio emission will
main be due to shock ionization, so such observations may be more
challenging to interpret to help break SED fit degeneracies. On the
other hand, measurements of protostellar outflow properties, including
cavity opening angle and mass and momentum fluxes may provide more
diagnostic power.
%jct4 - add the Rosero and Fedriani (Nature Communications) references to the paper.

In contrast with the high-mass protostars in Papers I and II, the best
models ($\chi^{2} - \chi^{2}_{\rm min} < 5$, within the red contours
shown in Figure~\ref{fig:primary}) of the intermediate-mass protostars
also occupy the region with lower $M_{c}$ at lower $\Sigma_{\rm
  cl}$. Another striking feature is that most sources have best models
with a core size larger than the aperture size, i.e., they appear
below the dashed line denoting when $R_{c} = R_{\rm ap}$ in
Figure~\ref{fig:primary}. To examine this matter further, we analyzed
the image profiles of the best 5 models of the sources and found that
the flux density at 37 $\mu$m usually decays to $10^{-3}$ of the peak
flux density within 5\arcsec\ from the center and the flux density at
70 $\mu$m usually decays to $10^{-3}$ of the peak flux density within
15\arcsec\ from the center. The typical aperture radius is $\sim$
10\arcsec\ (except for the three sources in IRAS 22172 where it is
$\sim$ 5\arcsec, but their best models have the flux density decaying
to $10^{-3}$ of the peak within 2\arcsec\ and 5\arcsec\ at 37 and 70
$\mu$m, respectively). This indicates that when the models have a core
size larger than the aperture used for measuring the SED, only a small
amount of the total flux from the model is being missed (however, the
proportion of missed flux would be larger at longer
wavelengths). Nevertheless, to better illustrate the importance of
this effect, in the following discussion we present two cases, i.e.,
with and without the constraint on the model core size needing to be
within a factor of two of the aperture size.

\section{Discussion}

We now discuss results of the global sample of 29 protostars that have
been derived from an uniform SED fitting analysis that always includes
{\it SOFIA}-FORCAST data, as presented in Papers I, II and III.

%jct3 - I moved text here, since it belongs in this section. And I re-wrote for clarity - check.

In general, we select the best five or fewer models that satisfy
$\chi^2 < \chi^2_{\rm min} +5$, where $\chi^2_{\rm min}$ is the value
of $\chi^2$ of the best model, and then present averages of model
properties. However, for G45.12+0.13, which was discussed in Paper II
as not being especially well fit by the ZT models because of its high
luminosity (it is likely to be multiple sources), there is only one
model with $\chi^2 < \chi^2_{\rm min} +5$. Thus for this source we
average all the best 5 models. The model properties are averaged in
log space, i.e., geometric averages, except for $A_V$, $\theta_{\rm
  view}$ and $\theta_{\rm w,esc}$, which are evaluated as arithmetic means.

Then, as explained at the end of the last section, we also consider
two cases, i.e., without and with the constraint of ``best-fit''
models having core sizes that are within a factor of two of the
aperture size. Without the core size constraint, the best five models
of all sources automatically satisfy $\chi^2 < \chi^2_{\rm min} +5$,
except for G45.12+0.13. With the core size constraint (which we regard
as our best, fiducial method), there can be cases, especially of
intermediate-mass sources from Paper III (i.e., this work), where
there are fewer than five models with $\chi^2 < \chi^2_{\rm min}
+5$. Still, G45.12+0.13 is kept as a special case, as above. Key
average source properties are listed in Table~\ref{tab:gmean}.

\subsection{The SOMA Sample Space}

\begin{figure*}
%\figurenum{}
\epsscale{0.97}
\plotone{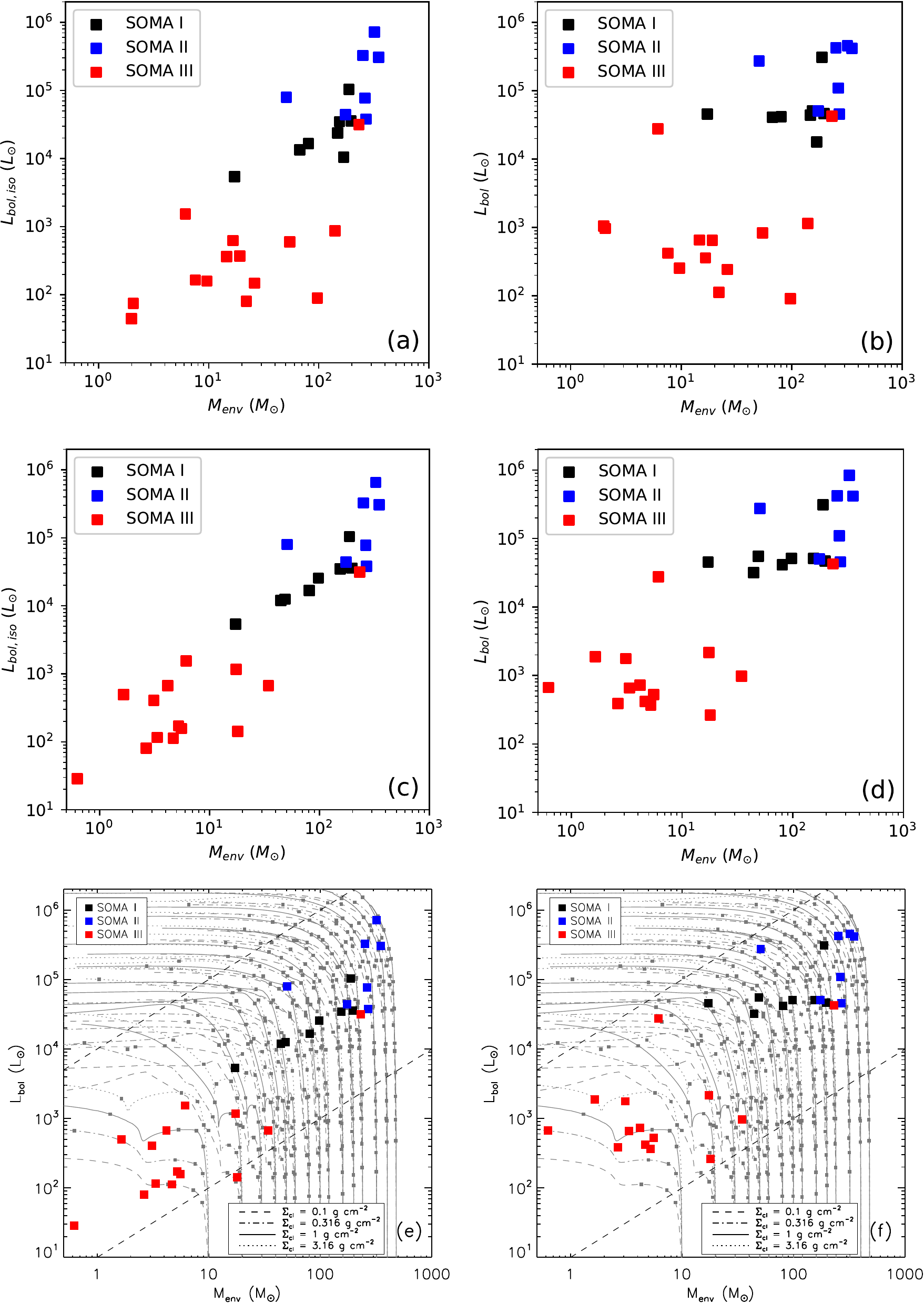}
\caption{
(a)
Average (geometric mean) isotropic bolometric luminosity versus
envelope mass returned by the best five (see text) ZT models for each
SOMA source from Papers I, II and III (this work), as labelled.
(b) Same as (a), but now with true bolometric luminosities plotted
versus envelope mass.
(c) Same as (a), but now using the average of the best five or fewer models with $R_c \la 2R_{\rm ap}$ and $\chi^2 < \chi^2_{\rm min} +5$.
%the best ZT models that also satisfy the size constraint of being within a factor of two of the aperture used for measuring the SED (see text).
%
(d) Same as (c), but now with true bolometric luminosities plotted
versus envelope mass.
(e) Same as (c), but now also showing the ZT18 protostar models (grey
squares), which are a collection of different evolutionary tracks
(grey lines) for different initial core masses and clump mass surface
densities (see legend). The two dashed black lines indicate $L_{\rm bol}/M_{\rm env} = 10$ and $10^4 \ L_{\odot}/M_{\odot}$, respectively.
(f) Same as (e), but now with true bolometric luminosities plotted
versus envelope mass.  
%
%Geometric mean isotropic luminosity versus geometric
%mean envelope mass returned by the best five or fewer models with core
%radius smaller than twice the aperture radius and $\chi^2$ smaller
%than $\chi^2_{\rm min} +5$. (d): Same with (c) except that true
%luminosities are plotted. (e): Evolutionary tracks of the ZT18 models
%plotted as grey curves. The true luminosities versus the envelope
%masses of the all the models in the model grid are plotted as grey
%squares with all the data points in (c) representing the isotropic
%luminosity plotted above. (f): Same with (e) except that the data
%points in (d) representing the true luminosity are plotted above. The
%red line indicates the relationship of $\rm log(L_{\rm bol}) = 1.120
%\times log(M_{\rm env}) + 2.420$. The two dashed black lines indicate
%$L_{\rm bol}/M_{\rm env} = 10$ and $10^4 \ L_{\odot}/M_{\odot}$,
%respectively. Note that in both (e) and (f), the true luminosities of
%the model grid are plotted.
}\label{fig:lm}
\end{figure*}

Figure~\ref{fig:lm}a shows $L_{\rm bol,iso}$ versus $M_{\rm env}$ for
the SOMA protostar sample from Papers I, II and this work, i.e., Paper
III. Figure~\ref{fig:lm}b shows $L_{\rm bol}$ versus $M_{\rm env}$ of
the same sample. This is the more fundamental property of the
protstar, since $L_{\rm bol,iso}$ is affected by the orientation of
protostellar geometry to our line of sight and the flashlight effect.
Compared with the sources presented in Papers I and II, which were
exclusively high-mass protostars, $L_{\rm bol,iso}$, $L_{\rm bol}$ and
$M_{\rm env}$ all extend down to lower values. When we apply the
constraint on model core sizes, i.e., radii of the models need to be
no larger than twice the radius of the aperture used to define the
SED, then we see from Figures~\ref{fig:lm}c and d that there is an
apparent tightening of the correlations between $L_{\rm bol,iso}$ or
$L_{\rm bol}$ with $M_{\rm env}$.
%becomes tighter. A similar trend is seen in $L_{\rm bol}$ as shown in
%Figure~\ref{fig:lm}d compared with Figure~\ref{fig:lm}b.
Note that the highest-mass, highest-luminosity YSOs
%(i.e., those SOMA I \& II)
usually have best models with $R_{c} \la R_{\rm ap}$ and are thus less
influenced by this constraint.

%We will discuss this point in more detail below.

Figures~\ref{fig:lm}e and f show the sample distribution in the
context of the whole ZT model grid, where lines indicate evolutionary
tracks, i.e., from low luminosity and high envelope mass to high
luminosity and low envelope mass, for different clump environment mass
surface densities, $\Sigma_{\rm cl}$.

The SOMA sample spans a relatively broad range of evolutionary stages
with $L_{\rm bol}/M_{\rm env}$ extending from $\sim$
10~$L_{\odot}/M_{\odot}$ up to almost $10^4 \: L_{\odot}/M_{\odot}$,
indicated by the dashed lines in Figure~\ref{fig:lm}f.  As a result of
this broad range and given the even wider range that is expected from
the theoretical models, we do not fit the observed $L_{\rm bol}$
versus $M_{\rm env}$ distribution with a power law relation (c.f.,
Molinari et al. 2008; Urquart et al. 2018). Rather, we simply note
that the sources that have so far been analyzed in the SOMA sample
span this wide range of evolutionary stages, but the expected very
late stages and very early stages are not especially well
represented.

\begin{figure*}
%\figurenum{}
\epsscale{1.18}
\plotone{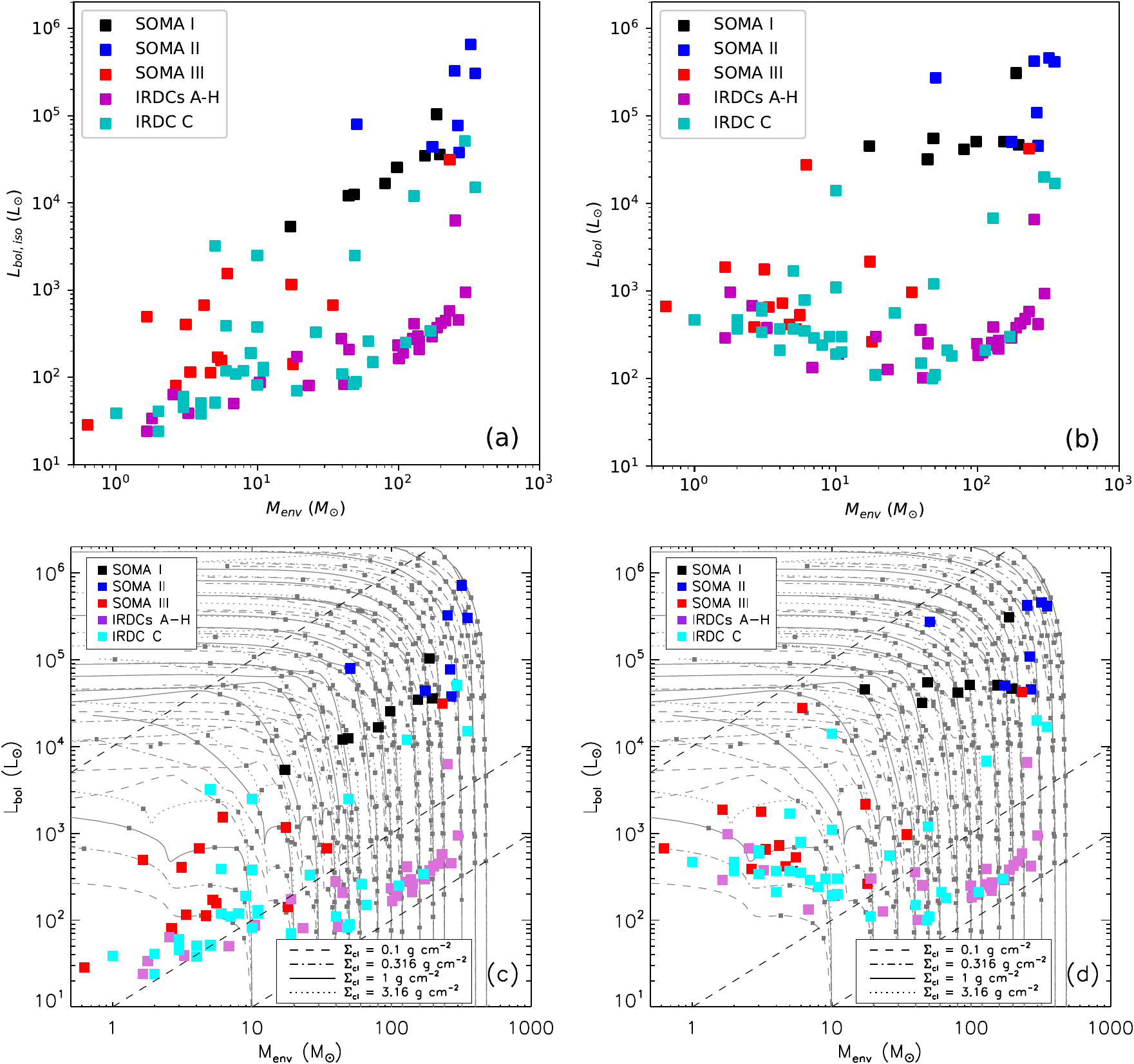}
\caption{
Protostellar evolutionary stages probed by the SOMA sample and IRDC
protostar samples: ``IRDC A-H'' (Liu et al. 2018; Liu et al., in
prep.); ``IRDC C'' (Moser et al. 2020). The format of the figures is
otherwise the same as Figures~\ref{fig:lm}c, d, e, f, respectively,
but with the average (geometric mean) results of the valid models of
IRDC sources added. The three dashed black lines in panels c and d
indicate $L_{\rm bol}/M_{\rm env} =1, 10$ and $10^4
\ L_{\odot}/M_{\odot}$.}\label{fig:lm2}
\end{figure*}

To further explore the evolutionary context of the SOMA protostars, in
Figure~\ref{fig:lm2} we show the SOMA sample in the luminosity versus
envelope mass plane, together with protostellar sources identified in
Infrared Dark Clouds (IRDCs), which are expected to be at earlier
stages of evolution. Two samples of protostars selected from IRDC
environments are shown, with the source SED construction and ZT model
fitting following the same methods as have been used for the SOMA
sample. The first, labelled ``IRDCs A-H'', is the sample of 28 sources
from Liu et al. (2018) and Liu et al., in prep., based on {\it ALMA}
observations of 32 clumps in IRDCs A to H from the sample of Butler \&
Tan (2009, 2012). The second, labelled ``IRDC C'', is a complete
census of the protostellar sources in IRDC C carried out by Moser et
al. (2020), based on sources identified in the region by {\it
  Herschel} 70~$\rm \mu m$ emission from the Hi-GAL point source
catalog (Molinari et al. 2016). After allowing for a few poorly
resolved sources that are treated as a single protostar in the SED
modeling, a total of 35 protostars have been analyzed by Moser et
al. (2020). The IRDC sources include protostars with intrinsic
bolometric luminosities down to about $100\:L_\odot$, including within
relatively massive core envelopes, so that the sampled values of
$L_{\rm bol}/M_{\rm env}$ now extend down to $\sim
1\:L_\odot/M_\odot$.

Various biases in the input catalog for the SOMA survey likely account
for the lack of sources at the final evolutionary stages of high
$L_{\rm bol}$ and low $M_{\rm env}$. For example, these sources will
have relatively weak MIR to FIR emission, which was used as a
consideration to target SOMA protostars. Such sources may also be
embedded within ultracompact H\,II regions, which we have tended to
avoid, so far for analysis, even if they are within our fields of
view: here the challenge is to isolate emission from any remaining
protostellar core from the thermal emission from hot dust in the large
scale H\,II region. Finally, this later phase of evolution may be
relatively short, so objects here may be intrinsically rare. Future
studies will attempt to identify such sources.

Finally, we note that a future goal is to extend complete surveys of
high- and intermediate-mass protostars across their full range of
evolutionary stages and across larger regions so that the samples can
be used for demographic analyses that will inform about topics such as
the duration of formation timescales. Previous work in this area,
e.g., Davies et al. (2011), which covered large regions of the
Galactic plane, focused only on high-mass protostars and have been
relatively restricted in their coverage of earlier evolutionary
stages.

%Compared with the massive star forming sources in Urquhart et
%al. (2018), which are found within a $L_{\rm bol}/M_{\rm env}$ range
%of 1 and 100 $L_{\odot}/M_{\odot}$, our sample reveals a higher upper
%bound of $L_{\rm bol}/M_{\rm env}$, indicating the possibility of
%generally higher star formation efficiency. We find a relationship of
%$\rm log(L_{\rm bol}) = 1.120 \times log(M_{\rm env}) + 2.420$,
%slightly shallower than the values reported by Urquhart et al. (2018)
%($\sim$ 1.314) and Molinari et al. (2008) ($\sim$ 1.27).

%and to achieve a certain $L_{\rm bol}$, models with relatively large
%$M_{\rm env}$ are excluded.
%Small $R_{c}$ results from low $M_{c}$ and high $\Sigma_{\rm cl}$.
%While also depending on the evolutionary stage, $M_{\rm env}$
%tend to be small if $M_{\rm c}$ is small.

\subsection{The Shapes of SEDs}

\begin{figure}
\epsscale{1.15}
\plotone{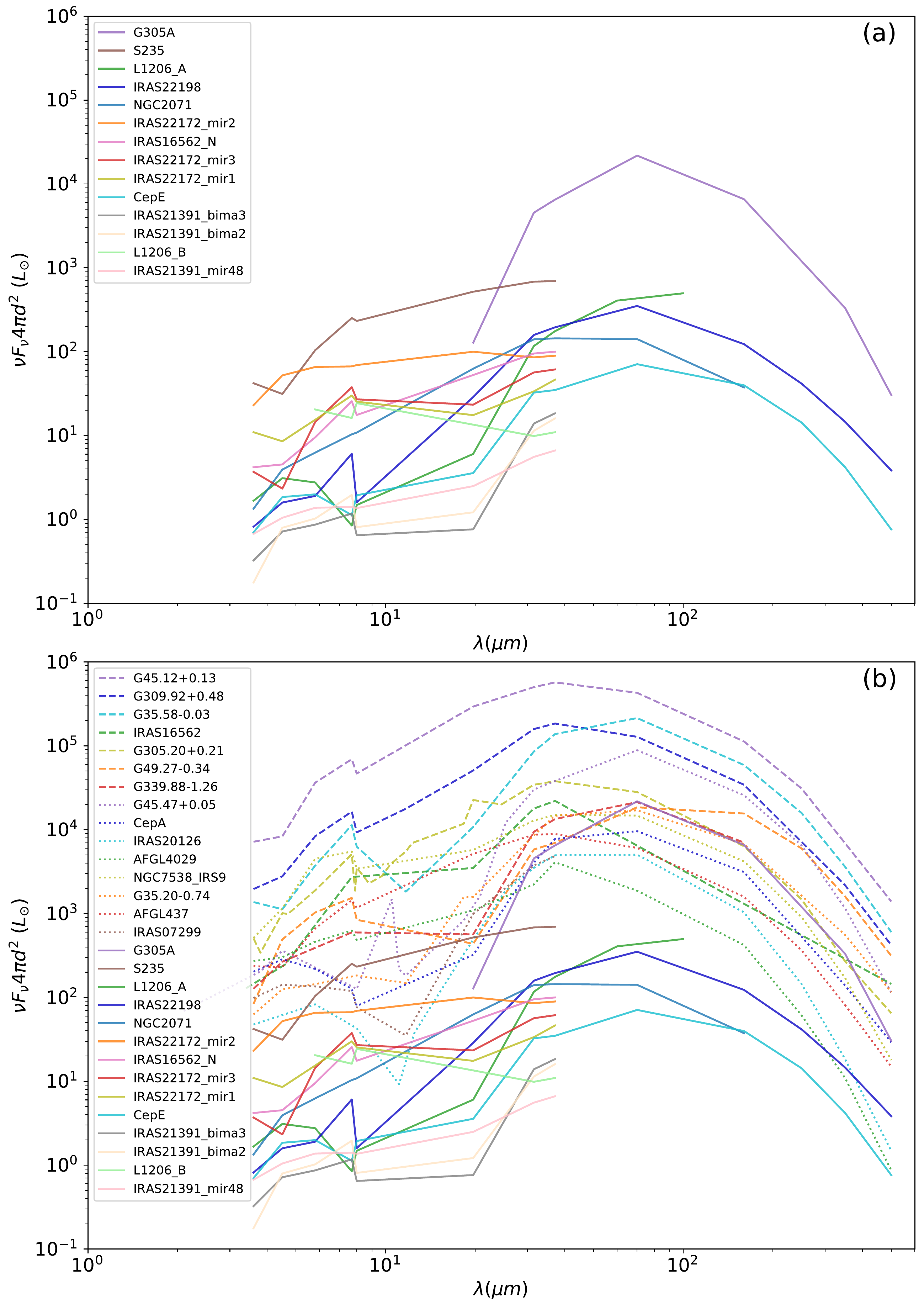}
\caption{
{\it a) Top panel:} Bolometric luminosity weighted SEDs of the 14 SOMA
protostars analyzed in this paper. The ordering of the legend is from
high to low ZT best fit model isotropic luminosity (top to
bottom). {\it b) Bottom panel:} Same as (a), but now with addition of dashed lines
denoting the sample of 15 sources from Papers I and
II.}\label{fig:Lbol}
\end{figure}

\begin{figure*}
\epsscale{1.0}
\plotone{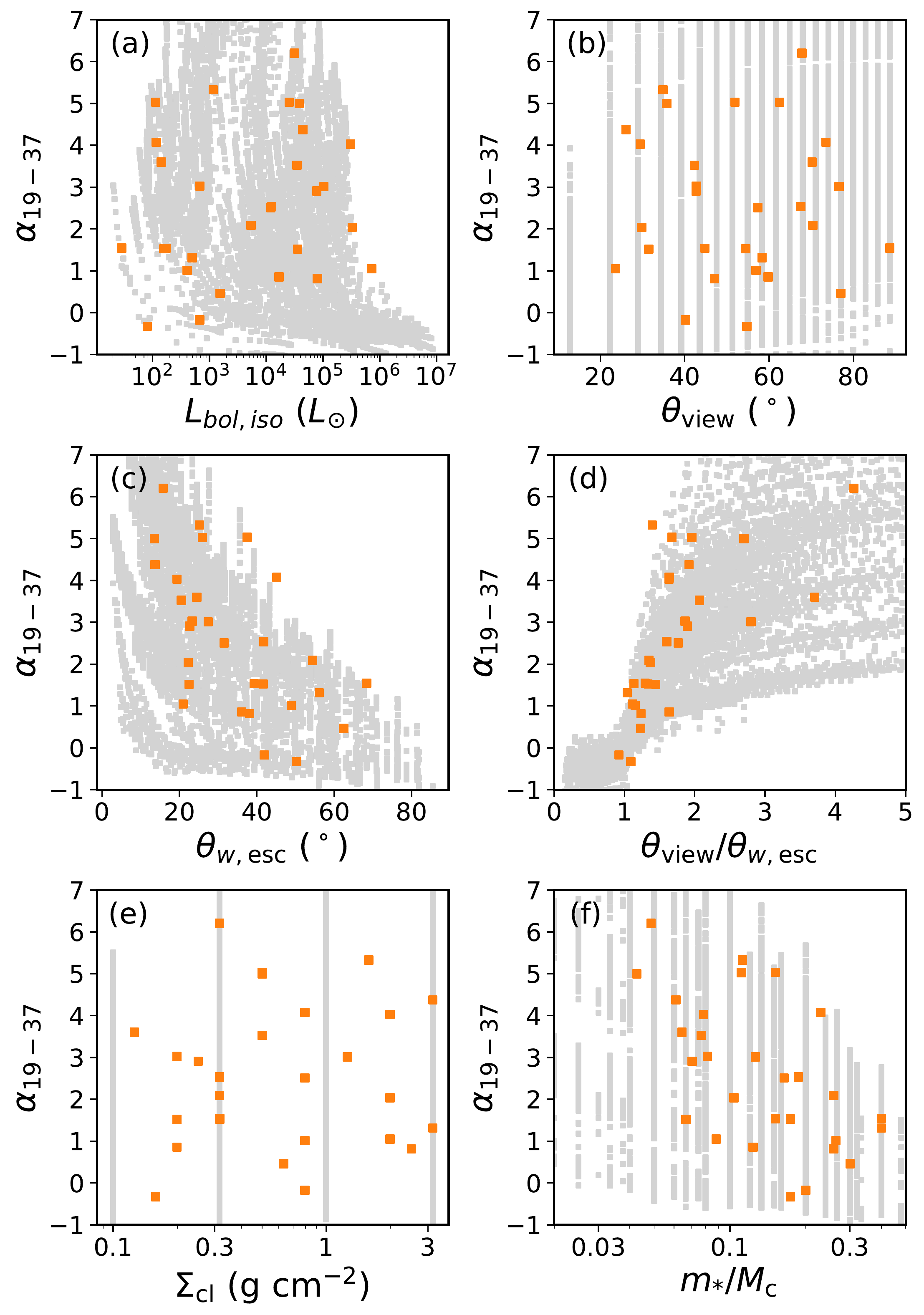}
\caption{
Spectral index, $\alpha_{19-37}$
%$= \frac{\nu_{37\mu m}F_{\nu, 37\mu m} - \nu_{19\mu m}F_{\nu, 19\mu
%    m}}{\lambda_{37\mu m} - \lambda_{19\mu m}}$,
between 19 $\mu$m and 37 $\mu$m (see text) versus: the geometric mean
isotropic luminosity $L_{\rm bol, iso}$ (a: top left); the arithmetic
mean inclination of viewing angle $\theta_{\rm view}$ (b: top right);
the arithmetic mean opening angle $\theta_{w, \rm esc}$ (c: middle
left); arithmetic mean $\theta_{\rm view} / \theta_{w, \rm esc}$ (d:
middle right); the geometric mean clump surface density $\Sigma_{\rm
  cl}$ (e: bottom left); and geometric mean $m_{*}/M_{\rm c}$ (f:
bottom right) returned by the best five or fewer models with $R_c \la 2R_{\rm ap}$ and $\chi^2 < \chi^2_{\rm min} +5$. The grey squares represents the ZT18 protostar models. Note that the spectral index of the models are calculated without foreground extinction and thus could be different from observations. }\label{fig:flatness}
\end{figure*}

In Figure~\ref{fig:Lbol} we show the bolometric luminosity spectral
energy distributions of the 14 protostars of this paper, together with
the sample of 15 generally higher luminosity sources from Papers I and
II. Here the $\nu F_{\nu}$ SEDs have been scaled by $4 \pi d^{2}$ so
that the height of the curves gives an indication of the luminosity of
the sources assuming isotropic emission. The ordering of the vertical
height of these distributions is largely consistent with the rank
ordering of the predicted isotropic luminosity of the protostars from
the best-fit ZT models (the legend in Figure~\ref{fig:Lbol} lists the
sources in order of decreasing ZT best model isotropic luminosity).

We define a 19--37 $\mu$m spectral index via
\begin{equation}
\alpha_{19-37} = \frac{\nu_{37\mu m}F_{\nu,37\mu m} - \nu_{19\mu m}F_{\nu,19\mu m}}{\lambda_{37\mu m} -  \lambda_{19\mu m}}.
\end{equation}
In general, we expect that this index may vary systematically with
protostellar source properties. Figure~\ref{fig:flatness} shows the
dependence of $\alpha_{19-37}$ of the SEDs on luminosity, inclination
of viewing angle, outflow cavity opening angle, ratio of inclination
of viewing angle to outflow cavity opening angle, $\Sigma_{\rm cl}$,
and $m_{*}/M_{c}$, respectively.  In all these panels, the results have
been averaged over those of the best 5 or fewer models with core radii
smaller than twice the aperture radius and $\chi^2 < \chi^2_{\rm min}
+5$ (except for G45.12+0.13, see above).
%jctnew - check this
%jct2 - the numbers of these averages should be
%shown somewhere. I guess we need a new table that includes the Paper
%I and II sources. Basically all the data plotted in Fig. 14 onwards
%needs to be presented.
We see that the outflow cavity opening angle has a strong influence on
the 19--37 $\mu$m index, following the expectation that a relatively
greater flux of shorter wavelength photons are able to escape from the
protostellar core if the outflow cavity opening angle is larger. Also
a viewing angle inclination that is relatively small compared to the
outflow cavity opening angle will result in a flatter shorter
wavelength SED, as also discussed in Paper II.

In Figure~\ref{fig:flatness}, we also plot the ZT18 models as grey squares beneath the observations to illustrate the model coverage. Note that the range shown here serves to best show the observations and does not represent the full parameter space of the ZT18 models. We note that while
the observed correlations are in general built in the ZT models, the results
of Figure~\ref{fig:flatness} show how tight (or loose) the
correlations are in practice of the observed SED spectral index in the
{\it SOFIA}-FORCAST bands with best average protostellar parameters
derived from the fitting the entire available MIR to FIR SED. This
information gives an idea of how much information can be derived from
only an observed value of $\alpha_{19-37}$.

Finally, and along the same lines, another important feature that is
revealed by $\alpha_{19-37}$ is the protostellar evolutionary stage,
as measured by $m_{*}/M_{c}$ (Figure~\ref{fig:flatness}f). Again, this
general trend is expected in the context of the ZT models, since the
outflow cavity systematically opens up during the course of the
evolution and the envelope mass is depleted, resulting in lower
overall extinction. There is also generally lower levels of extinction
in protostellar cores in lower $\Sigma_{\rm cl}$ environments, but
little correlation is seen here between $\alpha_{19-37}$ and
$\Sigma_{\rm cl}$ (Figure~\ref{fig:flatness}e), indicating other
factors have a more important influence.

%Zhang \& Tan (2018) discussed that from the model grid the 19-37
%$\mu$m index can be affected not only by the inclination of the
%viewing angle, but also other factors, such as $\Sigma_{\rm cl}$ and
%the evolutionary stage indicated by $m_{*}/M_{c}$. From our sample,
%the dependence on $m_{*}/M_{\rm c}$ is quite clear as shown in
%Figure~\ref{fig:flatness}f, which makes sense because the model
%involves an increasing opening angle with increasing evolutionary
%stage.

\subsection{Dependence of Massive Star Formation on Environment}

\begin{figure*}
%\figurenum{}
\epsscale{1.2}
\plotone{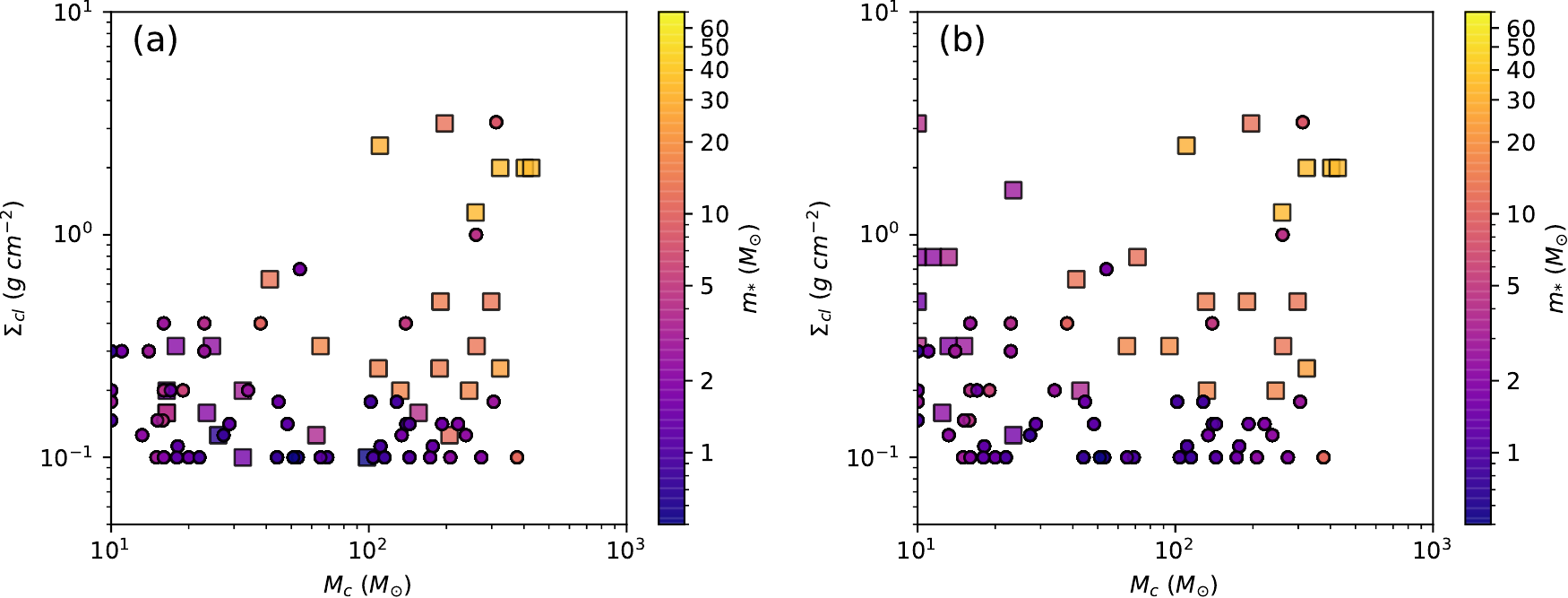}
\caption{
{\it a) Left:} Average clump mass surface density, $\Sigma_{\rm cl}$,
versus average initial core mass, $M_c$, of the SOMA sources (squares)
and IRDC sources (circles, Liu et al. 2018; Moser et al. 2020; Liu et
al., in prep.), based on ZT model fits: the average is made for the
best five selected models.
{\it b) Right:} Same as (a), but with the average made for the best
five or fewer models with $R_c \la 2R_{\rm ap}$ and $\chi^2 < \chi^2_{\rm min} +5$.
%(c): Distribution of models in the model grid. Note that for illustration $\Sigma_{\rm cl}$ was plotted timed by a random factor uniformly distributed between 0.67 and 1.5 in logarithmic space.
}\label{fig:m_sigma_ms}
\end{figure*}

\begin{figure*}
\epsscale{1.2}
\plotone{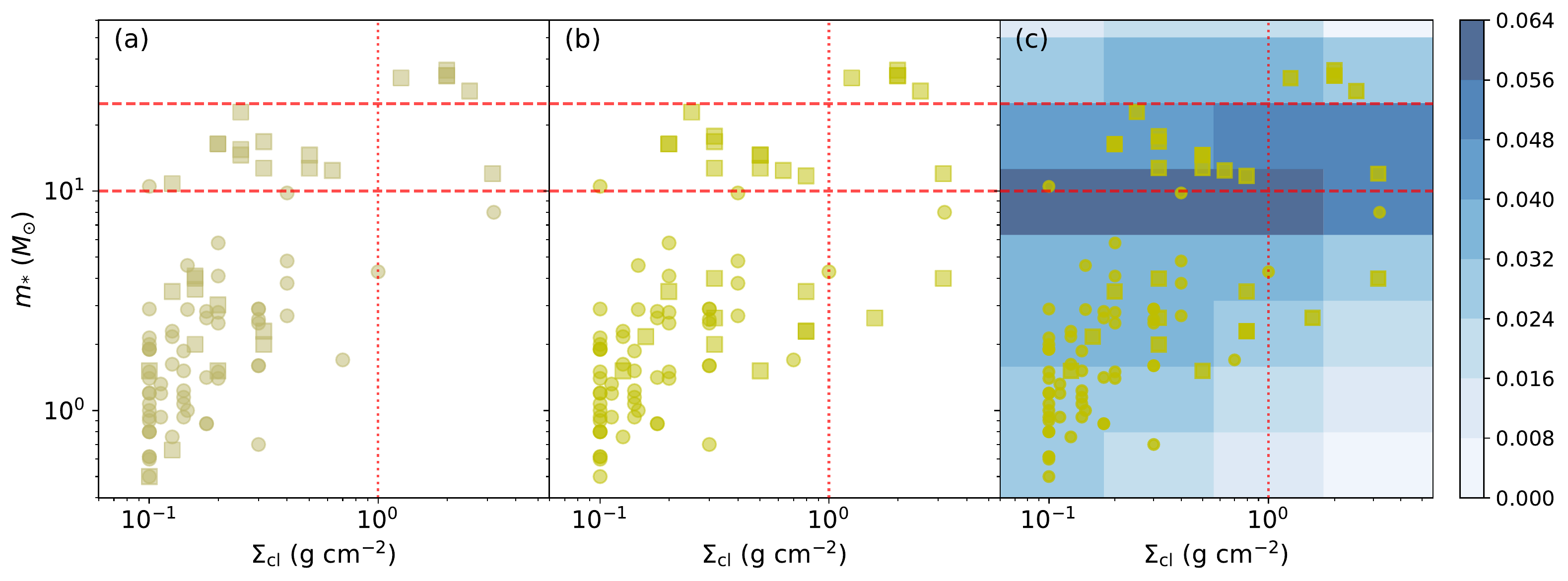}
\caption{
{\it a) Left:} Average protostellar mass, $m_*$, versus average clump
mass surface density, $\Sigma_{\rm cl}$, of SOMA sources (squares) and
IRDC sources (circles, Liu et al. 2018; Moser et al. 2020; Liu et al.,
in prep.), based on ZT model fits: the average is made for the best
five selected models. The red dotted and dashed lines indicate
fiducial threshold values of $m_*$ (10 and 25$\:M_{\odot}$) and $\Sigma_{\rm cl}$ (1$\:{\rm g\:cm}^{-2}$, see text).
{\it b) Middle:} Same as (a), but with the average made for best five
or fewer models with $R_c \la 2R_{\rm ap}$ and $\chi^2 < \chi^2_{\rm min} +5$.
{\it c) Right:} Same as (b), but now also showing the distribution of
models in the ZT model grid (shading indicates
the density of models).}\label{fig:sigma_ms}
\end{figure*}

\begin{figure*}
%\epsscale{0.8}
\centering
\includegraphics[width=13cm]{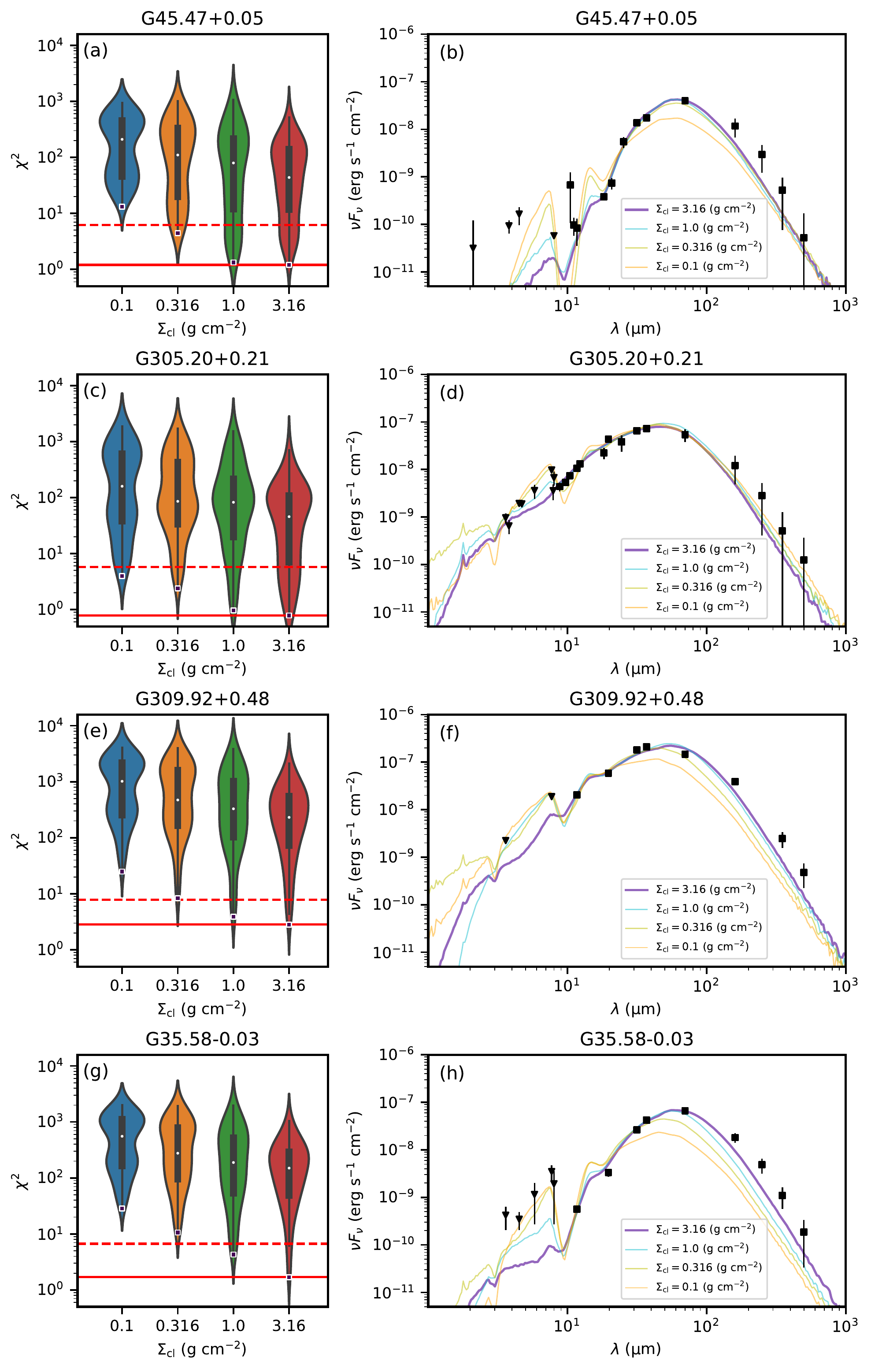}
\caption{
{\it Left column}: Violin plots of $\chi^{2}$ versus $\Sigma_{\rm cl}$
of all the models for several of the most massive protostars:
G45.47+0.05, G305.20+0.21, G309.92+0.48 and G35.58-0.03. For the
violin of each $\Sigma_{\rm cl}$, the white dot denotes the median
$\chi^{2}$. The black bar in the center of the violin denotes the
interquartile range (IQR). The black lines stretched from the bar
denote the lower/upper adjacent values -- defined as the furthest
observation within 1.5 IQR of the lower/upper end of the bar. The
width of the violin represents the probability density of the data
value smoothed by a kernel density estimator. The squares at the
bottom of each violin denote the smallest $\chi^{2}$ achieved by that
$\Sigma_{\rm cl}$. The red solid line denotes $\chi^{2}_{\rm min}$ for
the source. The red dashed line denotes $\chi^{2}_{\rm min}+5$. {\it
  Right column}: SEDs of the best model of each $\Sigma_{\rm cl}$ for
each source (thickest line is the overall best model). The black
triangles and squares with error bars denote the
observations.}\label{fig:lowsigma}
\end{figure*}

\begin{figure}
\epsscale{1.2}
\plotone{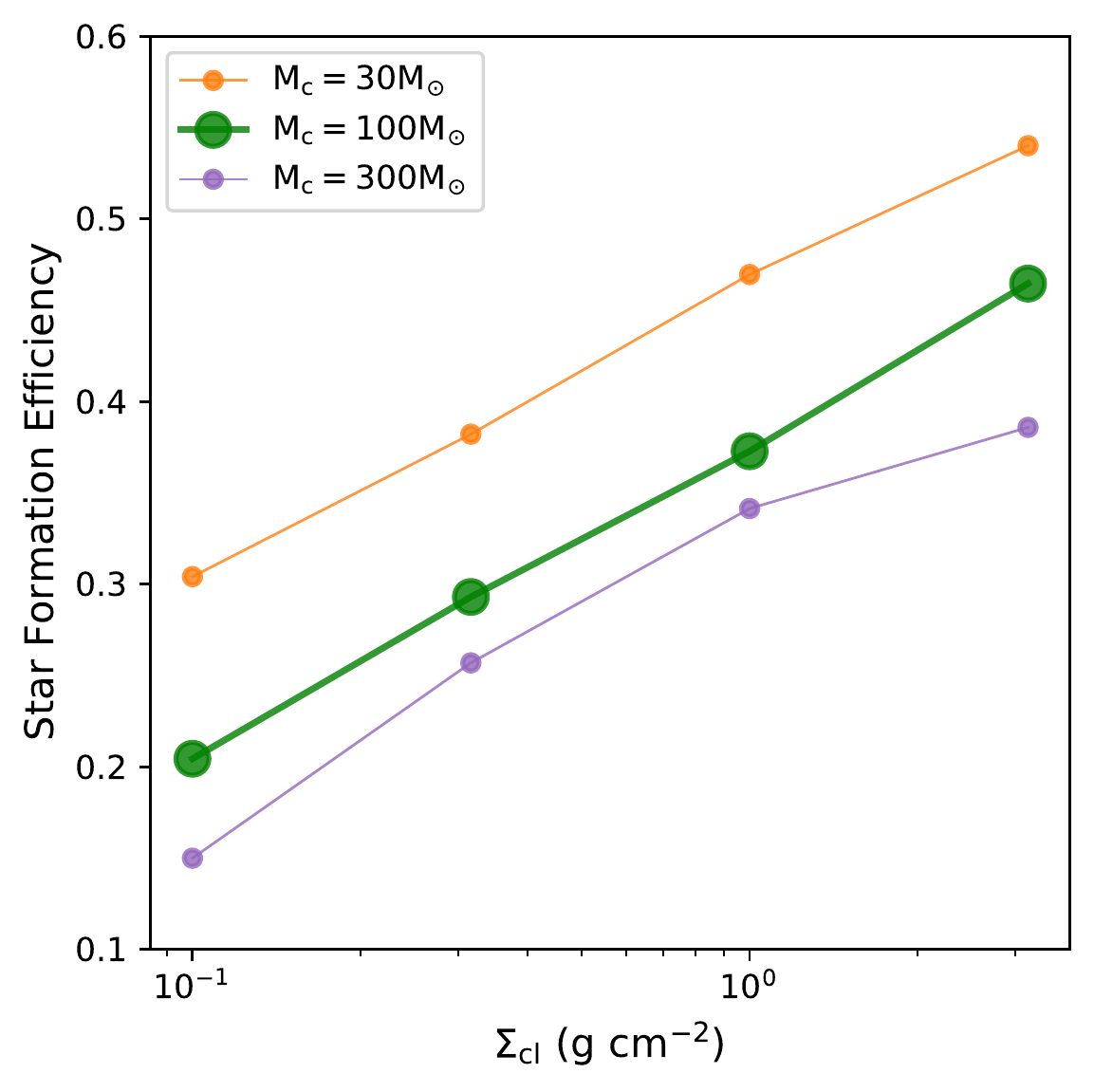}
\caption{
Star formation efficiency as a function of clump mass surface
density, $\Sigma_{\rm cl}$, from model calculations of Tanaka et
al. (2017). Models for initial core masses of $M_c=30,~100$, and
$300{\rm\:M_\odot}$ are shown, as labelled.}\label{fig:efficiency}
\end{figure}

\begin{figure*}
\epsscale{1.2}
\plotone{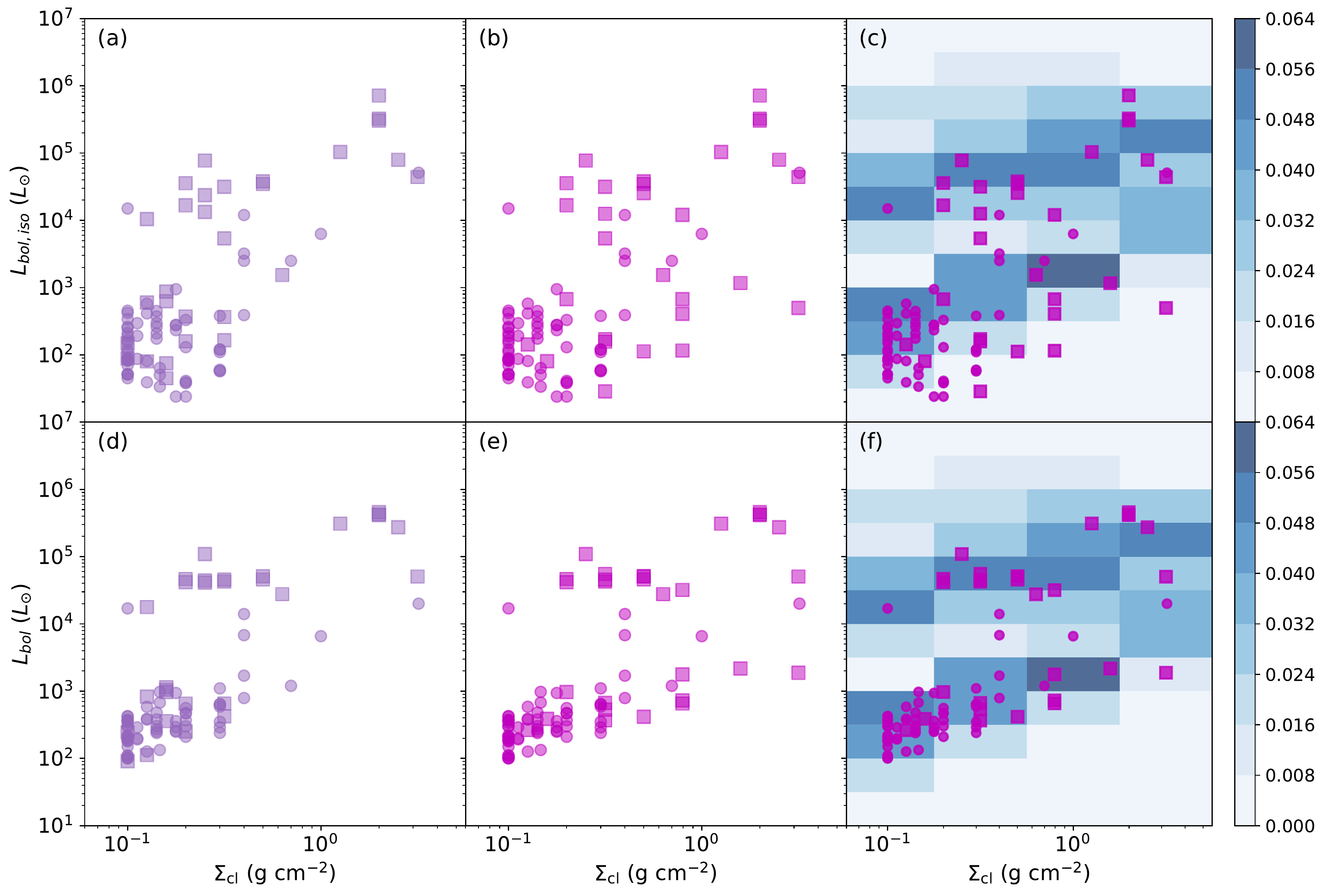}
\caption{
%jctnew - I would keep same color of points as used in previous figure.  - I think using different colors can help avoid confusion.
{\it a) Top Left:} Average protostellar isotropic bolometric
luminosity, $L_{\rm bol,iso}$, versus average clump mass surface
density, $\Sigma_{\rm cl}$, of SOMA sources (squares) and IRDC sources
(circles, Liu et al. 2018; Moser et al. 2020; Liu et al., in prep.),
based on ZT model fits: the average is made for the best five selected
models.
{\it b) Top Middle:} Same as (a), but with the average made for best five
or fewer models with $R_c \la 2R_{\rm ap}$ and $\chi^2 < \chi^2_{\rm min} +5$.
{\it c) Top Right:} Same as (b), but now also showing the distribution of
models in the ZT model grid (shading indicates
the density of models).
{\it d) Bottom Left:} Same as (a), but now for intrinsic bolometric
luminosity, $L_{\rm bol}$.
{\it e) Bottom Middle:} Same as (b), but now for intrinsic bolometric
luminosity, $L_{\rm bol}$.
{\it f) Bottom Right:} Same as (c), but now for intrinsic bolometric
luminosity, $L_{\rm bol}$.
%(a): Diagram of the geometric mean clump surface density versus the
%geometric mean isotropic luminosity of selected ZT models for each
%source in Paper I, II and this work. The best five ZT models are
%selected. (b): Same with (a) but with the best five or fewer models
%with core radius smaller than twice the aperture radius and $\chi^2$
%smaller than $\chi^2_{\rm min} +5$. (c): Distribution of models in the
%model grid. The colorscale of the shade indicates the density of
%models. The magenta squares represent the SOMA sample in
%(b).
}\label{fig:lSigma}
\end{figure*}

Figure~\ref{fig:m_sigma_ms} shows the distribution of values of
$M_{c}$ (i.e., initial core mass), $\Sigma_{\rm cl}$ and $m_{*}$ of
the 29 sources of the SOMA sample to date. With no constraint on the
model core size, there appears to be an absence of protostars with low
$M_{c}$ in high $\Sigma_{\rm cl}$ environments. However, this feature
is not seen after applying the core size constraint, which we regard
as the best method. Thus, the SOMA sample appears to contain
protostars that have a range of initial core masses that can be
present in the full range of protocluster clump mass surface density
environments. However, note that these properties of $M_c$ and
$\Sigma_{\rm cl}$ are not measured directly, but are inferred from the
SED fitting.

We next examine if current protostellar properties depend on
protocluster clump environment mass surface
density. Figure~\ref{fig:sigma_ms} shows $m_*$ versus $\Sigma_{\rm
  cl}$. Figure~\ref{fig:sigma_ms}a, similar to the results shown in
Figure~\ref{fig:m_sigma_ms}a, appears to show a lack of lower-mass
sources in high-$\Sigma_{\rm cl}$ environments. However, this changes
once the core size versus SED aperture constraint is applied
(Fig.~\ref{fig:sigma_ms}b), so we do not consider this to be a real
effect. From the data shown in Fig.~\ref{fig:sigma_ms}b, one potential
trend that we notice is a lack of highest mass
($m_*\gtrsim25\:M_\odot$) protostars in lower mass surface density
environments ($\Sigma_{\rm cl}\lesssim 1\:{\rm g\:cm}^{-2}$). All of
the five protostars with $m_*>25\:M_\odot$ (G45.47+0.05, G45.12+0.13, G305.20+0.21, G309.92+0.48, G35.58-0.03) are inferred to be in
$\Sigma_{\rm cl}>1\:{\rm g\:cm}^{-2}$ environments. In
Fig.~\ref{fig:sigma_ms}c, we see that this trend is not a direct
result of ZT model parameter space sampling, with density of models in
the grid shown by the blue shading. High $m_*$ protostars forming from
cores in low $\Sigma_{\rm cl}$ environments are present among the ZT
models. We note that these models include protostellar outflow
feedback, which sets star formation efficiencies close to 50\%, but do
not include radiative feedback, which would reduce the efficiency (see
below).
%(i.e., such cores do have relatively high star formation
%efficiencies and can reach such protostellar masses).

We further examine how low $\Sigma_{\rm cl}$ models fail for
  high $m_*$ sources in Figure~\ref{fig:lowsigma}. Here we exclude
  G45.12+0.13 because none of the models fit particularly well for
  this source (see Paper II). We can see that the median $\chi^{2}$
  and the smallest $\chi^{2}$ achieved generally decrease with
  $\Sigma_{\rm cl}$. Compared with high $\Sigma_{\rm cl}$ models, low
  $\Sigma_{\rm cl}$ models usually have higher fluxes at shorter
  wavelengths, i.e., $\la 8 \mu$m. These can be higher than the
  observational upper limits, which leads to a significant penalty in
  the fitting. Low $\Sigma_{\rm cl}$ models also tend to have lower
  fluxes at longer wavelength, i.e., $\ga 20 \mu$m. Therefore, they
  deviate away from the shape of the observed SEDs. We also tried
  manually adjusting $A_V$ or $L_{\rm bol}$ of the low $\Sigma_{\rm
    cl}$ models (not shown here), but such changes do not lead to
  significant improvement in model SED shape in comparison to the
  data.

Thus, we conclude there is tentative evidence from the SOMA sample
analyzed so far that the most massive protostars require their cores
to be in $\Sigma_{\rm cl}>1\ {\rm g\:cm}^{-2}$ environments, but
larger further testing with a larger number of sources is clearly
needed to confirm this.

Krumholz \& McKee (2008) proposed that a minimum mass surface density
of $1\:{\rm g\:cm}^{-2}$ is needed for massive star formation, based
on protostellar heating suppression of fragmentation of massive cores
by a population of surrounding lower-mass protostars (these protostars
have higher accretion rates and thus luminosities in higher
$\Sigma_{\rm cl}$ environments). While our result appears to confirm
this prediction, we caution that the Krumholz \& McKee model also
predicts that $10\:M_\odot$ protostars would not be able to form in
$\Sigma_{\rm cl}\lesssim 0.3\:{\rm g\:cm}^{-2}$ environments, which is
inconsistent with the SOMA data. As an alternative, magnetic
suppression of fragmentation to allow the existence of massive,
early-stage cores has been discussed by, e.g., Butler \& Tan (2012),
with evidence of strong, $\sim 1$~mG $B$-fields inferred several cores
in the IRDC 18310-4 region (Beuther et al. 2018).

The assembly of the highest mass pre-stellar cores, e.g., via a
bottom-up process of merging smaller pre-stellar cores together or by
general accumulation of clump gas, is expected to be more efficient in
denser regions and this could provide an explanation, in the context
of core accretion models, of the trends seen in
Figure~\ref{fig:sigma_ms}.

Once cores initiate star formation, then their accretion rates would
also be higher in high surface density environments and this is
expected to allow higher protostellar masses to be formed. Tanaka et
al. (2017) assessed the expected star formation efficiency from cores
due to both radiative and mechamical (i.e., outflow) feedback as a
function of $\Sigma_{\rm cl}$ and found it can decrease by more than a
factor of two for a given initial core as $\Sigma_{\rm cl}$ decreases
from 3.2 to 0.1~$\rm g\:cm^{-2}$ (see
Figure~\ref{fig:efficiency}). The decrease is greatest for more
massive cores, since once they start forming stars with
$m_*\gtrsim20\:M_\odot$, radiative feedback becomes powerful enough to
truncate further accretion. For example, the $\Sigma_{\rm
  cl}=0.1\:{\rm g\:cm^{-2}}$ models shown in
Figure~\ref{fig:efficiency} reach $m_*\simeq10\:M_\odot$ starting from
a $30\:M_\odot$ core, $m_*\simeq20\:M_\odot$ starting from a
$100\:M_\odot$ core, and $m_*\simeq45\:M_\odot$ starting from a
$300\:M_\odot$ core. However, the equivalent $\Sigma_{\rm cl}=1\:{\rm
  g\:cm^{-2}}$ models reach values of $m_*\simeq15, 40,$ and
$100\:M_\odot$, respectively. Thus, in the context of these models, it
is much more difficult to produce, e.g., $30\:M_\odot$ protostars in
low-$\Sigma_{\rm cl}$ environments due to feedback effects, especially
since the pre-stellar core mass function is expected to decline
rapidly with increasing mass.

For competitive accretion models (Bonnell et al. 2001; Wang et
al. 2010),
higher mass surface density environments are also expected to lead to
higher accretion rates and thus will probably also allow formation of
higher-mass stars. However, the equivalent calculations for the effect
of feedback have not yet been carried out for these models.

From an observational analysis of three clouds that are forming
  massive stars compared to several others that are not, Kauffmann et
  al. (2010) proposed a criterion for massive star formation
  equivalent to $\Sigma_{\rm cl} \geq 0.054 (M_{\rm
    cl}/1000\:M_\odot)^{-1/2}\:{\rm g\:cm}^{-2}$, which is relatively
  low compared to the thresholds discussed above. Also, this is a
  value smaller than the minimum of the range probed in the ZT18
  protostellar model grid of $\Sigma_{\rm cl}=0.1 g\:cm^{-2}$.
%criterion, $ \Sigma_{\rm cl} > 780 \ {\rm M_{\odot}\:pc}^{-2} = 0.163\: {\rm g\:cm}^{-2}$,
  Recently, Retes-Romero et al. (2020) studied 128 IRDCs to
  investigate if the Kauffmann et al. criterion predicts which of
  these IRDCs contains massive stars. They found that among the IRDCs
  satisfying this criterion, only one third of them currently contain
  massive YSOs. This may indicate that a higher, more localised value
  of $\Sigma_{\rm cl}$ is needed to form a massive star. For further
  progress on the general question of massive star formation
  thresholds, more direct measures of $\Sigma_{\rm cl}$, e.g., from
  dust continuum emission (in contrast to our indirect methods based
  on model fitting), on scales immediately surrounding the massive
  protostars and comparison to protostellar properties, e.g., as
  derived from SED fitting in the SOMA sample, are needed. However,
  such an analysis, which we defer to a future study, will inevitably
  be sensitive to how and where the protostellar core boundary is
  defined and such sensitivity will also need to be explored.

%the mass-size criterion, or equivalently a threshold density, for
%IRDCs to harbor high-mass YSOs. They found that among those IRDCs
%satisfying the Kauffmann et al. (2010) criterion, $ \Sigma_{\rm cl} >
%780 \ {\rm M_{\odot}\:pc}^{-2} = 0.163 \ {\rm g\:cm}^{-2}$, only one
%third contain massive YSOs.}

In summary, our results indicate, tentatively, that to form the most
massive, $\gtrsim25\:M_\odot$ protostars requires $\gtrsim1\:{\rm
  g\:cm}^{-2}$ protocluster clump environments, although this is based
on a relatively small number of (five) of protostellar sources
that are in this mass range. We have a larger number (about
10) of protostars with $10\:M_\odot\lesssim m_*\lesssim 25\:M_\odot$
that are best fitted by models with $\Sigma_{\rm cl}\lesssim 0.3\:{\rm
  g\:cm^{-2}}$, so that there does not appear to be a particular mass
surface density threshold, in this range, needed to form $10\:M_\odot$
protostars. These environmental dependencies on massive star formation
need confirmation with larger numbers of sources. Such trends are
consistent with several different theoretical expectations from core
accretion models, including that due to decreasing star formation
efficiency due to self-feedback for massive protostars in lower mass
surface density environments.

Finally, we investigate the dependence of $L_{\rm bol, iso}$ and
$L_{\rm bol}$ on $\Sigma_{\rm cl}$ in Figure~\ref{fig:lSigma}. Once
model core size to aperture constraints are applied (panels b and e),
there is no strong correlation present in the overall
distribution. The highest luminosity sources, which have the highest
protostellar masses, are preferentially found in high mass surface
density environments. This is not due to the sources having higher
current accretion rates, since for these high $m_*$ sources, the
accretion luminosity is only a relatively minor component of the total
luminosity. Thus this trend is simply a reflection of those seen in
the mass distribution of the sources.

%There is a seemingly proportional relation as shown in
%Figure~\ref{fig:lSigma}a with a Pearson correlation coefficient of
%$\sim$ 0.73. But once the constraint on the core size is applied as
%shown in Figure~\ref{fig:lSigma}b, the relation becomes much weaker
%with a Pearson correlation coefficient of $\sim$ 0.37. Also the models
%with $\Sigma_{\rm cl} \sim$ 0.1 g\,cm$^{-2}$ are much diminished as
%expected (see previous text). The empty space in the lower right
%corner in both cases is consistent with the empty space in the upper
%left corner in Figure~\ref{fig:sigma_ms}, which if not limited by
%sample incompleteness, indicates that to form a very high mass star
%does require an environment with very high $\Sigma_{\rm cl}$. Like
%$m_{*}$, in Figure~\ref{fig:lSigma}c we also see a slight trend of the
%average $L_{\rm bol}$ at each $\Sigma_{\rm cl}$ increasing with
%$\Sigma_{\rm cl}$. In addition to the same effects mentioned above,
%even for the same $M_{\rm c}$ and $m_{*}$, higher $\Sigma_{\rm cl}$
%will result in higher accretion rates and thus $L_{\rm bol}$ is
%higher.

%\clearpage
\renewcommand{\arraystretch}{0.9}
\begin{deluxetable*}{cccccccccccc}
\tabletypesize{\scriptsize}
\tablecaption{Average Parameters of SOMA Protostars\label{tab:gmean}} 
\tablewidth{18pt}
\tablehead{
\colhead{Source} & \colhead{$M_{\rm c}$} & \colhead{$\Sigma_{\rm cl}$} &\colhead{$m_{*}$} &\colhead{$m_{*}/M_{\rm c}$} &  \colhead{$M_{\rm env}$} & \colhead{$L_{\rm bol, iso}$} & \colhead{$L_{\rm bol}$}  & \colhead{$\theta_{\rm view}$} &\colhead{$\theta_{w,\rm esc}$} &\colhead{$\theta_{\rm view}/\theta_{w,\rm esc}$} & \colhead{$\alpha_{19-37}$}\\
\colhead{} &  \colhead{($M_\odot$)} & \colhead{(g $\rm cm^{-2}$)} & \colhead{($M_{\odot}$)} & \colhead{} & \colhead{($M_{\odot}$)} & \colhead{($L_{\odot}$)} & \colhead{($L_{\odot}$)} & \colhead{($\arcdeg$)} & \colhead{($\arcdeg$)} & \colhead{} & \colhead{} \\
 \vspace{-0.4cm}
}
\startdata
G45.12+0.13 & 403 & 2.0 & 35.5 & 0.09 & 319 & 7.2e+05 & 4.6e+05 & 24 & 21 & 1.12 & 1.05 \\
& 403 & 2.0 & 35.5 & 0.09 & 319 & 7.2e+05 & 4.6e+05 & 24 & 21 & 1.12 & 1.05 \\
\hline\noalign{\smallskip}
G309.92+0.48 & 323 & 2.0 & 33.5 & 0.10 & 251 & 3.3e+05 & 4.2e+05 & 30 & 22 & 1.37 & 2.04 \\
& 323 & 2.0 & 33.5 & 0.10 & 251 & 3.3e+05 & 4.2e+05 & 30 & 22 & 1.37 & 2.04 \\
\hline\noalign{\smallskip}
G35.58-0.03 & 427 & 2.0 & 33.5 & 0.08 & 350 & 3.1e+05 & 4.2e+05 & 29 & 19 & 1.63 & 4.03 \\
& 427 & 2.0 & 33.5 & 0.08 & 350 & 3.1e+05 & 4.2e+05 & 29 & 19 & 1.63 & 4.03 \\
\hline\noalign{\smallskip}
IRAS16562 & 323 & 0.3 & 22.9 & 0.07 & 263 & 7.7e+04 & 1.1e+05 & 43 & 23 & 1.90 & 2.91 \\
& 323 & 0.3 & 22.9 & 0.07 & 263 & 7.7e+04 & 1.1e+05 & 43 & 23 & 1.90 & 2.91 \\
\hline\noalign{\smallskip}
G305.20+0.21 & 110 & 2.5 & 28.5 & 0.26 & 51 & 7.9e+04 & 2.7e+05 & 47 & 38 & 1.24 & 0.82 \\
& 110 & 2.5 & 28.5 & 0.26 & 51 & 7.9e+04 & 2.7e+05 & 47 & 38 & 1.24 & 0.82 \\
\hline\noalign{\smallskip}
G49.27-0.34 & 197 & 3.2 & 12.0 & 0.06 & 174 & 4.4e+04 & 5.1e+04 & 26 & 14 & 1.92 & 4.38 \\
& 197 & 3.2 & 12.0 & 0.06 & 174 & 4.4e+04 & 5.1e+04 & 26 & 14 & 1.92 & 4.38 \\
\hline\noalign{\smallskip}
G339.88-1.26 & 298 & 0.5 & 12.7 & 0.04 & 269 & 3.8e+04 & 4.6e+04 & 36 & 14 & 2.70 & 5.00 \\
& 298 & 0.5 & 12.7 & 0.04 & 269 & 3.8e+04 & 4.6e+04 & 36 & 14 & 2.70 & 5.00 \\
\hline\noalign{\smallskip}
G45.47+0.05 & 260 & 1.3 & 32.8 & 0.13 & 187 & 1.0e+05 & 3.1e+05 & 77 & 27 & 2.80 & 3.01 \\
& 260 & 1.3 & 32.8 & 0.13 & 187 & 1.0e+05 & 3.1e+05 & 77 & 27 & 2.80 & 3.01 \\
\hline\noalign{\smallskip}
CepA & 188 & 0.3 & 14.6 & 0.08 & 148 & 2.4e+04 & 4.4e+04 & 62 & 24 & 3.05 & 5.03 \\
& 132 & 0.5 & 14.6 & 0.11 & 98 & 2.6e+04 & 5.1e+04 & 52 & 26 & 1.96 & 5.03 \\
\hline\noalign{\smallskip}
IRAS20126 & 109 & 0.3 & 15.5 & 0.14 & 67 & 1.3e+04 & 4.1e+04 & 67 & 35 & 2.14 & 2.54 \\
& 95 & 0.3 & 17.8 & 0.19 & 49 & 1.2e+04 & 5.5e+04 & 67 & 42 & 1.60 & 2.54 \\
\hline\noalign{\smallskip}
AFGL4029 & 65 & 0.3 & 16.8 & 0.26 & 17 & 5.4e+03 & 4.5e+04 & 70 & 54 & 1.35 & 2.09 \\
& 65 & 0.3 & 16.8 & 0.26 & 17 & 5.4e+03 & 4.5e+04 & 70 & 54 & 1.35 & 2.09 \\
\hline\noalign{\smallskip}
NGC7538\_IRS9 & 245 & 0.2 & 16.4 & 0.07 & 196 & 3.6e+04 & 4.7e+04 & 31 & 22 & 1.44 & 1.52 \\
& 245 & 0.2 & 16.4 & 0.07 & 196 & 3.6e+04 & 4.7e+04 & 31 & 22 & 1.44 & 1.52 \\
\hline\noalign{\smallskip}
G35.20-0.74 & 190 & 0.5 & 14.6 & 0.08 & 154 & 3.5e+04 & 5.1e+04 & 42 & 20 & 2.07 & 3.53 \\
& 190 & 0.5 & 14.6 & 0.08 & 154 & 3.5e+04 & 5.1e+04 & 42 & 20 & 2.07 & 3.53 \\
\hline\noalign{\smallskip}
AFGL437 & 133 & 0.2 & 16.4 & 0.12 & 80 & 1.7e+04 & 4.2e+04 & 60 & 36 & 1.64 & 0.86 \\
& 133 & 0.2 & 16.4 & 0.12 & 80 & 1.7e+04 & 4.2e+04 & 60 & 36 & 1.64 & 0.86 \\
\hline\noalign{\smallskip}
IRAS07299 & 206 & 0.1 & 10.8 & 0.05 & 168 & 1.0e+04 & 1.8e+04 & 83 & 21 & 4.85 & 2.51 \\
& 71 & 0.8 & 11.7 & 0.16 & 44 & 1.2e+04 & 3.2e+04 & 57 & 32 & 1.77 & 2.51 \\
\hline\noalign{\smallskip}
S235 & 41 & 0.6 & 12.4 & 0.30 & 6 & 1.5e+03 & 2.8e+04 & 77 & 62 & 1.23 & 0.46 \\
& 41 & 0.6 & 12.4 & 0.30 & 6 & 1.5e+03 & 2.8e+04 & 77 & 62 & 1.23 & 0.46 \\
\hline\noalign{\smallskip}
IRAS22198 & 63 & 0.1 & 3.5 & 0.06 & 55 & 6.0e+02 & 8.3e+02 & 65 & 19 & 3.52 & 3.03 \\
& 43 & 0.2 & 3.5 & 0.08 & 34 & 6.7e+02 & 9.7e+02 & 43 & 23 & 1.86 & 3.03 \\
\hline\noalign{\smallskip}
NGC2071 & 32 & 0.2 & 3.0 & 0.09 & 19 & 3.7e+02 & 6.5e+02 & 49 & 29 & 1.80 & 1.32 \\
& 10 & 3.2 & 4.0 & 0.40 & 2 & 5.0e+02 & 1.9e+03 & 58 & 56 & 1.04 & 1.32 \\
\hline\noalign{\smallskip}
CepE & 32 & 0.1 & 1.5 & 0.05 & 26 & 1.5e+02 & 2.4e+02 & 79 & 21 & 5.05 & 3.60 \\
& 24 & 0.1 & 1.5 & 0.06 & 18 & 1.4e+02 & 2.6e+02 & 70 & 24 & 3.70 & 3.60 \\
\hline\noalign{\smallskip}
L1206\_A & 156 & 0.2 & 4.0 & 0.03 & 140 & 8.7e+02 & 1.1e+03 & 81 & 14 & 8.64 & 5.33 \\
& 24 & 1.6 & 2.6 & 0.11 & 17 & 1.2e+03 & 2.2e+03 & 35 & 25 & 1.40 & 5.33 \\
\hline\noalign{\smallskip}
L1206\_B & 16 & 0.2 & 3.6 & 0.22 & 2 & 7.5e+01 & 9.7e+02 & 66 & 60 & 1.09 & -0.33 \\
& 12 & 0.2 & 2.2 & 0.17 & 3 & 8.0e+01 & 3.9e+02 & 55 & 50 & 1.09 & -0.33 \\
\hline\noalign{\smallskip}
IRAS22172\_mir2 & 24 & 0.2 & 2.0 & 0.09 & 17 & 6.3e+02 & 3.6e+02 & 29 & 28 & 1.02 & -0.17 \\
& 11 & 0.8 & 2.3 & 0.20 & 4 & 6.7e+02 & 7.3e+02 & 40 & 42 & 0.92 & -0.17 \\
\hline\noalign{\smallskip}
IRAS22172\_mir3 & 18 & 0.3 & 2.0 & 0.11 & 8 & 1.6e+02 & 4.2e+02 & 44 & 35 & 1.35 & 1.53 \\
& 15 & 0.3 & 2.6 & 0.17 & 6 & 1.6e+02 & 5.3e+02 & 54 & 42 & 1.34 & 1.53 \\
\hline\noalign{\smallskip}
IRAS22172\_mir1 & 16 & 0.2 & 1.5 & 0.09 & 10 & 1.6e+02 & 2.5e+02 & 37 & 31 & 1.22 & 1.54 \\
& 13 & 0.3 & 2.0 & 0.15 & 5 & 1.7e+02 & 3.7e+02 & 45 & 39 & 1.13 & 1.54 \\
\hline\noalign{\smallskip}
IRAS21391\_bima2 & 26 & 0.1 & 0.7 & 0.03 & 22 & 8.0e+01 & 1.1e+02 & 34 & 16 & 2.52 & 4.07 \\
& 10 & 0.8 & 2.3 & 0.23 & 3 & 1.2e+02 & 6.6e+02 & 73 & 45 & 1.64 & 4.07 \\
\hline\noalign{\smallskip}
IRAS21391\_bima3 & 98 & 0.1 & 0.5 & 0.01 & 97 & 8.9e+01 & 9.1e+01 & 54 & 5 & 11.10 & 5.03 \\
& 10 & 0.5 & 1.5 & 0.15 & 5 & 1.1e+02 & 4.2e+02 & 62 & 38 & 1.68 & 5.03 \\
\hline\noalign{\smallskip}
IRAS21391\_mir48 & 16 & 0.2 & 4.1 & 0.25 & 2 & 4.5e+01 & 1.0e+03 & 75 & 63 & 1.22 & 1.54 \\
& 10 & 0.3 & 4.0 & 0.40 & 1 & 2.9e+01 & 6.7e+02 & 89 & 68 & 1.30 & 1.54 \\
\hline\noalign{\smallskip}
G305A & 262 & 0.3 & 12.7 & 0.05 & 231 & 3.1e+04 & 4.3e+04 & 68 & 16 & 4.26 & 6.20 \\
& 262 & 0.3 & 12.7 & 0.05 & 231 & 3.1e+04 & 4.3e+04 & 68 & 16 & 4.26 & 6.20 \\
\hline\noalign{\smallskip}
IRAS16562\_N & 25 & 0.3 & 2.3 & 0.09 & 15 & 3.7e+02 & 6.5e+02 & 39 & 29 & 1.40 & 1.01 \\
& 13 & 0.8 & 3.5 & 0.26 & 3 & 4.1e+02 & 1.8e+03 & 57 & 49 & 1.15 & 1.01 \\
\enddata
\tablecomments{
The first line of each source shows the average (geometric mean, except for $\theta_{\rm view}$, $\theta_{\rm w,esc}$ and $\theta_{\rm view}/\theta_{w,\rm esc}$ for which arithmetic means are evaluated) of
the values of the best five models without any core size versus
aperture constraint applied. The second line shows the results of the
best five or fewer models with $R_c \leq 2R_{\rm ap}$ and $\chi^2 \leq \chi^2_{\rm min} +5$.}
\end{deluxetable*}

\section{Conclusions}

We have presented the results of MIR and FIR observations carried out
towards 14 protostars in the SOMA survey, with most of them being
intermediate-mass protostars. Following our standard methods developed
in Papers I \& II, we have built their SEDs with additional archival
{\it Spitzer}, {\it Herschel} and {\it IRAS} data and fit them with
Zhang \& Tan (2018) RT models of massive star formation via the
Turbulent Core Accretion paradigm.
%Our goal is to compare intermediate-mass star formation with high-mass
%star formation together with the sample of 15 protostars in Paper I \&
%II.
We have also supplemented the sample with protostars identified in
Infrared Dark Clouds (IRDCs) and expected to be at very early stages
in their evolution. By these methods we have extended the range of
masses, luminosities and evolutionary stages of protostellar sources
that have been analyzed in an uniform manner to test core accretion
theory.
%Through the MIR to FIR imaging and SED fitting, we test the star
%formation models in a large mass range and investigate properties and
%trends revealed by the models.
Our main results and conclusions are:
%summarized as follows.

1. The intermediate-mass protostars presented in this paper appear
relatively compact at 20 -- 40 $\mu$m, compared to the high-mass
protostars in Papers I \& II, whose 20 -- 40 $\mu$m images more
clearly show extension along their outflow axes. The protostars
presented here are forming in a variety of protocluster environments,
as revealed by NIR images. Higher resolution sub-mm images often
reveal presence of secondary dense gas cores within 0.1~pc (in
projection).

2. The SEDs of the 14 protostars of this paper are generally fit quite
well by the ZT models, but there are significant degeneracies among
acceptable models. These degeneracies in key model parameters, i.e.,
initial core mass, $M_c$, clump mass surface density, $\Sigma_{\rm
  cl}$, and current protostellar mass, $m_*$, are typically larger
than for the higher mass protostars, but this is often a reflection of
the more limited wavelength coverage of the intermediate-mass sources,
which are often away from the Galactic plane and thus lacking, e.g.,
longer wavelength {\it Herschel} data.
%In five of the protostars, a few millimeter sources are revealed
%within 0.04 pc in projection of the MIR source and so star formation
%may be leading to multiple star formation from within intermediate-
%and low-mass protostars may form with companions.
%good, though for the intermediate- and low-mass sources the best
%
For the sources analyzed here, we find that well-fitting models can
often have $R_{\rm c} > R_{\rm ap}$. Thus we have applied a further
constraint that model core radii should not exceed the aperture radius
used to define the SED by more than a factor of two.

%We have examined the model images and found the emission predicted by
%the models is highly concentrated. We have investigated the
%distributions and relations of model parameters with different
%constraints on $R_{\rm c}$. Models with high $M_{\rm env}$ and low
%$\Sigma_{\rm cl}$ tend to be excluded with more stringent constraints
%on $R_{\rm c}$. Another discrepancy lies in the estimates of $M_{\rm
%  env}$, $\dot{m}_{*}$ in several sources with previous results
%derived from IR and especially millimeter observations in
%literature. We have examined the whole model grid space and found that
%usually the discrepancy is either due to limited number of models in
%the grid available for the values of the previously estimated
%quantities, or due to the extremely high $\chi^2$ of the models
%associated with those values.

3. The SOMA sources analyzed in this paper and Papers I \& II
span a range of bolometric luminosities of $\sim10^{2}\: L_{\odot}$ to
$\sim10^{6} \: L_{\odot}$. The isotropic luminosity can be quite
different from the intrinsic luminosity, indicating a significant
flashlight effect in the sources.

4. The presented SOMA sample spans a range of light to mass ratios of
$L_{\rm bol}/M_{\rm env}$ from $\sim10\:L_\odot/M_\odot$ to
$\sim10^4\:L_\odot/M_\odot$. The addition of IRDC protostars extends
this range down to $\sim1\:L_\odot/M_\odot$, which is expected to be
near the very earliest phases of the star formation
process. Relatively late stages of evolution are currently missing
from the sample.

5. The SED shape, as measured by the spectral index from 19 to
  37 microns, shows trends with outflow opening angle, ratio of
  viewing angle to outflow opening angle, and evolutionary stage,
  i.e., $m_*/M_c$. However, such trends are features that are inherent
  in the ZT18 models and independent confirmation, e.g., from high
  resolution continuum and line studies of outflows and outflow
  cavities, is needed.

6. Protostars from low masses up to $\sim25\:M_\odot$ are
  inferred to be forming at all the clump mass surface densities
  probed by the models, i.e., from 0.1 to 3~$\rm g\:cm^{-2}$. However,
  to form protostars with $>25\:M_\odot$ appears to require
  $\Sigma_{\rm cl}\gtrsim 1\:{\rm g\:cm}^{-2}$ clump
  environments. Larger numbers of sources in this mass range are
  needed to confirm this result. While this finding is consistent with
  several possible theoretical expectations, we favor one based on
  internal feedback in the protostellar core, which becomes less
  effective for the denser cores that are associated with higher
  $\Sigma_{\rm cl}$ environments (Tanaka et al. 2017).

%Based on the sample in this paper and Paper I \& II, no matter what
%the constraint on $R_{\rm c}$ is, it seems $\Sigma_{\rm cl}$ can be as
%low as 0.2 g\,cm$^{-2}$ to form a protostar of $\sim$ 10
%$M_{\odot}$. But to form a very massive protostar for instance above
%$\sim$ 25 $M_{\odot}$, $\Sigma_{\rm cl}$ has to be at least as high as
%1 g\,cm$^{-2}$.

%5. If we assume the MIR to FIR emission of the protostars are mostly
%concentrated, which removes, the constraint on $R_{\rm c}$, we see a
%seemingly proportional trend of $m_{*}$ and $L_{\rm bol,iso}$ with
%$\Sigma_{\rm cl}$.

%7. Two protostars are shown to form in cluster from high-resolution
%NIR images, with one located at the center and the other located at
%the edge. The other protostars appear more isolated, but there is also
%possibility that their NIR image quality is not high enough to clearly
%reveal their environments.

\acknowledgments We thank the anonymous referee for helping improve the paper. M.L. acknowledges funding from the Jefferson Scholars
Foundation. M.L. and J.C.T. acknowledge funding from
NASA/USRA/SOFIA. J.C.T. acknowledges support from NSF grant
AST1411527, VR grant 2017-04522 and ERC project 788829 -
MSTAR. Y.Z. acknowledges support from JSPS KAKENHI grant JP19K14774.
K.E.I.T. acknowledges support from NAOJ ALMA Scientific Research grant
No. 2017-05A, and JSPS KAKENHI Grant Nos. JP19H05080, JP19K14760.
Y.L.Y. acknowledges support from a Virginia Initiative on Cosmic
Origins (VICO) postdoctoral fellowship. R.F. acknowledges support from
a Chalmers Initiative on Cosmic Origins (CICO) postdoctoral
fellowship.


\begin{thebibliography}{}
\bibitem[Andre et al.(1993)]{1993ApJ...406..122A} Andre, P., Ward-Thompson, D., \& Barsony, M.\ 1993, \apj, 406, 122
\bibitem[Anthony-Twarog(1982)]{1982AJ.....87.1213A} Anthony-Twarog, B.~J.\ 1982, \aj, 87, 1213
\bibitem[Ayala et al.(2000)]{2000AJ....120..909A} Ayala, S., Noriega-Crespo, A., Garnavich, P.~M., et al.\ 2000, \aj, 120, 909
\bibitem[Bally(1982)]{1982ApJ...261..558B} Bally, J.\ 1982, \apj, 261, 558
\bibitem[Beltr{\'a}n et al.(2002)]{2002ApJ...573..246B} Beltr{\'a}n, M.~T., Girart, J.~M., Estalella, R., et al.\ 2002, \apj, 573, 246
\bibitem[Beltr{\'a}n et al.(2006)]{2006A&A...457..865B} Beltr{\'a}n, M.~T., Girart, J.~M., \& Estalella, R.\ 2006, \aap, 457, 865
\bibitem[Beltr{\'a}n et al.(2008)]{2008A&A...481...93B} Beltr{\'a}n, M.~T., Estalella, R., Girart, J.~M., et al.\ 2008, \aap, 481, 93
\bibitem[Beltr{\'a}n et al.(2009)]{2009A&A...504...97B} Beltr{\'a}n, M.~T., Massi, F., L{\'o}pez, R., et al.\ 2009, \aap, 504, 97
\bibitem[Beltr{\'a}n(2015)]{2015Ap&SS.355..283B} Beltr{\'a}n, M.~T.\ 2015, \apss, 355, 283
\bibitem[Beuther et al.(2018)]{2018A&A...614A..64B} Beuther, H., Soler, J.~D., Vlemmings, W., et al.\ 2018, \aap, 614, A64
\bibitem[Boley et al.(2009)]{2009MNRAS.399..778B} Boley, P.~A., Sobolev, A.~M., Krushinsky, V.~V., et al.\ 2009, \mnras, 399, 778
\bibitem[Bonnell et al.(2001)]{2001MNRAS.323..785B} Bonnell, I.~A., Bate, M.~R., Clarke, C.~J., \& Pringle, J.~E.\ 2001, \mnras, 323, 785
\bibitem[Bontemps et al.(1996)]{1996A&A...311..858B} Bontemps, S., Andre, P., Terebey, S., et al.\ 1996, \aap, 311, 858
\bibitem[Burns et al.(2015)]{2015MNRAS.453.3163B} Burns, R.~A., Imai, H., Handa, T., et al.\ 2015, \mnras, 453, 3163
\bibitem[Butler \& Tan(2009)]{2009ApJ...696..484B} Butler, M.~J., \& Tan, J.~C.\ 2009, \apj, 696, 484
\bibitem[Butler, \& Tan(2012)]{2012ApJ...754....5B} Butler, M.~J., \& Tan, J.~C.\ 2012, \apj, 754, 5.
\bibitem[Cabrit \& Bertout(1992)]{1992A&A...261..274C} Cabrit, S., \& Bertout, C.\ 1992, \aap, 261, 274
\bibitem[Cesaroni et al.(1999)]{1999A&AS..136..333C} Cesaroni, R., Felli, M., \& Walmsley, C.~M.\ 1999, \aaps, 136, 333
\bibitem[Choudhury et al.(2010)]{2010ApJ...717.1067C} Choudhury, R., Mookerjea, B., \& Bhatt, H.~C.\ 2010, \apj, 717, 1067
\bibitem[Codella et al.(2001)]{2001A&A...376..271C} Codella, C., Bachiller, R., Nisini, B., et al.\ 2001, \aap, 376, 271
\bibitem[Crimier et al.(2010)]{2010A&A...516A.102C} Crimier, N., Ceccarelli, C., Alonso-Albi, T., et al.\ 2010, \aap, 516, A102
\bibitem[Davies et al.(2011)]{2011MNRAS.416..972D} Davies, B., Hoare, M.~G., Lumsden, S.~L., et al.\ 2011, \mnras, 416, 972
\bibitem[De Buizer et al.(2017)]{2017ApJ...843...33D} De Buizer, J.~M., Liu, M., Tan, J.~C., et al.\ 2017, \apj, 843, 33
\bibitem[Dewangan \& Anandarao(2011)]{2011MNRAS.414.1526D} Dewangan, L.~K., \& Anandarao, B.~G.\ 2011, \mnras, 414, 1526
\bibitem[Dewangan et al.(2016)]{2016ApJ...819...66D} Dewangan, L.~K., Ojha, D.~K., Luna, A., et al.\ 2016, \apj, 819, 66
\bibitem[Dewangan \& Ojha(2017)]{2017ApJ...849...65D} Dewangan, L.~K., \& Ojha, D.~K.\ 2017, \apj, 849, 65
\bibitem[Eisl{\"o}ffel(2000)]{2000A&A...354..236E} Eisl{\"o}ffel, J.\ 2000, \aap, 354, 236
\bibitem[Evans \& Blair(1981)]{1981ApJ...246..394E} Evans, N.~J., \& Blair, G.~N.\ 1981, \apj, 246, 394
\bibitem[Fazio et al.(2004)]{2004ApJS..154...10F} Fazio, G.~G., Hora, J.~L., Allen, L.~E., et al.\ 2004, \apjs, 154, 10
\bibitem[Fedriani et al.(2019)]{2019NatCo..10.3630F} Fedriani, R., Caratti o Garatti, A., Purser, S.~J.~D., et al.\ 2019, Nature Communications, 10, 3630
\bibitem[Felli et al.(1992)]{1992A&A...255..293F} Felli, M., Palagi, F., \& Tofani, G.\ 1992, \aap, 255, 293
\bibitem[Felli et al.(1997)]{1997A&A...320..594F} Felli, M., Testi, L., Valdettaro, R., et al.\ 1997, \aap, 320, 594
\bibitem[Felli et al.(2004)]{2004A&A...420..553F} Felli, M., Massi, F., Navarrini, A., et al.\ 2004, \aap, 420, 553
\bibitem[Felli et al.(2006)]{2006A&A...453..911F} Felli, M., Massi, F., Robberto, M., et al.\ 2006, \aap, 453, 911
\bibitem[Fontani et al.(2004)]{2004A&A...424..179F} Fontani, F., Cesaroni, R., Testi, L., et al.\ 2004, \aap, 424, 179
\bibitem[Fuente et al.(2005)]{2005A&A...433..535F} Fuente, A., Rizzo, J.~R., Caselli, P., et al.\ 2005, \aap, 433, 535
\bibitem[Fuente et al.(2009)]{2009A&A...507.1475F} Fuente, A., Castro-Carrizo, A., Alonso-Albi, T., et al.\ 2009, \aap, 507, 1475
\bibitem[Fujisawa et al.(2014)]{2014PASJ...66...78F} Fujisawa, K., Takase, G., Kimura, S., et al.\ 2014, \pasj, 66, 78
\bibitem[Griffin et al.(2010)]{2010A&A...518L...3G} Griffin, M.~J., Abergel, A., Abreu, A., et al.\ 2010, \aap, 518, L3
\bibitem[Gueth et al.(2001)]{2001A&A...375.1018G} Gueth, F., Schilke, P., \& McCaughrean, M.~J.\ 2001, \aap, 375, 1018
\bibitem[Gusdorf et al.(2017)]{2017A&A...602A...8G} Gusdorf, A., Anderl, S., Lefloch, B., et al.\ 2017, \aap, 602, A8
\bibitem[Hatchell et al.(2007)]{2007A&A...472..187H} Hatchell, J., Fuller, G.~A., \& Richer, J.~S.\ 2007, \aap, 472, 187
\bibitem[Herter et al.(2013)]{2013PASP..125.1393H} Herter, T.~L., Vacca, W.~D., Adams, J.~D., et al.\ 2013, \pasp, 125, 1393
\bibitem[Hirota et al.(2008)]{2008PASJ...60..961H} Hirota, T., Ando, K., Bushimata, T., et al.\ 2008, \pasj, 60, 961
\bibitem[Israel \& Felli(1978)]{1978A&A....63..325I} Israel, F.~P., \& Felli, M.\ 1978, \aap, 63, 325
\bibitem[Kauffmann et al.(2010)]{2010ApJ...716..433K} Kauffmann, J., Pillai, T., Shetty, R., et al.\ 2010, \apj, 716, 433
\bibitem[Klein et al.(2005)]{2005ApJS..161..361K} Klein, R., Posselt, B., Schreyer, K., et al.\ 2005, \apjs, 161, 361
\bibitem[Krassner et al.(1982)]{1982A&A...109..223K} Krassner, J., Pipher, J.~L., Sharpless, S., et al.\ 1982, \aap, 109, 223
\bibitem[Krumholz \& McKee(2008)]{2008Natur.451.1082K} Krumholz, M.~R., \& McKee, C.~F.\ 2008, \nat, 451, 1082
\bibitem[Kurtz et al.(2004)]{2004ApJS..155..149K} Kurtz, S., Hofner, P., \& {\'A}lvarez, C.~V.\ 2004, \apjs, 155, 149
\bibitem[Lawrence et al.(2007)]{2007MNRAS.379.1599L} Lawrence, A., Warren, S.~J., Almaini, O., et al.\ 2007, \mnras, 379, 1599.
\bibitem[Lefloch et al.(1996)]{1996A&A...313L..17L} Lefloch, B., Eisloeffel, J., \& Lazareff, B.\ 1996, \aap, 313, L17
\bibitem[Lefloch et al.(2011)]{2011A&A...527L...3L} Lefloch, B., Cernicharo, J., Pacheco, S., et al.\ 2011, \aap, 527, L3
\bibitem[Lefloch et al.(2015)]{2015A&A...581A...4L} Lefloch, B., Gusdorf, A., Codella, C., et al.\ 2015, \aap, 581, A4
\bibitem[Liu et al.(2018)]{2018ApJ...862..105L} Liu, M., Tan, J.~C., Cheng, Y., et al.\ 2018, \apj, 862, 105
\bibitem[Liu et al.(2019)]{2019ApJ...874...16L} Liu, M., Tan, J.~C., De Buizer, J.~M., et al.\ 2019, \apj, 874, 16
\bibitem[Matthews(1979)]{1979A&A....75..345M} Matthews, H.~I.\ 1979, \aap, 75, 345
\bibitem[McCutcheon et al.(1991)]{1991AJ....101.1435M} McCutcheon, W.~H., Dewdney, P.~E., Purton, C.~R., et al.\ 1991, \aj, 101, 1435
\bibitem[McKee \& Tan(2003)]{2003ApJ...585..850M} McKee, C.~F., \& Tan, J.~C.\ 2003, \apj, 585, 850
\bibitem[Molinari et al.(1996)]{1996A&A...308..573M} Molinari, S., Brand, J., Cesaroni, R., et al.\ 1996, \aap, 308, 573
\bibitem[Molinari et al.(2002)]{2002ApJ...570..758M} Molinari, S., Testi, L., Rodr{\'\i}guez, L.~F., et al.\ 2002, \apj, 570, 758
\bibitem[Molinari et al.(2008)]{2008A&A...481..345M} Molinari, S., Pezzuto, S., Cesaroni, R., et al.\ 2008, \aap, 481, 345
\bibitem[Molinari et al.(2016)]{2016A&A...591A.149M} Molinari, S., Schisano, E., Elia, D., et al.\ 2016, \aap, 591, A149
\bibitem[Moro-Mart{\'\i}n et al.(2001)]{2001ApJ...555..146M} Moro-Mart{\'\i}n, A., Noriega-Crespo, A., Molinari, S., et al.\ 2001, \apj, 555, 146
\bibitem[Moser et al.(2020)]{2019arXiv190712560M} Moser, E., Liu, M., Tan, J.~C., et al.\ 2019, arXiv e-prints, arXiv:1907.12560, accepted to ApJ.
\bibitem[Neri et al.(2007)]{2007A&A...468L..33N} Neri, R., Fuente, A., Ceccarelli, C., et al.\ 2007, \aap, 468, L33
\bibitem[Neugebauer et al.(1984)]{1984ApJ...278L...1N} Neugebauer, G., Habing, H.~J., van Duinen, R., et al.\ 1984, \apjl, 278, L1
\bibitem[Nisini et al.(2001)]{2001A&A...376..553N} Nisini, B., Massi, F., Vitali, F., et al.\ 2001, \aap, 376, 553
\bibitem[Ospina-Zamudio et al.(2018)]{2018A&A...618A.145O} Ospina-Zamudio, J., Lefloch, B., Ceccarelli, C., et al.\ 2018, \aap, 618, A145
\bibitem[Palau et al.(2010)]{2010A&A...510A...5P} Palau, A., S{\'a}nchez-Monge, {\'A}., Busquet, G., et al.\ 2010, \aap, 510, A5
\bibitem[Palau et al.(2013)]{2013ApJ...762..120P} Palau, A., Fuente, A., Girart, J.~M., et al.\ 2013, \apj, 762, 120
\bibitem[Palla \& Stahler(1993)]{1993ApJ...418..414P} Palla, F., \& Stahler, S.~W.\ 1993, \apj, 418, 414
\bibitem[Palla et al.(1993)]{1993A&A...280..599P} Palla, F., Cesaroni, R., Brand, J., et al.\ 1993, \aap, 280, 599
\bibitem[Patel et al.(2000)]{2000ApJ...538..268P} Patel, N.~A., Greenhill, L.~J., Herrnstein, J., et al.\ 2000, \apj, 538, 268
\bibitem[Ressler \& Shure(1991)]{1991AJ....102.1398R} Ressler, M.~E., \& Shure, M.\ 1991, \aj, 102, 1398
\bibitem[Retes-Romero et al.(2020)]{2020ApJ...897...53R} Retes-Romero, R., Mayya, Y.~D., Luna, A., et al.\ 2020, \apj, 897, 53
\bibitem[Robitaille et al.(2006)]{2006ApJS..167..256R} Robitaille, T.~P., Whitney, B.~A., Indebetouw, R., et al.\ 2006, \apjs, 167, 256
\bibitem[Robitaille et al.(2007)]{2007ApJS..169..328R} Robitaille, T.~P., Whitney, B.~A., Indebetouw, R., et al.\ 2007, The Astrophysical Journal Supplement Series, 169, 328.
\bibitem[Rosero et al.(2019)]{2019ApJ...873...20R} Rosero, V., Tanaka, K.~E.~I., Tan, J.~C., et al.\ 2019, \apj, 873, 20
\bibitem[Rygl et al.(2010)]{2010A&A...511A...2R} Rygl, K.~L.~J., Brunthaler, A., Reid, M.~J., et al.\ 2010, \aap, 511, A2
\bibitem[Saito et al.(2007)]{2007ApJ...659..459S} Saito, H., Saito, M., Sunada, K., et al.\ 2007, \apj, 659, 459
\bibitem[S{\'a}nchez-Monge et al.(2008)]{2008A&A...485..497S} S{\'a}nchez-Monge, {\'A}., Palau, A., Estalella, R., et al.\ 2008, \aap, 485, 497
\bibitem[S{\'a}nchez-Monge et al.(2010)]{2010ApJ...721L.107S} S{\'a}nchez-Monge, {\'A}., Palau, A., Estalella, R., et al.\ 2010, \apjl, 721, L107
\bibitem[Saraceno et al.(1996)]{1996A&A...315L.293S} Saraceno, P., Ceccarelli, C., Clegg, P., et al.\ 1996, \aap, 315, L293
\bibitem[Sargent(1977)]{1977ApJ...218..736S} Sargent, A.~I.\ 1977, \apj, 218, 736
\bibitem[Serabyn et al.(1993)]{1993ApJ...404..247S} Serabyn, E., Guesten, R., \& Mundy, L.\ 1993, \apj, 404, 247
\bibitem[Seth et al.(2002)]{2002ApJ...581..325S} Seth, A.~C., Greenhill, L.~J., \& Holder, B.~P.\ 2002, \apj, 581, 325
\bibitem[Shepherd \& Watson(2002)]{2002ApJ...566..966S} Shepherd, D.~S., \& Watson, A.~M.\ 2002, \apj, 566, 966
\bibitem[Shimoikura et al.(2016)]{2016ApJ...832..205S} Shimoikura, T., Dobashi, K., Matsumoto, T., et al.\ 2016, \apj, 832, 205
\bibitem[Skrutskie et al.(2006)]{2006AJ....131.1163S} Skrutskie, M.~F., Cutri, R.~M., Stiening, R., et al.\ 2006, \aj, 131, 1163
\bibitem[Smith \& Beck(1994)]{1994ApJ...420..643S} Smith, H.~A., \& Beck, S.~C.\ 1994, \apj, 420, 643
\bibitem[Snell \& Bally(1986)]{1986ApJ...303..683S} Snell, R.~L., \& Bally, J.\ 1986, \apj, 303, 683
\bibitem[Staff et al.(2019)]{2019ApJ...882..123S} Staff, J.~E., Tanaka, K.~E.~I., \& Tan, J.~C.\ 2019, \apj, 882, 123
\bibitem[Stojimirovi{\'c} et al.(2008)]{2008ApJ...679..557S} Stojimirovi{\'c}, I., Snell, R.~L., \& Narayanan, G.\ 2008, \apj, 679, 557
\bibitem[Sugitani et al.(1989)]{1989ApJ...342L..87S} Sugitani, K., Fukui, Y., Mizuni, A., et al.\ 1989, \apjl, 342, L87
\bibitem[Sugitani et al.(2000)]{2000AJ....119..323S} Sugitani, K., Matsuo, H., Nakano, M., et al.\ 2000, \aj, 119, 323
\bibitem[Takahashi et al.(2012)]{2012ApJ...752...10T} Takahashi, S., Saigo, K., Ho, P.~T.~P., et al.\ 2012, \apj, 752, 10
\bibitem[Tanaka et al.(2017)]{2017ApJ...835...32T} Tanaka, K.~E.~I., Tan, J.~C., \& Zhang, Y.\ 2017, \apj, 835, 32
\bibitem[Tofani et al.(1995)]{1995A&AS..112..299T} Tofani, G., Felli, M., Taylor, G.~B., et al.\ 1995, \aaps, 112, 299
\bibitem[Torrelles et al.(1998)]{1998ApJ...505..756T} Torrelles, J.~M., G{\'o}mez, J.~F., Rodr{\'\i}guez, L.~F., et al.\ 1998, \apj, 505, 756
\bibitem[Trinidad et al.(2009)]{2009ApJ...706..244T} Trinidad, M.~A., Rodr{\'\i}guez, T., \& Rodr{\'\i}guez, L.~F.\ 2009, \apj, 706, 244
\bibitem[Urquhart et al.(2018)]{2018MNRAS.473.1059U} Urquhart, J.~S., K{\"o}nig, C., Giannetti, A., et al.\ 2018, \mnras, 473, 1059
\bibitem[Valdettaro et al.(2005)]{2005A&A...443..535V} Valdettaro, R., Palla, F., Brand, J., et al.\ 2005, \aap, 443, 535
\bibitem[van Kempen et al.(2012)]{2012ApJ...751..137V} van Kempen, T.~A., Longmore, S.~N., Johnstone, D., et al.\ 2012, \apj, 751, 137
\bibitem[Velusamy et al.(2011)]{2011ApJ...741...60V} Velusamy, T., Langer, W.~D., Kumar, M.~S.~N., et al.\ 2011, \apj, 741, 60
\bibitem[Wang et al.(2010)]{2010ApJ...709...27W} Wang, P., Li, Z.-Y., Abel, T., \& Nakamura, F.\ 2010, \apj, 709, 27
\bibitem[Wilking et al.(1989)]{1989ApJ...345..257W} Wilking, B.~A., Mundy, L.~G., Blackwell, J.~H., et al.\ 1989, \apj, 345, 257
\bibitem[Wilking et al.(1993)]{1993AJ....106..250W} Wilking, B., Mundy, L., McMullin, J., et al.\ 1993, \aj, 106, 250
\bibitem[Wouterloot \& Walmsley(1986)]{1986A&A...168..237W} Wouterloot, J.~G.~A., \& Walmsley, C.~M.\ 1986, \aap, 168, 237
\bibitem[Wu et al.(2004)]{2004A&A...426..503W} Wu, Y., Wei, Y., Zhao, M., et al.\ 2004, \aap, 426, 503
\bibitem[Young et al.(2012)]{2012SPIE.8444E..10Y} Young, E.~T., Herter, T.~L., G{\"u}sten, R., et al.\ 2012, \procspie, 8444, 844410
\bibitem[Zapata et al.(2007)]{2007A&A...471L..59Z} Zapata, L.~A., Ho, P.~T.~P., Rodr{\'\i}guez, L.~F., et al.\ 2007, \aap, 471, L59
\bibitem[Zhang, \& Tan(2018)]{2018ApJ...853...18Z} Zhang, Y., \& Tan, J.~C.\ 2018, \apj, 853, 18
\end{thebibliography}
\end{document}